\documentclass[12pt]{article}
\usepackage{amssymb,amsmath}
\makeatletter

\newcount\comment@nesting

\begingroup
\xdef\comment@begincomment{\string\\begin\string\{comment\string\}}
\xdef\comment@endcomment{\string\\end\string\{comment\string\}}
\endgroup

\begingroup
\catcode`\^^M=\active
\def\@temp{\endgroup\def\comment@processline##1^^M}%
\@temp{%
    \def\comment@curline{#1}%
    \let\@next=\comment@processline
    \ifx\comment@curline\comment@endcomment
        \ifnum\comment@nesting=0
            \def\@next{\end{comment}}%
        \else
            \global\advance\comment@nesting by -1
        \fi
    \else
        \ifx\comment@curline\comment@begincomment
            \global\advance\comment@nesting by 1
        \fi
    \fi
    \@next
}

\makeatother

\usepackage[dvipdfmx]{graphicx}
\usepackage{bm}
\usepackage{color}
\usepackage{here}
\usepackage{enumerate}
\usepackage{here}
\usepackage{subfigure}
\usepackage{tikz}
\usepackage{hyperref}

\newcommand{\Zb}{\mathbb{Z}}

\newcommand{\Acal}{\mathcal{A}}
\newcommand{\Ccal}{\mathcal{C}}
\newcommand{\Dcal}{\mathcal{D}}
\newcommand{\Mcal}{\mathcal{M}}
\newcommand{\Ncal}{\mathcal{N}}
\newcommand{\Wcal}{\mathcal{W}}

\DeclareMathOperator*{\Tr}{{\rm Tr}}

\newcommand{\II}{\mathbb{II}}

\numberwithin{equation}{section}

\definecolor{mygreen}{rgb}{0,0.714,0.286}

\begin{document}

\thispagestyle{empty}
\begin{flushright}
KIAS-P21059

\end{flushright}
\vskip1.5cm
\begin{center}
{\Large \bf Web of Seiberg-like dualities for\\
\vskip0.75cm
3d $\mathcal{N}=2$ quivers}

\vskip1.5cm
Tadashi Okazaki\footnote{tokazaki@kias.re.kr}

\bigskip
{\it School of Physics, Korea Institute for Advanced Study,\\
85 Hoegi-ro, Cheongnyangri-dong, Dongdaemun-gu, Seoul 02455, Republic of Korea}

\bigskip
and
\\
\bigskip
Douglas J. Smith\footnote{douglas.smith@durham.ac.uk}

\bigskip
{\it Department of Mathematical Sciences, Durham University,\\
Upper Mountjoy, Stockton Road, Durham DH1 3LE, UK}

\end{center}

\vskip1cm
\begin{abstract}
We construct Seiberg-like dualities of 3d $\mathcal{N}=2$ general quiver gauge theories with unitary, symplectic and orthogonal gauge groups coupled to fundamental and bifundamental matter fields. We illustrate this with several examples of linear, circular and star-shaped quiver gauge theories. 
We examine the local operators in the theories by computing supersymmetric indices and also find precise matching for the proposed dualities as strong evidence. 
We also generalize the dualities in the presence of a boundary on which the theories obey $\mathcal{N}=(0,2)$ chiral half-BPS boundary conditions and check the matching of half-indices. 
\end{abstract}

\newpage
\setcounter{tocdepth}{3}
\tableofcontents

\section{Introduction and conclusions}
\label{sec_Intro}
There exist various dualities in 3d $\mathcal{N}=2$ supersymmetric field theories. 
One of the dualities which is called Seiberg-like duality states that 
two or more different ultraviolet (UV) gauge theories flow to an exactly same infrared (IR) fixed point. 
Unlike Seiberg duality \cite{Seiberg:1994pq} of 4d $\mathcal{N}=1$ gauge theories, 
Seiberg-like duality of 3d $\mathcal{N}=2$ gauge theories is available even for the cases without matter field if big enough Chern-Simons coupling is turned on. 
This adds spice to the 3d dualities. 
Over the past quarter century numerous Seiberg-like dualities have been reported (see e.g.\ \cite{Aharony:1997gp, Giveon:2008zn, Niarchos:2008jb,Benini:2011mf,Kapustin:2011gh,Hwang:2011ht,Aharony:2011ci,Kim:2013cma,Aharony:2013dha,Park:2013wta,Aharony:2014uya,Nii:2014jsa,Hwang:2015wna,Benini:2017dud,Amariti:2018wht,Hwang:2018uyj,Nii:2018bgf,Nii:2019qdx,Nii:2019wjz,Nii:2020eui,Nii:2020xgd,Nii:2020ikd,Amariti:2020xqm,Amariti:2021snj,Benvenuti:2020gvy,Kubo:2021ecs}). 
Although most of the investigations of Seiberg-like duality had been centered on gauge theories with a single factor of gauge group, 
those of quiver gauge theories with product of two gauge nodes consisting of classical groups coupled to matter in the vector and rank-2 representations have been recently proposed in \cite{Benvenuti:2020wpc}
by performing Seiberg-like duality on one of the two gauge
nodes.
Similar techniques of dualizing a node within a quiver have been performed \cite{Giacomelli:2017vgk, Pasquetti:2019uop, Pasquetti:2019tix, Garozzo:2019xzi, Benvenuti:2020gvy} for 
the related monopole duality on nodes within quivers.
It has also recently been demonstrated that various 3d IR dualities can be
understood through reduction of 5d SCFTs \cite{Sacchi:2021afk, Sacchi:2021wvg}.

In addition, the story becomes more interesting in the presence of a boundary. 
The 3d $\mathcal{N}=2$ supersymmetric field theories with a Lagrangian description can preserve supersymmetry at a boundary by specifying appropriate UV boundary conditions of bulk fields \cite{Okazaki:2013kaa}.
Half-BPS boundary conditions can preserve either $\mathcal{N}=(1,1)$ which
have been described in \cite{Okazaki:2013kaa,Aprile:2016gvn}, or
$\mathcal{N}=(0,2)$ which we focus here.
In particular, $\mathcal{N}=(0,2)$ half-BPS chiral UV boundary conditions preserving the $U(1)_R$ symmetry lead to various applications \cite{Gadde:2013wq,Okazaki:2013kaa,Gadde:2013sca,Yoshida:2014ssa,Dimofte:2017tpi,Brunner:2019qyf,Costello:2020ndc,Sugiyama:2020uqh,Jockers:2021omw,Dedushenko:2021mds,Bullimore:2021rnr,Zeng:2021zef}.
For pairs of UV boundary conditions, there exist dual boundary conditions which flow to the same IR boundary conditions. Abelian dualities of $\mathcal{N}=(0,2)$ half-BPS boundary conditions in 3d $\mathcal{N}=2$ gauge theories were proposed in \cite{Okazaki:2013kaa}. Seiberg-like dualities of the $\mathcal{N}=(0,2)$ half-BPS boundary conditions in 3d $\mathcal{N}=2$ gauge theories are found in \cite{Dimofte:2017tpi} for the unitary gauge theories coupled to chiral multiplets in the vector representation and in \cite{Okazaki:2021pnc} for the orthogonal and symplectic gauge theories with chiral multiplets in the vector representation. 

In this paper we obtain Seiberg-like dualities of generic linear, circular and star-shaped quiver gauge theories with symplectic, orthogonal and unitary gauge groups coupled to chiral multiplets in the fundamental and bifundamental representations by generalizing the dualities of quivers with two gauge nodes in \cite{Benvenuti:2020wpc}. 
As strong evidence we find that supersymmetric full-indices \cite{Bhattacharya:2008zy,Bhattacharya:2008bja,Kim:2009wb,Imamura:2011su, Kapustin:2011jm,Dimofte:2011py} computed as  $S^1\times S^2$ partition functions 
precisely agree with each other for the proposed dual theories. 
Also we propose dualities of the $\mathcal{N}=(0,2)$ half-BPS boundary conditions in these quiver gauge theories. 
We support our claim by checking that half-indices \cite{Gadde:2013wq, Gadde:2013sca, Yoshida:2014ssa,Dimofte:2017tpi} computed as $S^1\times HS^2$ partition functions (also see \cite{Pasquetti:2011fj,Beem:2012mb}) which count boundary local operators precisely coincide for the proposed dualities where $HS^2$ is a hemisphere. 
By expanding the full- and half-indices we check the operator mapping across the bulk and boundary dualities. 

\subsection{Structure}
\label{sec_structure}
The structure of this paper is as follows. 
In section \ref{sec_symplectic} we generalize the dualities of quiver gauge theories with two nodes of symplectic gauge groups by introducing Chern-Simons couplings and analyzing three gauge nodes. We also propose the dualities of the $\mathcal{N}=(0,2)$ half-BPS boundary conditions in these theories. 
In section \ref{sec_orthogonal} we extend the dualities of quivers of orthogonal gauge groups to the case with more than two $SO$ gauge nodes and with other orthogonal gauge groups, $O_{\pm}$, $Spin$ and $Pin_{\pm}$. In section \ref{sec_orthosym} we further extend the dualities to the quivers consisting of both orthogonal and symplectic gauge nodes. 
In section \ref{sec_linear} we argue for Seiberg-like dualities of general linear quivers including circular quivers and linear quivers of arbitrary lengths. 
In section \ref{sec_star} we propose that a star-shaped quiver gauge theory is dual to a polygonal bipyramid quiver gauge theory. We also argue that the nodes
we are not dualizing can have arbitrary additional matter including coupling to other gauge nodes. In this way the star-shaped quiver can be embedded in an
arbitrary quiver and this shows that in such cases we can dualize any gauge node which has only fundamental and bifundamental matter.

\subsection{Open questions}
\label{sec_future}
There are several interesting open problems which we leave for future works. 
\begin{itemize}

\item It would be nice to elucidate general rules of mapping of the
(bare and dressed) BPS monopole operators under the proposed Seiberg-like dualities, 
to study the moduli spaces of supersymmetric vacua and to provide further tests of the dualities by computing other protected quantities, including a sphere partition function \cite{Kapustin:2009kz,Jafferis:2010un,Hama:2010av}, a squashed sphere partition function \cite{Hama:2011ea} and a twisted index \cite{Nekrasov:2014xaa,Benini:2015noa,Benini:2016hjo,Closset:2016arn,Bullimore:2019qnt}. 

\item It would be interesting to study in more detail quivers with Chern-Simons couplings, including the monopole operators, the relevant level-rank dualities and emergence or enhancement of supersymmetry in the bulk \cite{Gang:2018huc} and at the boundary \cite{Bae:2021lvk}. This could include examples with $\Ncal > 2$ supersymmetry \cite{Aharony:2008ug, Aharony:2008gk, Gaiotto:2008sd, Hosomichi:2008jb}. 

\item While the proposed Seiberg-like dualities are field theory phenomena, 
the brane constructions of 3d $\mathcal{N}=2$ gauge theories \cite{deBoer:1997ka, Aharony:1997ju, Giveon:2008zn} would be useful to generalize our results. 
The $\mathcal{N}=(0,2)$ boundary conditions should be studied by tilting the 5-branes of the brane configurations in \cite{Chung:2016pgt} with the cross-determinant Fermi at the NS5-NS5 junction \cite{Hanany:2018hlz, Gaiotto:2019jvo} or taking the T-dual of the brane configurations in \cite{Okazaki:2020pbb}. 

\item In the presence of the BPS Wilson and vortex line operators, the full- and half-indices can be evaluated \cite{Drukker:2012sr, Dimofte:2011py, Dimofte:2017tpi}. 
It would be interesting to compute the indices to extend the dualities by introducing the line operators. 

\item Mutations on quivers in the cluster algebras \cite{MR1887642, MR2004457} have been argued to play several roles in Seiberg-like dualities \cite{Berenstein:2002fi, Xie:2012mr, Franco:2012mm, Xie:2013lya, Benini:2014mia, Kim:2015fba, Yamazaki:2016wnu} (See also \cite{Dimofte:2013iv,Terashima:2013fg} for the relation between certain 3d $\mathcal{N}=2$ Abelian Chern-Simons matter theories and the cluster algebra). 
It would be intriguing to explore the relation between the proposed dualities and the mutations. 

\item Some identities of indices resulting from the Abelian dualities of 3d $\mathcal{N}=2$ theories are demonstrated e.g.\ in \cite{Krattenthaler:2011da, Dimofte:2017tpi},
and \cite{Hwang:2017kmk} showed that many exact results, including
for $U(N)$ Aharony dualities, could be proven using a vortex partition
function identity.
While we have checked the matching of indices as strong evidence of the dualities, analytic proof of the various identities of indices presented in this paper is desirable. 
Also the half-index which involves the 3d bulk degrees of freedom can exhibit interesting properties under modular transformations \cite{Cheng:2018vpl, Jockers:2021omw}. 

\item The Seiberg-like dualities in 3d $\mathcal{N}=2$ gauge theories are obtained from 4d dualities by compactifying 4d $\mathcal{N}=1$ gauge theories on a circle 
\cite{Aharony:2013dha, Aharony:2013kma}. It would be interesting to explore 4d uplifts of the proposed dualities. 

\item Inclusion of boundary $\mathcal{N}=(0,2)$ charged chiral multiplets and gauge multiplets should generalize the web of dualities 
of the $\mathcal{N}=(0,2)$ boundary conditions. 
In the presence of 2d bosonic fields the Neumann half-index would be computed by following the Jeffrey-Kirwan (JK) residue prescription \cite{MR1318878, Benini:2013nda, Benini:2013xpa} which picks the contour around poles associated to the bosonic fields of positive (or negative) charge as evaluated for $\mathcal{N}=(0,4)$ boundary conditions in \cite{Okazaki:2019bok}. 

\item The $\mathcal{N}=(0,2)$ half-BPS boundary conditions in 3d $\mathcal{N}=2$ field theories are compatible with the holomorphic twist. They define the boundary Vertex Operator Algebra (VOA) \cite{Costello:2020ndc} (see also \cite{Zeng:2021zef}) which conjecturally reproduces the bulk operator algebra. It would be nice to categorify the proposed dualities to equivalences of algebras or modules. 

\item The 3d $\mathcal{N}=2$ $U(N)$ gauge theory coupled to a chiral multiplet in the adjoint representation plays an important role in the 3d-3d correspondence \cite{Dimofte:2011ju, Dimofte:2011py, Chung:2014qpa, Gukov:2015sna, Gukov:2016gkn, Gukov:2017kmk}. As our proposed duality can provide us with dual description of the theory $T[L(k,1)]$ \cite{Gukov:2015sna} as a special case, it would be interesting to explore further applications and to extend the description of dualities to include quivers with rank-$2$ tensor matter, especially as such matter appears in the quivers after dualizing one gauge node. Further generalizations to include multiple adjoint chirals for a gauge node are also interesting \cite{Hwang:2018uyj}.

\item We proposed the dualities of star-shaped quiver gauge theories. The 3d $\mathcal{N}=4$ star-shaped quiver gauge theories arise from the compactifications of the 6d $(2,0)$ theory \cite{Benini:2010uu, Cremonesi:2014vla, Dimofte:2018abu} which are related to 4d $\mathcal{N}=2$ theories of class $\mathcal{S}$ \cite{Gaiotto:2010okc,Gaiotto:2009we}. 
In addition, the 3d $\mathcal{N}=2$ star-shaped quiver-type theory has been argued to appear from certain compactifications of 6d or 5d SCFTs \cite{Razamat:2019sea}. It would be interesting to further study the details of theories including moduli spaces and the boundary conditions in the star-shaped quiver gauge theories. 

\end{itemize}

\section{Symplectic linear quivers}
\label{sec_symplectic}
We describe dualities with symplectic gauge groups. We present examples of
linear quivers with two and three gauge nodes. The description in the case
of two gauge nodes without boundary was first presented in
\cite{Benvenuti:2020wpc}.

\subsection{$USp(2N_1) \times USp(2N_2)-[2N_f]$}
\label{sec_USpUSp}
We start by reviewing the triality of 3d $\mathcal{N}=2$ quiver gauge theories with two gauge nodes and one flavor node proposed in \cite{Benvenuti:2020wpc}. 
Theory A is a $USp(2N_1) \times USp(2N_2)$ gauge theory with bifundamental chirals $B$ in the $({\bf 2N_1}, {\bf 2N_2})$
representation, as well as $2N_f$ chirals $Q$ in the fundamental representation of $USp(2N_2)$. 
Theory A has three kinds of bare monopole operators $v_A^{*,0}$, $v_A^{0,*}$ and $v_A^{*,*}$ 
where the $i$th superscript stands for the GNO flux for the $i$th gauge group
and we denote by $*$ an arbitrary non-zero flux. 
The bare monopole $v_A^{*,*}$ with fluxes on both nodes can be dressed by polynomials in $\Tr(BB)$, resulting in the dressed monopole operators $v_A^{*,*}\Tr((BB)^k)$, $k=1,\cdots, N_1$. 
The Chern-Simons levels and superpotential vanish. 

Provided $\tilde{N_1} \equiv N_2 - N_1 - 1 \ge 0$ there exists a dual theory B given by Seiberg-like duality on the $USp(2N_1)$ gauge factor. 
The result is a $USp(2\tilde{N_1}) \times USp(2N_2)$ gauge theory with
bifundamental chirals $b$ in the $({\bf 2\tilde{N_1}}, {\bf 2N_2})$
representation, an antisymmetric rank-$2$ chiral $\phi_b$ of $USp(2N_2)$, as well as
$2N_f$ chirals $Q$ in the
fundamental representation of $USp(2N_2)$.
We also have a gauge singlet $\sigma_B$ which is dual to a monopole $v_A^{*,0}$ in theory A. 
While the Chern-Simons levels vanish, we have a superpotential
\begin{align}
\mathcal{W}_B&=\sigma_B v_{B}^{\pm,0}+\Tr (b\phi_b b). 
\end{align}
The first term requires the monopole operators $v_B^{*,0}$ to vanish and 
the second term sets $bb$ to zero so that the bare monopoles $v_B^{*,*}$ are not dressed by $\Tr(bb)$. 
On the other hand, the bare monopoles $v_B^{0,*}$ can be dressed by monomials in the antisymmetric chiral $\phi_b^k$, $k=1, \cdots, N_2$ \cite{Benvenuti:2018bav}, as explained in general in \cite{Cremonesi:2015dja}.
Consequently theory B contains two kinds of monopole operators $v_B^{0,*}\phi_b^k$ and $v_B^{*,*}$ in the chiral ring. 

Provided $\tilde{N_2} \equiv N_1 - N_2 + N_f - 1 \ge 0$ there can be a dual theory C given by 
Seiberg-like duality on the $USp(2N_2)$ gauge factor. The result is a
$USp(2N_1) \times USp(2\tilde{N_2})$ gauge theory with
bifundamental chirals $c$ in the $({\bf 2N_1}, {\bf 2\tilde{N_2}})$
representation, an antisymmetric rank-$2$ chiral $\phi_c$ of $USp(2N_1)$, as well as 
$2N_f$ chirals $p$ and $q$ in the
fundamental representations of $USp(2N_1)$ and $USp(2\tilde{N_2})$.
Again we have a gauge singlet $\sigma_C$ which is dual to a monopole $v_A^{0,*}$ in 
theory A, but now also additional singlets $M$ in the rank-2 antisymmetric
representation of the flavor symmetry group $SU(2N_f)$. 
Whereas the Chern-Simons levels is turned off, theory C has a superpotential 
\begin{align}
\mathcal{W}_C&=\sigma_C v_C^{0,\pm}+\Tr(c\phi_c c)+\Tr(cpq)+\Tr(qMq). 
\end{align}
The first term sets the monopoles $v_C^{0,*}$ to zero  
and the second term gets rid of $cc$. 
As a consequence, there are two types of monopole operators $v_C^{*,0}\phi_c^k$ and $v_C^{*,*}$ 
in the chiral ring. 

\begin{figure}
\centering
\scalebox{0.7}{
\begin{tikzpicture}
\path (0,0) node[circle, minimum size=64, fill=yellow!20, draw](AG1) {$USp(2N_1)$}
(4,0) node[circle, minimum size=64, fill=yellow!20, draw](AG2) {$USp(2N_2)$}
(4,4) node[minimum size=50, fill=cyan!10, draw](AF2) {$2N_f$};
\draw (AG1) -- (AG2) node [midway, above] {$B$};
\draw (AG2) -- (AF2) node [midway, right] {$Q$};
\path (-6,-8) node[circle, minimum size=64, fill=yellow!20, draw](BG1) {$USp(2\tilde{N_1})$}
(-2,-8) node[circle, minimum size=64, fill=yellow!20, draw](BG2) {$USp(2N_2)$}
(-2,-4) node[minimum size=50, fill=cyan!10, draw](BF2) {$2N_f$};
\draw (BG1) -- node[above]{$b$} (BG2);
\draw (BG2) -- node[right]{$Q$}(BF2);
\draw (BG2.south west) .. controls ++(-1, -2) and ++(1, -2) .. node[below]{$\phi_b$} (BG2.south east);
%
\path (6,-8) node[circle, minimum size=64, fill=yellow!20, draw](CG1) {$USp(2N_1)$}
(10,-8) node[circle, minimum size=64, fill=yellow!20, draw](CG2) {$USp(2\tilde{N_2})$}
(10,-4) node[minimum size=50, fill=cyan!10, draw](CF2) {$2N_f$};
\draw (CG1) -- node[above]{$c$} (CG2);
\draw (CG1) -- node[right]{$p$}(CF2);
\draw (CG2) -- node[right]{$q$}(CF2);
\draw (CG1.south west) .. controls ++(-1, -2) and ++(1, -2) .. node[below]{$\phi_c$} (CG1.south east);
\draw (CF2.north east) .. controls ++(1, 2) and ++(-1, 2) .. node[below]{$M$} (CF2.north west);
\end{tikzpicture}
}
\caption{Triality of $USp \times USp$ quivers where $\tilde{N_1}=N_2-N_1-1$ and $\tilde{N_2}=N_1+N_f-N_2-1$.
Note that $\phi_b$, $\phi_c$ and $M$ are in antisymmetric rank-$2$
representations.} \label{fig:USp_USp}
\end{figure}
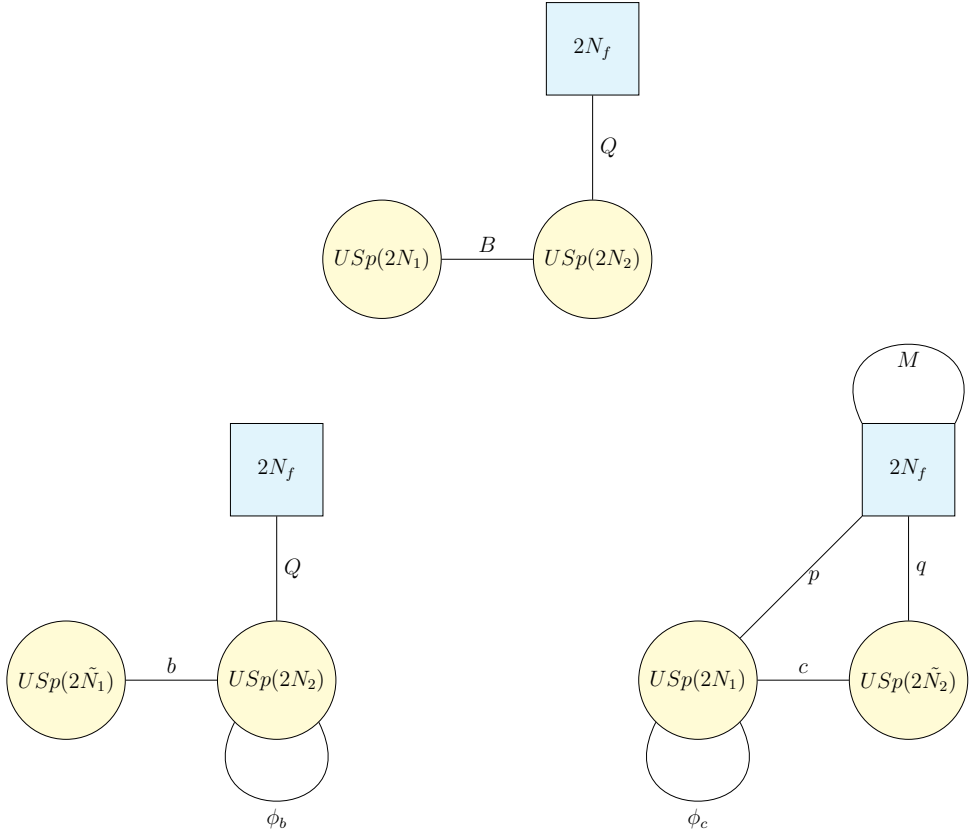

The triality of the $USp\times USp$ quivers is shown in Figure \ref{fig:USp_USp}. 

The continuous global symmetry groups of the theories are given by a $SU(2N_f)$ flavor symmetry, two axial symmetries $U(1)_{a_1} \times U(1)_{a_2}$, and the $U(1)_R$ R-symmetry. 
The following table lists the charges and representations of the fields in each theory.
\begin{align}
\label{USp_charges}
\begin{array}{c|c|c|c|c}
& SU(2N_f) & U(1)_{a_1} & U(1)_{a_2} & U(1)_R \\ \hline
B & {\bf 1} & 1 & 0 & r_B \\
Q & {\bf \overline{2N}_f} & 0 & 1 & r_Q \\ \hline
b & {\bf 1} & -1 & 0 & 1 - r_B \\
\phi_b & {\bf 1} & 2 & 0 & 2r_B \\
Q & {\bf \overline{2N}_f} & 0 & 1 & r_Q \\
\sigma_B & {\bf 1} & -2N_2 & 0 & 2N_2(1 - r_B) - 2N_1 \\ \hline
c & {\bf 1} & -1 & 0 & 1 - r_B \\
\phi_c & {\bf 1} & 2 & 0 & 2r_B \\
p & {\bf \overline{2N}_f} & 1 & 1 & r_B + r_Q \\
M & {\bf N_f(2N_f - 1)} & 0 & 2 & 2r_Q \\
q & {\bf \overline{2N}_f} & 0 & -1 & 1 - r_Q \\ 
\sigma_C & {\bf 1} & -2N_1 & -2N_f & 2N_1(1 - r_B) + 2N_f(1 - r_Q) - 2N_2
\end{array}
\end{align}
The operator map across the triality is summarized as
\begin{align}
\label{USpUSp_map1}
\begin{array}{c|c|c}
A&B&C\\ \hline 
\Tr(QQ)&\Tr(QQ)&M \\
\Tr(Q(BB)^k Q)&\Tr(Q\phi_b^k Q)&\Tr(p\phi_c^{k-1}p)\\\
\Tr((BB)^k)&\Tr(\phi_b^k)&\Tr(\phi_c^k)\\
v_A^{\pm,0}&\sigma_B&v_C^{\pm,\pm} \\
v_A^{0,\pm}&v_B^{\pm,\pm}&\sigma_C\\
v_A^{\pm,\pm} & v_B^{0,\pm} & v_C^{\pm,0} \\
\end{array}
\end{align}

The monopole operators above are gauge invariant so can be dressed by gauge
invariant combinations of the chiral multiplets, with the mapping between theories following the pattern indicated in the first 3 lines of the table.
Note that there can be restrictions on the allowed dressed monopoles
as noted
in this context in \cite{Benvenuti:2020wpc}. These follow from the condition
that the fields dressing the monopole must be `massless' \cite{Cremonesi:2015dja},
as well as requiring gauge invariance and that all F-term constraints are
satisfied. This massless condition is that $\rho(\vec{m}) = 0$ where
$\rho$ is the weight of the field under the gauge group and $\vec{m}$
represents all fluxes of the monopole. Since the monopole flux breaks the gauge
group, each field must be decomposed into irreducible representations of the
broken gauge group and then we can determine which (if any) of these components remain massless in the given monopole background.
We comment on this in some of the examples we give later.

\subsection{Supersymmetric indices for symplectic gauge theories}
\label{sec_USpUSpindices}
The supersymmetric full-indices \cite{Bhattacharya:2008zy,Bhattacharya:2008bja,Kim:2009wb,Imamura:2011su, Kapustin:2011jm, Dimofte:2011py, Bashkirov:2011vy} for theories A, B and C with symplectic gauge groups are
\begin{align}
I^A = & Z^{USp_1}_{gauge} Z^{USp_2}_{gauge} Z_{matter \; A} \\
I^B = & Z^{\widetilde{USp}_1}_{gauge} Z^{USp_2}_{gauge} Z_{matter \; B} \\
I^C = & Z^{USp_1}_{gauge} Z^{\widetilde{USp}_2}_{gauge} Z_{matter \; C}
\end{align}
where $Z_{gauge}$ and $Z_{matter *}$ are the contributions from the vector and
chiral multiplets with $USp_I$ referring to gauge node $USp(N_I)$ and
$\widetilde{USp}_I$ referring to gauge node $USp(\tilde{N}_I)$.
Above we have included Chern-Simons levels $k_1$ for $USp(2N_*)$ but for now
we take these levels to all vanish.

The gauge contributions are given by
\begin{align}
Z^{USp_I}_{gauge} = &
 \sum_{m_i^{(I)} \in \Zb} \frac{1}{2^{N_I} N_I!}
\oint \left( \prod_{i=1}^{N_I} \frac{ds^{(I)}_i}{2\pi i s_i^{(I)}} (-s_i^{(I)})^{k_I m_i^{(I)}} \right) \nonumber \\
 & \times q^{-\sum_{i=1}^{N_I} |m_i^{(I)}| - \sum_{i < j}^{N_I} |m_i^{(I)} \pm m_j^{(I)}|/2}
 \left( \prod_{i=1}^{N_I} (1 - q^{|m_i^{(I)}|}s_i^{(I)\pm 2} \right) \nonumber \\
 & \times \prod_{i < j}^{N_I}
 ( 1 - q^{|m_i^{(I)} \pm m_j^{(I)}|/2} s_{j}^{(I)} s_j^{(I)\pm}) ( 1 - q^{|-m_i^{(I)} \pm m_j^{(I)}|/2} s_i^{(I)-1} s_j^{(I)\pm}) \; .
 \end{align}
 
The contributions from the chiral multiplet are 
\begin{align}
Z_{matter \; A} = & Z_B Z_Q^{N_f} \\
Z_{matter \; B} = & Z_b Z_{\phi_b} Z_Q^{N_f} Z_{\sigma_B} \\
Z_{matter \; C} = & Z_c Z_{\phi_c} Z_p^{N_f} Z_M^{N_f} Z_q^{N_f} Z_{\sigma_C} \; .
\end{align}
These can be calculated using the general expressions
 \begingroup
\allowdisplaybreaks
\begin{align}
&Z_{BiFund \; USp_I-USp_J}(r, a) =  (q^{\frac{1-r}{2}} a^{-1})^{\sum_{i=1}^{N_I} \sum_{j=1}^{N_J} |m_i^{(I)} \pm m_j^{(J)}|} \nonumber \\
 & \times \prod_{i = 1}^{N_I} \prod_{j=1}^{N_J}
 \frac{(q^{1-\frac{r}{2}+\frac{|m_i^{(I)}-m_j^{(J)}|}{2}} a^{-1} s_i^{(I)\mp} s_j^{(J)\pm}; q)_{\infty} 
 (q^{1-\frac{r}{2}+\frac{|m_i^{(I)}+m_j^{(J)}|}{2}} a^{-1} s_i^{(I)\pm} s_j^{(J)\pm}; q)_{\infty}}
 {(q^{\frac{r}{2}+\frac{|m_i^{(I)} - m_j^{(J)}|}{2} } a s_i^{(I)\pm} s_j^{(J)\mp}; q)_{\infty} 
(q^{\frac{r}{2}+\frac{|m_i^{(I)} + m_j^{(J)}|}{2} } a s_i^{(I)\pm} s_j^{(J)\pm}; q)_{\infty}} \\
&Z_{N_F Fund \; USp_I}(r, a) =  (q^{\frac{1-r}{2}} a^{-1})^{N_F \sum_{i=1}^{N_I} |m_i^{(I)}|} \prod_{\alpha = 1}^{N_F} \prod_{i=1}^{N_I}
 \frac{(q^{1-\frac{r}{2}+ \frac{|{m_i^{(I)}}|}{2} } a^{-1} {s_j^{(I)}}^{\mp} x_{\alpha}^{-1}; q)_{\infty}}
 {(q^{\frac{r}{2}+\frac{|{m_i^{(I)}}|}{2}} a {s_i^{(I)}}^{\pm} x_{\alpha}; q)_{\infty}} \\
&Z_{Antisym \; USp_I}(r, a) =  (q^{\frac{1 - 2r}{2}} a^{-1})^{\sum_{i < j}^{N_I} |{m_i^{(I)}} \pm {m_j^{(I)}}|} \left( \prod_{i, j = 1}^{N_I}
 \frac{(q^{1-r +\frac{|{m_i^{(I)}} - {m_j^{(I)}}|}{2}} a^{-1} {s_i^{(I)}}^{-1} {s_j^{(I)}}; q)_{\infty}}
 {(q^{r+ \frac{|{m_i^{(I)}} - {m_j^{(I)}}|}{2}} a {s_i^{(I)}} {s_j^{(I)}}^{-1}; q)_{\infty}} \right) \nonumber \\
 & \times \left( \prod_{i < j}^{N_I} 
 \frac{(q^{1-r + \frac{|{m_i^{(I)}} + {m_j^{(I)}}|}{2}} a^{-1} {s_i^{(I)}}^{\mp} {s_j^{(I)}}^{\mp}; q)_{\infty}}
 {(q^{r+\frac{|{m_i^{(I)}} + {m_j^{(I)}}|}{2}} a {s_i^{(I)}}^{\pm} {s_j^{(I)}}^{\pm}; q)_{\infty}} \right) \\
&Z_{Antisym \; SU(N_F)}(r, a)  = \prod_{\alpha < \beta}^{N_F} \frac{(q^{1 - r} a^{-1} x_{\alpha}^{-1} x_{\beta}^{-1}; q)_{\infty}}{(q^{r} a x_{\alpha} x_{\beta}; q)_{\infty}} \\
&Z_{Singlet}(r, a) = 
 \frac{(q^{1 - \frac{r}{2}} a^{-1}; q)_{\infty}}
 {(q^{\frac{r}{2}} a; q)_{\infty}} \; .
\end{align}
\endgroup
In particular, we have
\begingroup
\begin{align}
Z_B = & Z_{BiFund \; USp_I-USp_J}(r_B, a_1) \\
Z^{N_f}_Q = & Z_{2N_f Fund \; I}(r_Q, a_2) \\
Z_b = & Z_{BiFund \; USp_I-USp_J}(1 - r_B, a_1^{-1}) \\
Z_{\phi_b} = & Z_{Antisym \; USp_I}(2r_B, a_1^2) \\
Z_{\sigma_B} = & Z_{Singlet}(r_{\sigma_B}, a_1^{-2N_2}) \\
Z_c = & Z_{BiFund \; USp_I-USp_J}(1 - r_B, a_1^{-1}) \\
Z_{\phi_c} = & Z_{Antisym \; USp_I}(2r_B, a_1^2) \\
Z^{N_f}_p = & Z_{BiFund \; USp_I-USp_J}(r_B + r_Q, a_1 a_2) \\
Z^{N_f}_M = & Z_{Antisym \; SU(N_F)}(2r_Q, a_2^2) \\
Z^{N_f}_q = & Z_{BiFund \; USp_I-USp_J}(1 - r_Q, a_2^{-1}) \\
Z_{\sigma_C} = & Z_{Singlet}(r_{\sigma_C}, a_1^{-2N_1} a_2^{-2N_f})
\end{align}
\endgroup
where $r_{\sigma_B} \equiv 2(1-r_B)N_2 - 2N_1$ and
$r_{\sigma_C} \equiv 2(1-r_B)N_1 + 2(1-r_Q)N_f - 2N_2$.

\subsection{Boundary 't Hooft anomalies}
When we include $\mathcal{N}=(0,2)$ half-BPS boundary conditions, we need to calculate the anomaly polynomial
for the 2d boundary theory. We must ensure that the gauge anomalies cancel and
for proposed dualities we must have matching boundary 't Hooft anomaly polynomials. 
In the case of Dirichlet boundary conditions for gauge fields we also need the
anomaly polynomial to calculate the effective Chern-Simons coupling which enters the half-index \cite{Dimofte:2017tpi}. 

For the multiplets we have discussed the contributions to the anomaly polynomial
are given by the following expressions \cite{Dimofte:2017tpi, Okazaki:2021pnc}
if we have Dirichlet boundary conditions.
For Neumann boundary conditions we just take the opposite sign.
Taking the multiplets to have R-charge $q_R$ and a vector of $U(1)_{a_i}$
charges $\underline{q}$ we have
\begin{align}
\Acal_{VM \; USp_I} = & -(N_I+1) \Tr({\bf s_I}^2) - \frac{N_I(2N_I + 1)}{2} {\bf r}^2 \\
\Acal_{2N_f \; Fund \; USp_I}(q_R, \underline{q}) = & N_I \Tr({\bf x}^2) + N_f \Tr({\bf s_I}^2) + 2N_I N_f \left( \underline{q} \cdot \underline{{\bf a}} + (q_R - 1){\bf r} \right)^2 \\
\Acal_{BiFund \; USp_I-USp_J}(q_R, \underline{q}) = & N_I \Tr({\bf s_J}^2) + N_J \Tr({\bf s_I}^2) + 2N_I N_J \left( \underline{q} \cdot \underline{{\bf a}} + (q_R - 1){\bf r} \right)^2 \\
\Acal_{Antisym \; USp_I}(q_R, \underline{q}) = & (N_I - 1) \Tr({\bf s_I}^2) + \frac{N_I(2N_I - 1)}{2} \left( \underline{q} \cdot \underline{{\bf a}} + (q_R - 1){\bf r} \right)^2 \\
\Acal_{Antisym \; SU(N_f)}(q_R, \underline{q}) = & \frac{N_f - 2}{2} \Tr({\bf x}^2) + \frac{N_f(N_f - 1)}{4} \left( \underline{q} \cdot \underline{{\bf a}} + (q_R - 1){\bf r} \right)^2 \\
\Acal_{Singlet}(q_R, \underline{q}) = & \frac{1}{2} \left( \underline{q} \cdot \underline{{\bf a}} + (q_R - 1){\bf r} \right)^2 \; .
\end{align}
In this notation ${\bf s}_I$ is the $USp(2N_I)$ field strength,
$\underline{\bf a}$ is a vector of $U(1)_{a_i}$ field strengths and
${\bf r}$ is the $U(1)_R$ field strength. We use ${\bf x}$ for the $SU(2N_f)$
flavor symmetry field strength and the expressions for fundamental matter
and the antisymmetric multiplet $M$ are given by the above expressions for
bifundamental and antisymmetric multiplets by replacing an appropriate
${\bf s}_I$ with ${\bf x}$. Also, we use the notation $\tilde{I}$ in the above
expressions to refer to the dual $USp(2\tilde{N}_I)$ gauge group with
field strength $\tilde{\bf s}_I$.

We propose a triality of the following sets of boundary conditions:
\begin{itemize}
\item $(\Ncal, \Ncal, N, N)$ for
$(\mathrm{VM}_1, \mathrm{VM}_2, B, Q)$ in theory A.
\item $(\Dcal, \Ncal, D, N, N, D)$ for 
$(\tilde{\textrm{VM}_1}, \mathrm{VM}_2, b, \phi_b, Q, \sigma_B)$ in theory B.
\item $(\Ncal, \Dcal, D, N, N, N, D, D)$ for 
$(\mathrm{VM}_1, \tilde{\textrm{VM}_2}, c, \phi_c, p, M, q, \sigma_C)$ in theory C.
\end{itemize}
We will see that with suitable additional 2d boundary multiplets, we can
cancel the gauge anomalies for the gauge group factors with Neumann boundary
conditions, and match the anomalies. We can then check the matching of the
half-indices.

Explicitly, we find the following anomaly polynomials for the bulk fields.
\begin{align}
\Acal^{A \; Bulk}_{\Ncal, \Ncal, N, N} = &
  -2N_1 N_2 {\bf a}_1^2 - 2N_2 N_f {\bf a}_2^2 + 4N_1 N_2 (1 - r_B){\bf a}_1 {\bf r
} + 4N_2 N_f (1 - r_Q){\bf a}_2 {\bf r} \nonumber \\
 & + \left( (N_1 - N_2)^2 + 2N_1 N_2 r_B(2 - r_B) - 2N_2 N_f (1 - r_Q)^2 + \frac{1}{2}(N_1 + N_2) \right) {\bf r}^2 \nonumber \\ 
 & + (N_1 - N_2 + 1) \Tr({\bf s}_1^2) + (1 - N_1 + N_2 - N_f)\Tr({\bf s}_2^2) - N_2 \Tr({\bf x}^2) \\
\Acal^{B \; Bulk}_{\Dcal, \Ncal, D, N, N, D} = &
  -2N_1 N_2 {\bf a}_1^2 - 2N_2 N_f {\bf a}_2^2 + 4N_1 N_2 (1 - r_B){\bf a}_1 {\bf r
} + 4N_2 N_f (1 - r_Q){\bf a}_2 {\bf r} \nonumber \\
 & + \left( (N_1 - N_2)^2 + 2N_1 N_2 r_B(2 - r_B) - 2N_2 N_f (1 - r_Q)^2 + \frac{1}{2}(N_1 + N_2) \right) {\bf r}^2 \nonumber \\
 & - (N_1 + N_f - N_2 - 1) \Tr({\bf s}_2^2) + N_1 \Tr(\tilde{\bf s}_1^2) - N_2 \Tr({\bf x}^2) \\
\Acal^{C \; Bulk}_{\Ncal, \Dcal, D, N, N, N, D, D} = & 
  -2N_1 N_2 {\bf a}_1^2 - 2N_2 N_f {\bf a}_2^2 + 4N_1 N_2 (1 - r_B){\bf a}_1 {\bf r
} + 4N_2 N_f (1 - r_Q){\bf a}_2 {\bf r} \nonumber \\
 & + \left( (N_1 - N_2)^2 + 2N_1 N_2 r_B(2 - r_B) - 2N_2 N_f (1 - r_Q)^2 + \frac{1}{2}(N_1 + N_2) \right) {\bf r}^2 \nonumber \\
 & - (N_2 - N_1 - 1) \Tr({\bf s}_1^2) + N_2 \Tr(\tilde{\bf s}_2^2) - N_2 \Tr({\bf x}^2)
\end{align}

Recalling that $\tilde{N_1} = N_2 - N_1 - 1$ and $\tilde{N_2} = N_1 - N_2 + N_f - 1$, and that due to
the $\Dcal$ boundary conditions $USp(2\tilde{N_1})$ and $USp(2\tilde{N_2})$ are global symmetries on
the boundary, whereas $USp(2N_1)$ and $USp(2N_2)$ are gauge symmetries due to the
$\Ncal$ boundary conditions, we can cancel all gauge anomalies with the
following 2d multiplets.
\begin{align}
\label{USp_charges_2d}
\begin{array}{c|c|c|c|c|c}
 & & USp(2N_1) & USp(2N_2) & USp(2\tilde{N_1}) & USp(2\tilde{N_2}) \\ \hline
\mathrm{Fermi} & \Gamma_{1} & {\bf 2N_1} & {\bf 1} & {\bf 2\tilde{N_1}} & {\bf 1} \\
\mathrm{Fermi} & \Gamma_{2} & {\bf 1} & {\bf 2N_2} & {\bf 1} & {\bf 2\tilde{N_2}} \\
\end{array}
\end{align}
In particular, we need to include
\begin{itemize}
\item $\Gamma_{1}$, $\Gamma_{2}$ in theory A.
\item $\Gamma_{2}$ in theory B.
\item $\Gamma_{1}$ in theory C.
\end{itemize}

Including the contribution of those 2d multiplets in each theory all gauge anomalies are cancelled and the resulting anomaly polynomials match
\begin{align}
\Acal^{Total} = &
 -2N_1 N_2 {\bf a}_1^2 - 2N_2 N_f {\bf a}_2^2 + 4N_1 N_2 (1 - r_B){\bf a}_1 {\bf r
} + 4N_2 N_f (1 - r_Q){\bf a}_2 {\bf r} \nonumber \\
 & + \left( (N_1 - N_2)^2 + 2N_1 N_2 r_B(2 - r_B) - 2N_2 N_f (1 - r_Q)^2 + \frac{1}{2}(N_1 + N_2) \right) {\bf r}^2 \nonumber \\
 & + N_1 \Tr(\tilde{\bf s}_1^2) + N_2 \Tr(\tilde{\bf s}_2^2) - N_2 \Tr({\bf x}^2)
\end{align}

It is possible to consider other boundary conditions. The obvious case is to switch all boundary conditions so we have all Dirichlet in theory A. This will
simply change the sign of the bulk contribution to the anomaly polynomial and
it is easy to see that all gauge anomalies will be cancelled and the anomalies
will match if we include 2d Fermi multiplets
\begin{itemize}
\item None in theory A.
\item $\Gamma_{1}$ in theory B.
\item $\Gamma_{2}$ in theory C.
\end{itemize}

We will focus on examples with all Neumann boundary conditions for theory A, but we believe the same dualities will hold with all boundary conditions reversed along with the above 2d Fermi multiplets. Such dualities have been checked for 
the case of a single gauge node for unitary gauge groups \cite{Dimofte:2017tpi}, and for
symplectic and orthogonal gauge groups \cite{Okazaki:2021pnc}.

\subsection{Half-indices}
We now give the expressions for the half-indices with the previously chosen boundary conditions. For the three theories we have
\begin{align}
\II^A_{\Ncal, \Ncal, N, N} = & \II^{VM \; USp_1}_{\Ncal} \II^{VM \; USp_2}_{\Ncal} \II^B_N \II^Q_N
 \\
\II^B_{\Dcal, \Ncal, D, N, N, D} = & \II^{VM \; USp_{\tilde{1}}}_{\Dcal} \II^{VM \; USp_2}_{\Ncal} \II^b_D \II^{\phi_b}_N \II^Q_N \II^{\sigma_B}_D
 \\
\II^C_{\Ncal, \Dcal, D, N, N, N, D, D} = & \II^{VM \; USp_1}_{\Ncal} \II^{VM \; USp_{\tilde{2}}}_{\Dcal} \II^c_D \II^{\phi_c}_N \II^p_N \II^M_N \II^q_D \II^{\sigma_C}_D
\end{align}
where
\begin{align}
\II^{VM \; USp_I}_{\Ncal} = & \frac{(q)_{\infty}^{N_I}}{2^{N_I} N_I!} \left( \prod_{i=1}^{N_I} \oint \frac{ds_i^{(I)}}{2\pi i s_i^{(I)}} \right)
 \left( \prod_{i \ne j}^{N_I} (s_i^{(I)} s_j^{(I)-1}; q)_{\infty} \right)
 \left( \prod_{i \le j} (s_i^{(I)\pm} s_j^{(I)\pm}; q)_{\infty} \right) \\
\II^{VM \; USp_I}_{\Dcal} = & \frac{1}{(q)_{\infty}^{N_I}} \sum_{m_i^{(I)} \in \Zb^{N_I}} 
\frac{q^{\frac{k_{eff \; I}}{2} \sum_{i=1}^{N_I} {m_i^{(I)}}^2} (\prod_i^{N_I} {u_i^{(I)}}^{k_{eff \; I} m_i^{(I)}})}
{\prod_{i \ne j}^{\tilde{N_1}} (q^{1 + m_i^{(I)} - m_j^{(I)}} u_i^{(I)} {u_j^{(I)}}^{-1}; q)_{\infty} \prod_{i \le j} (q^{1 \pm (m_i^{(I)} + m_j^{(I)})} u_i^{(I) \pm} u_j^{(I) \pm}; q)_{\infty}} \\
k_{eff \; I} = & 2\tilde{N}_I
\end{align}
recalling that $N_{\tilde{I}} = \tilde{N}_I$ so $\tilde{N}_{\tilde{I}} = N_I$.
We also have the prescription that for Dirichlet boundary conditions for the
vector multiplet we make the replacement
$s_i^{(I)} \rightarrow q^{m_i^{(I)}} u_i^{(I)}$ in the following matter
contributions.

The 3d matter contributions are given, for either Dirichlet or Neumann boundary conditions by
\begingroup
\begin{align}
\II^B = & \II^{BiFund \; USp_I-USp_J}(r_B, a_1) \\
\II^Q = & \II^{2N_f Fund \; USp_I}(r_Q, a_2) \\
\II^b = & \II^{BiFund \; USp_I-USp_J}(1 - r_B, a_1^{-1}) \\
\II^{\phi_b} = & \II^{Antisym \; USp_I}(2r_B, a_1^2) \\
\II^{\sigma_B} = & \II^{Singlet}(r_{\sigma_B}, a_1^{-2N_2}) \\
\II^c = & \II^{BiFund \; USp_I-USp_J}(1 - r_B, a_1^{-1}) \\
\II^{\phi_c} = & \II^{Antisym \; USp_I}(2r_B, a_1^2) \\
\II^p = & \II^{BiFund \; USp_I-USp_J}(r_B + r_Q, a_1 a_2) \\
\II^M = & \II^{Antisym \; SU(N_F)}(2r_Q, a_2^2) \\
\II^q = & \II^{BiFund \; USp_I-USp_J}(1 - r_Q, a_2^{-1}) \\
\II^{\sigma_C} = & \II^{Singlet}(r_{\sigma_C}, a_1^{-2N_1} a_2^{-2N_f})
\end{align}
\endgroup

For Neumann boundary conditions the 3d matter contributions are
\begingroup
\allowdisplaybreaks
\begin{align}
\II^{BiFund \; USp_I-USp_J}_N(r,a) = & \prod_{i=1}^{N_I} \prod_{j=1}^{N_J} 
\frac{1}{(q^{\frac{r}{2}} a s_i^{(I)\pm} {s_j^{(J)}}^{\mp} ; q)_{\infty} 
(q^{\frac{r}{2}} a s_i^{(I)\pm} {s_j^{(J)}}^{\pm} ; q)_{\infty}} \\
\II^{N_F Fund \; USp_I}_{N}(r,a) = & \prod_{i=1}^{N_I} \prod_{\alpha=1}^{N_F}
\frac{1}{(q^{\frac{r}{2}} a {s_i^{(I)}}^{\pm} x_{\alpha} ; q)_{\infty}} \\
\II^{Antisym \; USp_I}_{N}(r,a) = & \left( \prod_{i,j=1}^{N_I} \frac{1}{(q^{\frac{r}{2}} a {s_i^{(I)}} {s_j^{(I)}}^{-1} ; q)_{\infty}} \right) 
 \left( \prod_{j < l}^{N_2} \frac{1}{(q^{\frac{r}{2}} a {s_i^{(I)}}^{\pm} {s_j^{(I)}}^{\pm} ; q)_{\infty}} \right) \\
\II^{Singlet}_{N}(r,a) = & \frac{1}{(q^{\frac{r}{2}} a; q)_{\infty}} \\
\II^{Antisym \; SU(N_F)}_{N}(r,a) = & \prod_{\alpha < \beta}^{N_F} 
\frac{1}{(q^{\frac{r}{2}} a x_{\alpha} x_{\beta} ; q)_{\infty}}
\end{align}
\endgroup
while for Dirichlet boundary conditions the 3d matter contributions are
\begingroup
\allowdisplaybreaks
\begin{align}
\II^{BiFund \; USp_I-USp_J}_D(r,a) = & \prod_{i=1}^{N_I} \prod_{j=1}^{N_J} 
(q^{ 1 - \frac{r}{2}} a^{-1} s_i^{(I)\pm} {s_j^{(J)}}^{\mp} ; q)_{\infty} 
(q^{1 - \frac{r}{2}} a^{-1} s_i^{(I)\pm} {s_j^{(J)}}^{\pm} ; q)_{\infty} \\
\II^{N_F Fund \; USp_I}_{D}(r,a) = & \prod_{i=1}^{N_I} \prod_{\alpha=1}^{N_F}
(q^{1 - \frac{r}{2}} a^{-1} {s_i^{(I)}}^{\pm} x_{\alpha}^{-1} ; q)_{\infty} \\
\II^{Antisym \; USp_I}_{D}(r,a) = & \left( \prod_{i,j=1}^{N_I} 
(q^{1 - \frac{r}{2}} a^{-1} {s_i^{(I)}} {s_j^{(I)}}^{-1} ; q)_{\infty} \right) 
\left( \prod_{i < j}^{N_I} (q^{1 - \frac{r}{2}} a^{-1} {s_i^{(I)}}^{\pm} {s_j^{(I)}}^{\pm} ; q)_{\infty} \right) \\
\II^{Singlet}_{D}(r,a) = & (q^{1 - \frac{r}{2}} a^{-1}; q)_{\infty} \\
\II^{Antisym \; SU(N_F)}_{D}(r,a) = & \prod_{\alpha < \beta}^{N_F}
(q^{1 - \frac{r}{2}} a^{-1} x_{\alpha} x_{\beta} ; q)_{\infty} \\
\end{align}
\endgroup

Finally, the contributions from the 2d matter multiplets are
\begin{align}
I^{\Gamma_{I}} = & \prod_{i=1}^{N_I} \prod_{j=1}^{\tilde{N}_I} 
(q^{\frac12} s_i^{(I) \pm} u_j^{(I) \pm}; q)_{\infty} (q^{\frac12} s_i^{(I) \pm} u_j^{(I) \mp}; q)_{\infty}
\end{align}

\subsection{$USp(2)\times USp(6)-[8]$ $(N_1 = 1$, $N_2 = 3$, $N_f = 4)$}
For simplicity we set the global $SU(2N_f)$ flavor fugacities $x_{\alpha} = 1$. 
Here we show the expansion of the indices for $\tilde{N_1} = 1$ and $\tilde{N_2} = 1$. 

For $r_B=2/5$, $r_Q=2/7$, the full-indices which perfectly match for theory A, B and C
are~\footnote{Here we fix the R-charges as in \cite{Benvenuti:2020wpc}. Note that in order to check the dualities, it is sufficient to pick any choice of the R-charges which gives a well defined (convergent) index for all theories.}
\begin{align}
&I^A = I^B = I^C 
\nonumber\\
=&1+\underbrace{28 a_2^2 q^{2/7}}_{\Tr(QQ)}+\underbrace{a_1^2 q^{2/5}}_{\Tr(BB)}
+\underbrace{\frac{q^{16/35}}{a_1^2 a_2^8}}_{v_A^{0,\pm}}+
\underbrace{406 a_2^4 q^{4/7}}_{\Tr(QQ)^2}
+\underbrace{\frac{q^{23/35}}{a_1^6 a_2^8}}_{v_A^{\pm,\pm}}+
\underbrace{56a_1^2 a_2^2 q^{24/35}}_{\begin{smallmatrix} \Tr(Q(BB)Q), \\ \Tr(QQ)\Tr(BB) \end{smallmatrix}}
+\underbrace{\frac{28 q^{26/35}}{a_1^2 a_2^6}}_{v_A^{0,\pm} \Tr(QQ)}
   \nonumber\\
   &
+
(\underbrace{a_1^4}_{\Tr (BB)^2}+\underbrace{\frac{1}{a_1^6}}_{v_A^{\pm,0}}) q^{4/5}
   +(
  \underbrace{4060
   a_2^6}_{\Tr (QQ)^3}+\underbrace{\frac{1}{a_2^8}}_{v^{0,\pm} \Tr(BB)}) q^{6/7}
   +\underbrace{\frac{q^{32/35}}{a_1^4 a_2^{16}}}_{v_A^{0,\pm 2}}
   +\underbrace{\frac{28 q^{33/35}}{a_1^6 a_2^6}}_{v_A^{\pm,\pm} \Tr(QQ)}
   +\underbrace{1190 a_1^2a_2^4 q^{34/35}}_{\begin{smallmatrix} \Tr(QQ)^2\Tr(BB), \\ \Tr(QQ)\Tr(Q(BB)Q) \end{smallmatrix}} -\underbrace{65q}_{\begin{smallmatrix} \Tr(Q \psi_Q), \\ \Tr(B \psi_B) \end{smallmatrix}}
      \nonumber\\
   &
   +\underbrace{\frac{406 q^{36/35}}{a_1^2 a_2^4}}_{v_A^{0,\pm} \Tr(QQ)^2}+\underbrace{\frac{q^{37/35}}{a_1^4 a_2^8}}_{v_A^{\pm,\pm}\Tr(BB)}
   + 28 (\underbrace{2 a_1^{4}}_{\Tr(Q(BB)^2Q)}+\underbrace{a_1^{-6}}_{v_A^{\pm,0}\Tr(QQ)} )a_2^2 q^{38/35} 
   +\underbrace{\frac{q^{39/35}}{a_1^8 a_2^{16}}}_{v_A^{\pm,\pm2}}
   \nonumber\\
   &+8 (\underbrace{3933 a_2^{8}}_{\Tr (QQ)^4}+
   \underbrace{7a_2^{-6}}_{\begin{smallmatrix} v_A^{0,\pm}\Tr(Q(BB)Q), \\ v_A^{0,\pm}\Tr(QQ)\Tr(BB) \end{smallmatrix}}) q^{8/7}
 + \Big( \underbrace{a_1^6}_{\Tr(BB)^3} + \underbrace{\frac{28}{a_1^4 a_2^{14}}}_{v_A^{0, \pm 2} \Tr(QQ)} \Big)q^{6/5}
 +\cdots
\end{align}
In theory A the term $28a_2^2 q^{2/7}$ counts the meson $\Tr (Q^iQ^j)$ whose coefficient $\frac{8\cdot 7}{2}=28$ implies the antisymmetric representation under the $SU(8)$ flavor symmetry, as expected due to the contraction of the $USp(6)$ indices.
Then the term $406a_2^4 q^{4/7}$ counts the operator $\Tr (Q^i Q^j) \Tr(Q^k Q^l)$ which gives 28 for each trace and the product of traces is symmetric so the
 number is $28 \cdot 29 / 2=406$. Similarly for
$\Tr (QQ)^3$ the counting is $28 \cdot 29 \cdot 30 / 3! = 4060$. For
$\Tr (QQ)^4$ it is naively $28 \cdot 29 \cdot 30 \cdot 31 / 4! = 31465$ 
but we see that 
the coefficient in the index is $31464$ indicating that there is one linear
relation amongst these $31465$ expressions. Indeed, there is
a single linear relation between the $105$ expressions formed by
contracting pairs of $USp(4)$ indices on $Q^1 Q^2 \cdots Q^8$.
The term $56a_1^2 a_2^2q^{26/35}$ corresponds to the dressed meson $\Tr(Q^i (BB) Q^j)$ whose number is $8\cdot 7 / 2 = 28$,
with another 28 from $\Tr(Q^i Q^j) \Tr(BB)$. 
The term $28a_1^{-2}a_2^{-6}q^{26/35}$ corresponds to the operator $v_A^{0,\pm}\Tr (Q^iQ^j)$. 
We note that the monopole operators can be dressed by $\Tr(QQ)$ or $\Tr(BB)$
provided at least one component of $Q$ or $B$ remains massless
\cite{Cremonesi:2015dja} in the monopole background. To the order of our
expansion we see that all monopoles can be dressed by $\Tr(QQ)$ or $\Tr(BB)$ except for
$v_a^{*,0}$ which cannot be dressed by $\Tr(BB)$ -- this would appear in the index as $q^{6/5}/a_1^4$ but we can see it is absent above. However, we note that
$v_a^{0,*}$ can be dressed by $\Tr(BB)$.
The asymmetry is because if we only turn on the flux for
$USp(2)$ there is no component of the
bifundamental (or any field in a non-trivial representation of $USp(2)$, although in the case of multiple fluxes some
components can be massless due to cancellations allowing $\rho(\vec{m}) = 0$) which can satisfy the massless condition $\rho(\vec{m}) = 0$
\cite{Cremonesi:2015dja}.
However, turning on only one of the fluxes for $USp(6)$ the component of
$B$ which is in the bifundamental representation of
$USp(2) \times USp(4)$ (with this being the original $USp(2)$ and with $USp(4) \subset USp(6)$) remains massless.

As we conjecture, the half-indices for theories A, B and C agree perfectly with each other. 
With the same R-charge assignment as the full-indices, they are given by
\begin{align}
&\II^A_{\mathcal{N},\mathcal{N}} = \II^B_{\mathcal{D},\mathcal{N}}=\II^C_{\mathcal{N},\mathcal{D}}
\nonumber\\
&=1+\underbrace{28 a_2^2 q^{2/7}}_{\Tr(QQ)}
+\underbrace{a_1^2 q^{2/5}}_{\Tr(BB)}
+\underbrace{406 a_2^4 q^{4/7}}_{\Tr (QQ)^2}
-\underbrace{\frac{8 q^{9/14} \left(a_2 \left(w^2+1\right)\right)}{w}}_{\Tr(Q \Gamma_2)}
+\underbrace{56 a_1^2 a_2^2
   q^{24/35}}_{\Tr(Q(BB)Q)}+\underbrace{a_1^4 q^{4/5}}_{\Tr (BB)^2}
\nonumber\\
&-\underbrace{\frac{8 q^{59/70} \left(a_1 a_2 \left(u^2+1\right)\right)}{u}}_{\Tr(Q B \Gamma_1)}
+\underbrace{4060 a_2^6 q^{6/7}}_{\Tr (QQ)^3}
-\underbrace{\frac{224 q^{13/14} \left(a_2^3
   \left(w^2+1\right)\right)}{w}}_{\Tr(QQ)\Tr(Q\Gamma_2)}+
  \underbrace{1190 a_1^2 a_2^4 q^{34/35}}_{\Tr(QQ)^2\Tr(BB)}
\nonumber\\
&+q \underbrace{\left(u^2+\frac{1}{u^2}+w^2+\frac{1}{w^2}+2\right)}_{\Tr(\Gamma_1\Gamma_1) , \; \Tr(\Gamma_2\Gamma_2)} +\cdots
\end{align}
where $u$ and $w$ are the fugacities for $USp(2\tilde{N}_1) = USp(2)$ and
$USp(2\tilde{N}_2) = USp(2)$. 
Since in theory A the both vector multiplets obey the Neumann b.c. there is no monopole operator at the boundary. 

The boundary operator map of across the proposed triality is given by
\begin{align}
\label{USpUSp_bdymap}
\begin{array}{c|c|c}
A&B&C\\ \hline 
\Tr(Q\Gamma_2)&\Tr(Q\Gamma_2)& \psi_{q} \\
\Tr(QB\Gamma_1)&\Tr(Q\psi_{b})&\Tr(qc\Gamma_1)\\
\end{array}
\end{align}
where $\psi_q$ is the fermion in the $q$ multiplet with Dirichlet boundary
conditions which is charged under the global $SU(2N_f)$ and $USp(2\tilde{N}_2)$
symmetries, and similarly $\psi_b$ is the fermion in the $b$ multiplet.

\subsection{Chern-Simons levels}
As is well known, we can also generate non-zero Chern-Simons levels by giving large positive or
negative masses to the chirals $Q$ as first demonstrated for symplectic groups
by taking a limit of the partition function in \cite{Willett:2011gp}. In the simplest case we
start with $2N_f = \hat{N}_f + 2|k|$ fundamental and anti-fundamental
chirals and give masses (of the same sign) to $2|k| \in \Zb$ of them to leave
$\hat{N}_f$ fundamental chirals. This generates
a Chern-Simons level $k$ for the $USp(2N_2)$ gauge node in theories A and B with
$k>0$ by sending the masses to $+\infty$ and $k<0$ by sending the masses to
$-\infty$. There are no other changes to theories A and B so the result is
that the duality holds for arbitrary Chern-Simons level $k$ for $USp(2N_2)$.
However, the effect on theory C is more involved. In the case where we send the
masses of $2|k|$ multiplets $Q$ to $\pm \infty$, in theory C we
must also send the masses of $2|k|$ multiplets $q$ and of
$\sigma_C$ to $\mp \infty$ while also sending the masses
of $2|k|$ multiplets $p$ to $\pm \infty$. This modifies theory
C by having $\tilde{N_2} = N_1 - N_2 + \hat{N}_f/2 + |k| - 1$ while having $\hat{N}_f$
flavors, removing the $\sigma_C$ multiplet, and giving gauge group with
Chern-Simons levels $USp(2N_1)_k \times USp(2\tilde{N_2})_{-k}$. 

\subsection{$USp(2)\times USp(6)_{2k}-[2(4-|k|)]$ $(N_1 = 1$, $N_2 = 3$, $N_f = 4 - |k|)$}
Here we have $\tilde{N_1} = 1$ and $\tilde{N_2} = 1$, with Chern-Simons level satisfying
$|k|$ $\in$ $\{\frac{1}{2}$, $1$, $\frac{3}{2}$, $2$, $\frac{5}{2}$, $3$, $\frac{7}{2}$, $4 \}$. 

We have checked the matching of supersymmetric indices and 
for $k=1/2$, $r_B=2/5$ and $r_Q=2/7$ the full indices are given by
\begin{align}
&I^A = I^B = I^C
\nonumber\\
=&1+\underbrace{21 a_2^2 q^{2/7}}_{\Tr(QQ)}+\underbrace{a_1^2 q^{2/5}}_{\Tr(BB)}
+\underbrace{231 a_2^4 q^{4/7}}_{\Tr(QQ)^2}
+\underbrace{42 a_1^2 a_2^2 q^{24/35}}_{\begin{smallmatrix} \Tr(QQ) \Tr(BB), \\ \Tr(Q(BB)Q) \end{smallmatrix}}
+(\underbrace{a_1^4}_{\Tr(BB)^2}+\underbrace{\frac{1}{a_1^6}}_{v_{A}^{\pm,0}}) q^{4/5}
+\underbrace{1771 a_2^6 q^{6/7}}_{\Tr(QQ)^3}
\nonumber\\
&+\underbrace{672 a_1^2 a_2^4 q^{34/35}}_{\begin{smallmatrix} \Tr(QQ)^2 \Tr(BB), \\ \Tr(QQ) \Tr(Q(BB)Q) \end{smallmatrix}}
- \underbrace{50q}_{\begin{smallmatrix} \Tr(Q \psi_Q), \\ \Tr(B \psi_B) \end{smallmatrix}}
+21 (\underbrace{2 a_1^{4}}_{\begin{smallmatrix} \Tr(QQ) \Tr(BB)^2, \\ \Tr(Q(BB)^2Q) \end{smallmatrix}} + \underbrace{a_1^{-6}}_{v_A^{\pm,0} \Tr(QQ)}) a_2^2 q^{38/35}
+\underbrace{10626 a_2^8 q^{8/7}}_{\Tr(QQ)^4}
 + \cdots
\end{align}
We see that the monopole operators with the charge $(0,*)$ are removed due to the non-zero Chern-Simons level for the second gauge node.
In fact, in \cite{Aharony:2015pla} it is argued~\footnote{The detailed analysis in \cite{Aharony:2015pla} is for $U(N)$ gauge groups, but we expect a similar results for other gauge groups -- see also \cite{Amariti:2015kha, Nii:2018bgf} for $SU(N)$ theories. It is an important issue to understand the spectrum of chiral monopole operators for symplectic and orthogonal groups, even in the case of a single gauge node. We leave this for future work and in this paper simply comment on `missing' monopole operator contributions in some examples.} that the Chern-Simons terms allow monopole operators to carry an electric charge given by the Chern-Simons level. In order to form a gauge invariant operator they must be dressed by charged fields. However, typically even the candidate minimal gauge invariant dressed monopole operator is not chiral so does not contribute to the index, with the lowest order dressed chiral monopole operators having even higher dimension \cite{Aharony:2015pla} if there even are any chiral monopoles in the theory. Therefore, not surprisingly, we do not see contributions from any monopole operators charged under $USp(6)$ to the order we expand. However, the charge $(*,0)$ monopole operators and dressed versions survive as the $USp(2)$ gauge node has vanishing Chern-Simons level.

Note that in this example there are some features specific to the chosen gauge
groups. E.g.\ $\Tr(BB)^2$ and $\Tr(BBBB)$ are not distinct operators since the first
gauge node is $USp(2)$.

The half-indices for $k = -\frac12$ are given by
\begin{align}
&\II^A_{\mathcal{N},\mathcal{N}} = 
\II^B_{\mathcal{D},\mathcal{N}} =
\II^C_{\mathcal{N},\mathcal{D}}
\nonumber\\
=&
1+\underbrace{21 a_2^2 q^{2/7}}_{\Tr(QQ)}
+\underbrace{a_1^2 q^{2/5}}_{\Tr(BB)}
+\underbrace{231 a_2^4 q^{4/7}}_{\Tr(QQ)^2}
-\underbrace{\frac{7 q^{9/14} \left(a_2 \left(w^2+1\right)\right)}{w}}_{\Tr(Q\Gamma_2)}
+\underbrace{42 a_1^2 a_2^2 q^{24/35}}_{\begin{smallmatrix} \Tr(QQ) \Tr(BB), \\ \Tr(Q(BB)Q) \end{smallmatrix} }
+\underbrace{a_1^4 q^{4/5}}_{\Tr (BB)^2}
\nonumber\\
&
-\underbrace{\frac{7 q^{59/70} \left(a_1 a_2 \left(u^2+1\right)\right)}{u}}_{\Tr(QB\Gamma_1)}
+\underbrace{1771 a_2^6 q^{6/7}}_{\Tr(QQ)^3}
-\underbrace{\frac{147 q^{13/14} \left(a_2^3
   \left(w^2+1\right)\right)}{w}}_{\Tr(QQ)\Tr(Q\Gamma_2)}
   +\underbrace{672 a_1^2 a_2^4 q^{34/35}}_{\begin{smallmatrix} \Tr(QQ)^2\Tr(BB), \\ \Tr(QQ) \Tr(Q(BB)Q) \end{smallmatrix}}
   \nonumber\\
   &
   + \underbrace{q \left(u^2+\frac{1}{u^2}+w^2+\frac{1}{w^2}+2\right)}_{\Tr(\Gamma_1\Gamma_1) , \; \Tr(\Gamma_2\Gamma_2)} + \cdots
\end{align}

\subsection{$USp(2N_1) \times USp(2N_2)-[2N_f] \times USp(2N_3)$}
\label{sec_USpUSpUSp}
Here we would like to propose a quadrality of quiver gauge theories with three nodes of symplectic gauge groups $USp(2N_1)$, $USp(2N_2)$ and $USp(2N_3)$.
Let us call the original theory A. 
Dualizing the $USp(2N_1)$ or $USp(N_3)$ lead to a theories B or D which are 
easily obtained from the previous discussion. 

Another interesting theory arises when we take the Seiberg-like dual of the $USp(2N_2)$ gauge node. 
We propose that it gives a dual theory C with $USp(2\tilde{N}_2)$ where $\tilde{N_2}=N_1 + N_3 + N_f - N_2 - 1$ gauge node. 
The three gauge nodes $USp(2N_1)$, $USp(2\tilde{N}_2)$ and $USp(2N_3)$ are coupled to $2N_f$ chiral multiplets $p_1$, $q$ and $p_2$. 
There are antisymmetric rank-$2$ chirals $\phi_c^1$ for $USp(2N_1)$ and $\phi_c^2$ for $USp(2N_3)$, 
and a singlet chiral $M$ in the antisymmetric rank-$2$ representation of the flavor symmetry $SU(2N_f)$. 
It has a superpotential
\begin{align}
\label{spot_3node}
\mathcal{W}_C&=
\sigma_C v_C^{0,\pm,0} +
\Tr(c_1 \phi_c^1 c_1)+\Tr(c_2 \phi_c^2 c_2)
\nonumber\\
&+\Tr (c_1 p_1 q)+\Tr (c_2 p_2 q)+ \Tr(c_1c_2c_3) + \Tr(qMq).
\end{align}
The first term replaces the bare monopole operator $v_C^{0,\pm,0}$ with the
singlet $\sigma_C$ while
the second and third terms replace $c_1c_1$ and $c_2 c_2$ with $\phi_c^1$ and
$\phi_c^2$. The final term replaces $qq$ with $M$. The other terms can be
viewed as replacing $c_1 q$ with $p_1$, $c_2 q$ with $p_2$ and $c_1c_2$ with $c_3$.

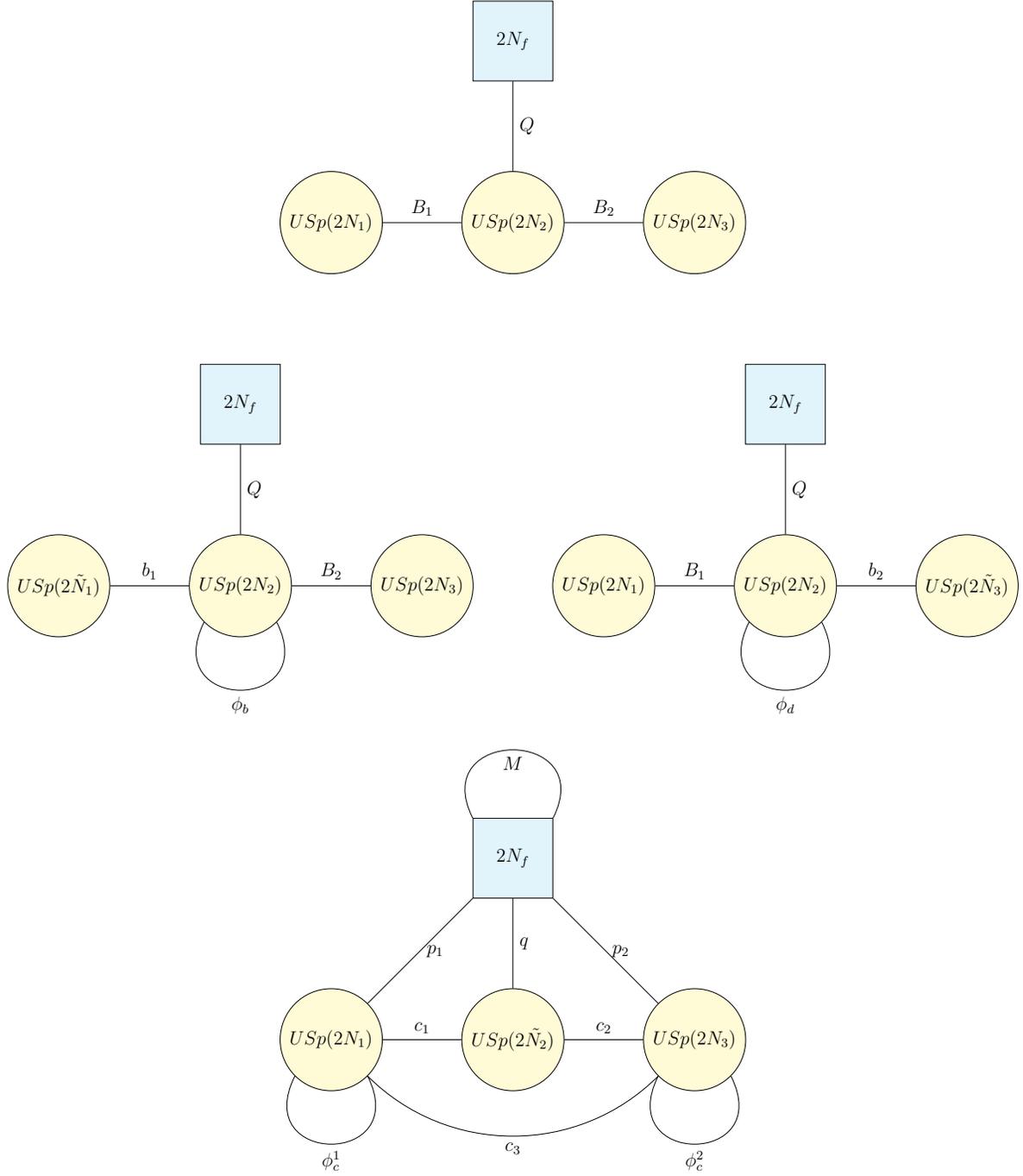
\begin{figure}
\centering
\scalebox{0.7}{
\begin{tikzpicture}
\path (-4,0) node[circle, minimum size=64, fill=yellow!20, draw](AG1) {$USp(2N_1)$}
(0,0) node[circle, minimum size=64, fill=yellow!20, draw](AG2) {$USp(2N_2)$}
(4,0) node[circle, minimum size=64, fill=yellow!20, draw](AG3) {$USp(2N_3)$}
(0,4) node[minimum size=50, fill=cyan!10, draw](AF2) {$2N_f$};
\draw (AG1) -- (AG2) node [midway, above] {$B_1$};
\draw (AG3) -- (AG2) node [midway, above] {$B_2$};
\draw (AG2) -- (AF2) node [midway, right] {$Q$};
\path (-10,-8) node[circle, minimum size=64, fill=yellow!20, draw](BG1) {$USp(2\tilde{N}_1)$}
(-6,-8) node[circle, minimum size=64, fill=yellow!20, draw](BG2) {$USp(2N_2)$}
(-2,-8) node[circle, minimum size=64, fill=yellow!20, draw](BG3) {$USp(2N_3)$}
(-6,-4) node[minimum size=50, fill=cyan!10, draw](BF2) {$2N_f$};
\draw (BG1) -- node[above]{$b_1$} (BG2);
\draw (BG3) -- node[above]{$B_2$} (BG2);
\draw (BG2) -- node[right]{$Q$}(BF2);
\draw (BG2.south west) .. controls ++(-1, -2) and ++(1, -2) .. node[below]{$\phi_b$} (BG2.south east);
%
%
\path (2,-8) node[circle, minimum size=64, fill=yellow!20, draw](DG1) {$USp(2N_1)$}
(6,-8) node[circle, minimum size=64, fill=yellow!20, draw](DG2) {$USp(2N_2)$}
(10,-8) node[circle, minimum size=64, fill=yellow!20, draw](DG3) {$USp(2\tilde{N}_3)$}
(6,-4) node[minimum size=50, fill=cyan!10, draw](DF2) {$2N_f$};
\draw (DG1) -- node[above]{$B_1$} (DG2);
\draw (DG3) -- node[above]{$b_2$} (DG2);
\draw (DG2) -- node[right]{$Q$}(DF2);
\draw (DG2.south west) .. controls ++(-1, -2) and ++(1, -2) .. node[below]{$\phi_d$} (DG2.south east);
%
\path (-4,-18) node[circle, minimum size=64, fill=yellow!20, draw](CG1) {$USp(2N_1)$}
(0,-18) node[circle, minimum size=64, fill=yellow!20, draw](CG2) {$USp(2\tilde{N_2})$}
(4,-18) node[circle, minimum size=64, fill=yellow!20, draw](CG3) {$USp(2N_3)$}
(0,-14) node[minimum size=50, fill=cyan!10, draw](CF2) {$2N_f$};
\draw (CG1) -- node[above]{$c_1$} (CG2);
\draw (CG3) -- node[above]{$c_2$} (CG2);
\draw (CG1) -- node[right]{$p_1$}(CF2);
\draw (CG3) -- node[right]{$p_2$}(CF2);
\draw (CG2) -- node[right]{$q$}(CF2);
\draw (CG1) to [out=-45, in=-135] node[below]{$c_3$}(CG3);
\draw (CG1.south west) .. controls ++(-1, -2) and ++(1, -2) .. node[below]{$\phi_c^1$} (CG1.south east);
\draw (CG3.south west) .. controls ++(-1, -2) and ++(1, -2) .. node[below]{$\phi_c^2$} (CG3.south east);
\draw (CF2.north east) .. controls ++(1, 2) and ++(-1, 2) .. node[below]{$M$} (CF2.north west);
\end{tikzpicture}
}
\caption{Quadrality of $USp(2N_1) \times USp(2N_2)-[2N_f] \times USp(2N_3)$ quiver 
where $\tilde{N_1}=N_2-N_1-1$, $\tilde{N_2}=N_1+\tilde{N}_1+N_f-N_2-1$ and $\tilde{N}_3=N_2-N_3-1$. All chirals $\phi_b$, $\phi_c^1$, $\phi_c^2$, $\phi_d$ and $M$ are in antisymmetric rank-$2$
representations.} \label{fig:USp_USp_USp}
\end{figure}

The proposed quadrality of the $USp\times USp\times USp$ gauge theories is shown in Figure \ref{fig:USp_USp_USp}. 

The operator map across the proposed quadrality is given by
\begin{align}
\label{USpUSpUSp_map1}
\begin{array}{c|c|c|c}
A&B&C&D\\ \hline 
\Tr(QQ)&\Tr(QQ)&M&\Tr(QQ) \\
\Tr(B_1(B_2B_2)^k B_1) & \Tr(\phi_b (B_2B_2)^k) & \Tr(c_3c_3 (\phi_c^2)^{k-1}) & \Tr((\phi_d)^k B_1B_1) \\
\Tr(B_2(B_1B_1)^k B_2) & \Tr((\phi_b)^k B_2B_2) & \Tr(c_3c_3 (\phi_c^1)^{k-1}) & \Tr(\phi_d (B_1B_1)^k) \\
\Tr(Q(B_1B_1)^k Q)&\Tr(Q\phi_b^k Q)&\Tr(p_1(\phi_c^1)^{k-1}p_1)&\Tr(Q(B_1B_1)^k Q) \\
\Tr(Q(B_2B_2)^k Q)&\Tr(Q(B_2B_2)^k Q)&\Tr(p_2(\phi_c^2)^{k-1}p_2)&\Tr(Q\phi_d^k Q) \\
\Tr(B_1B_1)^k&\Tr(\phi_b^k)&\Tr(\phi_c^1)^k&\Tr(B_1B_1)^k \\
\Tr(B_2B_2)^k&\Tr(B_2B_2)^k&\Tr(\phi_c^2)^k&\Tr(\phi_d^k) \\
v_A^{\pm,0,0}&\sigma_B&v_C^{\pm,\pm,0} &v_D^{\pm,0,0} \\
v_A^{0,0,\pm}&v_B^{0,0,\pm}&v_C^{0,\pm,\pm}&\sigma_D \\
v_A^{0,\pm,0}&v_B^{\pm,\pm,0}&\sigma_C&v_D^{0,\pm,\pm} \\
v_A^{\pm,\pm,0}& v_B^{0,\pm,0}&v_C^{\pm,0,0}&v_D^{\pm,\pm,\pm}\\
v_A^{0,\pm,\pm}& v_B^{\pm,\pm,\pm}&v_C^{0,0,\pm}&v_D^{0,\pm,0} \\
v_A^{\pm,0,\pm}& v_B^{\pm,0,\pm}&v_C^{\pm,0,\pm}&v_D^{\pm,0,\pm} \\
v_A^{\pm,\pm,\pm}&v_B^{0,\pm,\pm}&v_C^{\pm,\pm,\pm}&v_D^{\pm,\pm,0}
\end{array}
\end{align}

The monopole operators above are gauge invariant so can be dressed by gauge
invariant combinations of the chiral multiplets, with the mapping between theories following the pattern indicated in the first 7 lines of the table.
As for the case of two gauge nodes there will be restrictions on the
allowed dressed monopole operators \cite{Cremonesi:2015dja} and we comment on this in the example we
include next.

\subsection{$USp(2)\times USp(6)-[6]\times USp(2)$ $(N_1 = 1$, $N_2 = 3$, $N_3 = 1$, $N_f = 3)$}
\label{sec_USpUSpUSp_example}
We have confirmed that the four supersymmetric indices perfectly match. 
With $r_B = 2/5$ and $r_Q = 2/7$ the full indices are given by
\begin{align}
&I^A=I^B=I^C=I^D
\nonumber\\
&=1+\underbrace{15 a_2^2 q^{2/7}}_{\Tr(QQ)}
+\underbrace{\frac{q^{12/35}}{a_1^2 a_2^6  a_3^2}}_{v_A^{0,\pm,0}}
+q^{2/5} (\underbrace{a_1^2}_{\Tr(B_1 B_1)}+ \underbrace{a_3^2}_{\Tr(B_2 B_2)})
+\frac{q^{19/35}}{a_1^6
   a_2^6  a_3^6} (\underbrace{a_1^4}_{v_A^{0, \pm, \pm}}+ \underbrace{a_3^4}_{v_A^{\pm, \pm, 0}})
   +\underbrace{120 a_2^4 q^{4/7}}_{\Tr(QQ)^2}
\nonumber\\
&+\underbrace{\frac{15 q^{22/35}}{a_1^2 a_2^4  a_3^2}}_{v_A^{0,\pm,0} \Tr(QQ)}
+q^{24/35} (\underbrace{\frac{1}{a_1^4 a_2^{12}  a_3^4}}_{v_A^{0,\pm2, 0}}
+\underbrace{30 a_1^2a_2^2}_{\begin{smallmatrix} \Tr(QQ)\Tr(B_1 B_1), \\ \Tr(Q(B_1B_1)Q) \end{smallmatrix}}
+\underbrace{30 a_2^2  a_3^2)}_{\begin{smallmatrix} \Tr(QQ)\Tr(B_2 B_2), \\ \Tr(Q(B_2B_2)Q) \end{smallmatrix}}
\nonumber\\
&
+q^{26/35} (
\underbrace{\frac{1}{a_2^6 a_3^2}}_{v_A^{0,\pm,0}B_1B_1}
+\underbrace{\frac{1}{a_1^2 a_2^6}}_{v_A^{0,\pm,0}B_2B_2}
+ \underbrace{\frac{1}{a_1^6 a_2^6 a_3^6}}_{v_A^{\pm, \pm, \pm}})
+q^{4/5} (
\underbrace{2a_1^2  a_3^2}_{\begin{smallmatrix} \Tr(B_1 B_1)\Tr(B_2 B_2) \\ \Tr(B_1 B_2 B_2 B_1)\end{smallmatrix}}
+\underbrace{a_1^4}_{\Tr(B_1 B_1)^2}
+\underbrace{\frac{1}{a_1^6}}_{v_A^{\pm,0,0}}
+ \underbrace{a_3^4}_{\Tr(B_2 B_2)^2}
+\underbrace{\frac{1}{ a_3^6}}_{v_A^{0,0,\pm}})
   \nonumber\\
&
   +15q^{29/35} (\underbrace{\frac{1}{a_1^2 a_2^4 a_3^6}}_{v_A^{\pm,\pm,0}\Tr(QQ)} 
   + \underbrace{\frac{1}{a_1^6 a_2^4 a_3^2}}_{v_A^{0,\pm,\pm}\Tr(QQ)} ) 
    +\underbrace{680 a_2^6 q^{6/7}}_{\Tr(QQ)^3}
+ q^{31/35} ( 
\underbrace{\frac{1}{a_1^6 a_2^{12} a_3^6}}_{v_A^{\pm,\pm\pm,\pm}} + 
\underbrace{\frac{1}{a_1^4 a_2^{12} a_3^8}}_{v_A^{0,\pm2,\pm}} + 
\underbrace{\frac{1}{a_1^8 a_2^{12} a_3^4}}_{v_A^{\pm,\pm2,0}} )
   \nonumber\\
&
 + \underbrace{\frac{120
   q^{32/35}}{a_1^2 a_2^2  a_3^2}}_{v_A^{0, \pm, 0} \Tr(QQ)^2}
 + q^{33/35} (
\underbrace{\frac{1}{a_1^2 a_2^6 a_3^4}}_{v_A^{\pm,\pm,0}B_2 B_2} 
+ \underbrace{\frac{1}{a_1^4 a_2^6 a_3^2}}_{v_A^{0,\pm,\pm}B_1 B_1} 
 + \underbrace{\frac{1}{a_2^6 a_3^6}}_{v_A^{\pm,\pm,0}B_1 B_1} 
 + \underbrace{\frac{1}{a_1^6 a_2^6}}_{v_A^{0,\pm,\pm}B_2 B_2} )
\nonumber\\
&+q^{34/35} ( \underbrace{\frac{15}{a_1^4 a_2^{10}  a_3^4}}_{v_A^{0, \pm 2, 0} \Tr(QQ)} 
+\underbrace{345 a_1^2 a_2^4}_{\begin{smallmatrix} \Tr(QQ)^2 \Tr(B_1 B_1), \\ \Tr(QQ) \Tr(Q(B_1 B_1)Q) \end{smallmatrix}}
+\underbrace{345 a_2^4  a_3^2}_{\begin{smallmatrix} \Tr(QQ)^2 \Tr(B_2 B_2), \\ \Tr(QQ) \Tr(Q(B_2 B_2)Q) \end{smallmatrix}})
-\underbrace{38q}_{\begin{smallmatrix}\Tr (Q\psi_Q), \\ \Tr(B_1\psi_{B_1}), \; \Tr(B_2\psi_{B_2}) \end{smallmatrix}}+\cdots
\nonumber\\
&+
q^{8/7}
\Big(
\underbrace{\frac{1}{a_1^4 a_2^6 a_3^6}}_{v_A^{\pm,\pm,\pm}B_1B_1}
+\underbrace{\frac{1}{a_1^6 a_2^6 a_3^4}}_{v_A^{\pm,\pm,\pm}B_2B_2}
 + \underbrace{3060a_2^8}_{\Tr(QQ)^4}
 + \cdots \Big)
 + q^{6/5} \Big(
\underbrace{\frac{a_3^2}{a_1^6}}_{v_A^{\pm,0,0} B_2B_2}
+\underbrace{\frac{a_1^2}{a_3^6}}_{v_A^{0,0,\pm} B_1 B_1}
+ \cdots \Big)
+\cdots
\end{align} 

Note that unlike the $USp(2) \times USp(6)-[8]$ example, here we get the
naively expected coefficient for the $q^{8/7} a_2^8$ term, in this case
$3060 = 15 \cdot 16 \cdot 17 \cdot 18 / 4!$ for $\Tr(QQ)^4$.

We see several bare monopole operators and at higher order they are dressed with gauge invariant combinations of the other operators. However, some terms are missing. The first occurs at order $q^{6/5}$ where we do not see the terms $q^{6/5}/a_1^4$ or $q^{6/5}/a_3^4$ which would be the contributions from $v_A^{\pm,0,0}$ dressed by $B_1B_1$ and $v_A^{0,0,\pm}$ dressed by $B_2B_2$. We have checked the index to higher order than presented here and we see no contributions which would correspond to $v_A^{\pm,0,0}$ dressed by any operators containing $B_1$ operators (or, of course by symmetry, $v_A^{0,0,\pm}$ dressed by any operators containing $B_2$ operators).
This is entirely as expected since with only flux for one of
the $USp(2)$ gauge nodes, and fields in non-trivial representations of that
$USp(2)$ group cannot obey the massless condition $\rho(\vec{m}) = 0$, and this
clearly applies to $B_1$ and $B_2$ in these backgrounds.

One point to note is that for symplectic gauge groups we have no topological
fugacities in the index to keep track of monopole contributions.~\footnote{
Although we can of course calculate the contribution to the index for each
choice of magnetic fluxes.} This means that some coefficients can arise from
contributions from different sectors. One example at higher order is the
coefficient of $q^{8/5}/(a_1^6 a_3^6)$ which is found to be $680$. This can
receive contributions from the gauge invariant bare monopole operator
$v_A^{\pm, 0, \pm}$ and from $v_A^{\pm, \pm, \pm}$ dressed by $\Tr(QQ)^3$.
Naively, $\Tr(QQ)^3$ gives contribution $15 \cdot 16 \cdot 17 / 3! = 680$ so
at first it appears as though the $v_A^{\pm, 0, \pm}$ monopole is missing.
However, in the $v_A^{\pm, \pm, \pm}$ monopole background the massless part
of $Q$ transforms in the fundamental representation of the $USp(4)$
gauge group and a careful check reveals a linear relation amongst the $15$
naively independent gauge invariant contraction of $Q^1 Q^2 Q^3 Q^4 Q^5 Q^6$.

The half-indices are given by
\begin{align}
&\II^A_{\mathcal{N},\mathcal{N},\mathcal{N}} 
= \II^B_{\mathcal{D},\mathcal{N},\mathcal{N}}  
= \II^C_{\mathcal{N},\mathcal{D},\mathcal{N}} 
= \II^D_{\mathcal{N},\mathcal{N},\mathcal{D}} 
\nonumber\\
&=1+\underbrace{15 a_2^2 q^{2/7}}_{\Tr(QQ)}
+q^{2/5} (\underbrace{a_1^2}_{\Tr(B_1 B_1)}+ \underbrace{a_3^2}_{\Tr(B_2 B_2)})
+\underbrace{120 a_2^4 q^{4/7}}_{\Tr(QQ)^2}
-\underbrace{\frac{6 q^{9/14} \left(a_2 \left(w^2+1\right)\right)}{w}}_{\Tr(Q\Gamma_2)}
\nonumber\\
&
+30 a_2^2 q^{24/35}
   (\underbrace{a_1^2}_{\begin{smallmatrix} \Tr(QQ) \Tr(B_1 B_1), \\ \Tr(Q(B_1 B_1)Q) \end{smallmatrix}}+ \underbrace{a_3^2}_{\begin{smallmatrix} \Tr(QQ) \Tr(B_2 B_2), \\ \Tr(Q(B_2 B_2)Q) \end{smallmatrix}})
+q^{4/5} 
(
\underbrace{2a_1^2  a_3^2}_{\begin{smallmatrix} \Tr(B_1 B_1)\Tr(B_2 B_2) \\ \Tr(B_1 B_2 B_2 B_1)\end{smallmatrix}}
+\underbrace{a_1^4}_{\Tr(B_1B_1)^2}
+\underbrace{a_3^4}_{\Tr(B_2B_2)^2}
)
\nonumber\\
&
- \underbrace{\frac{6 q^{59/70} \left(a_2 \left(a_1 u \left(v^2+1\right)+ a_3
   \left(u^2+1\right) v\right)\right)}{u v}}_{\Tr(Q B_1 \Gamma_1), \; \Tr(Q B_3 \Gamma_3)}
   +\underbrace{680 a_2^6 q^{6/7}}_{\Tr(QQ)^3}
- \underbrace{\frac{90 q^{13/14} \left(a_2^3 \left(w^2+1\right)\right)}{w}}_{\Tr(Q \Gamma_2) \Tr(QQ)}
\nonumber\\
&
+345 a_2^4 q^{34/35}
   (\underbrace{a_1^2}_{\begin{smallmatrix} \Tr(QQ)^2 \Tr(B_1 B_1), \\ \Tr(QQ) \Tr(Q(B_1 B_1)Q) \end{smallmatrix}}
   +\underbrace{a_3^2}_{\begin{smallmatrix} \Tr(QQ)^2 \Tr(B_2 B_2), \\ \Tr(QQ) \Tr(Q(B_2 B_2)Q) \end{smallmatrix}} ) + \cdots
\end{align}
where $v, w, u$ are the fugacities for $USp(2\tilde{N}_1), USp(2\tilde{N}_2), USp(2\tilde{N}_3)$ and $\Gamma_I$ is the Fermi in the bifundamental representation of $USp(2N_I) \times USp(2\tilde{N}_I$ which is included in the theories for Neumann boundary conditions for the $USp(2N_I)$ vector multiplet.

\section{Orthogonal linear quivers}
\label{sec_orthogonal}
We now discuss similar dualities where we have orthogonal rather than symplectic gauge groups. We present examples with two and three gauge nodes.

\subsection{$SO(N_1) \times SO(N_2)-[N_f]$}
\label{sec_SOSO}
There are distinct orthogonal gauge groups 
$SO(N_c)$, $O(N_c)_{\pm}$, $Spin(N_c)$ and $Pin(N_c)_{\pm}$ 
for the Lie algebra $\mathfrak{so}(N_c)$. 
The minimal gauge group is $SO(N_c)$ and other gauge groups can be obtained by gauging two 0-form global symmetries, 
the charge conjugation symmetry $\mathbb{Z}_2^{\mathcal{C}}$ 
and the magnetic symmetry $\mathbb{Z}_{2}^{\mathcal{M}}$. 
In the indices and half-indices we have discrete fugacities $\chi$ for $\Zb_2^{\Ccal}$ and $\zeta$ for $\Zb_2^{\Mcal}$.
We first review the triality of the $SO\times SO$ quiver gauge theories proposed in \cite{Benvenuti:2020wpc} and mostly consider only the cases where these discrete fugacities are set to one, i.e.\ so the gauge groups are $SO$. However, we expect all dualities to be generalizable to other orthogonal groups as for the single gauge node cases without \cite{Aharony:2013kma} or with boundaries \cite{Okazaki:2021pnc}.

The structure of the $SO\times SO$ quivers take the similar form as the $USp\times USp$ quivers in section \ref{sec_USpUSp}. 
Theory A is a $SO(N_1)\times SO(N_2)$ gauge theory with bifundamental chirals $B$ in the $({\bf N_1}, {\bf N_2})$ representation, 
$N_f$ chirals $Q$ in the fundamental representation under the $SO(N_2)$. 
In these theories there can exist three types of baryonic operators
\begin{align}
\epsilon_1 \epsilon_2 B^{N_1} Q^{N_2-N_1}
&=\epsilon_{a_1 \cdots a_{N_1}}\epsilon^{b_1\cdots b_{N_2}}B_{b_1}^{a_1}\cdots B_{b_{N_1}}^{a_{N_1}}Q_{b_{N_1+1}}^{i_1}\cdots Q_{b_{N_2}}^{i_{N_2 -N_1}}, 
\nonumber\\
\epsilon_1 B^{N_1}Q^{N_1}
&=\epsilon^{a_1 \cdots a_{N_1}}B^{b_1}_{a_1}\cdots B^{b_{N_1}}_{a_{N_1}}Q^{i_1}_{b_{1}}\cdots Q^{i_{N_1}}_{b_{N_1}},
\nonumber\\
\epsilon_2 Q^{N_2}&=\epsilon^{b_1\cdots b_{N_2}} Q^{i_1}_{b_1}\cdots Q^{i_{N_2}}_{b_{N_2}},
\end{align}
where $a_i$, $b_j$ and $i_k$ are the $SO(N_1)$, $SO(N_2)$ and $SU(N_f)$ indices.
Since they are antisymmetric in the flavor indices, the
second and third types exist only when $N_f \ge N_1$ or $N_f \ge N_2$
respectively.
The first type exists when $N_f \ge N_2 - N_1 \ge 0$.
The dualities we discuss are also valid for $N_2 < N_1$~\footnote{Specifically, for both dualities discussed in this section to be valid
we need $N_1 - 2 \le N_2 \le N_1 + N_f + 2$.} but in those cases we
have no baryonic operators of the first type since the
analogous constructions contracting bifundamentals using the antisymmetric
tensors would have an excess of $SO(N_1)$ indices and we have no flavors in
the fundamental of $SO(N_1)$ to contract these indices.

For a single gauge node
with $N_c \ge 2$~\footnote{There are additional monopoles which we do not
discuss here in the case where $N_c \ge 4$ \cite{Aharony:2013kma} while for
$N_c < 2$ we have no monopoles.
}
there are two types of monopole operators $v^{+}$ and $v^{-}$, 
which are even and odd under the charge conjugation symmetry  $\mathbb{Z}_2^{\mathcal{C}}$ \cite{Aharony:2013kma}. 
The monopole operators break the gauge group $SO(N_c)$ down to $S(O(N_c - 2)\times O(2))$. 
For the product gauge group $SO(N_1)\times SO(N_2)$ the even monopole operators are 
\begin{align}
v_{A}^{+,0},\qquad v_{A}^{0,+}, \qquad v_{A}^{+,+}. 
\end{align}
They are gauge invariant and can be dressed with
gauge invariant combinations of $B$ and $Q$.

The odd monopole operators are not gauge invariant for $N_c > 2$. However, 
they can form gauge invariant operators dressed by $B$ and $Q$ in baryonic
combinations \cite{Benvenuti:2020wpc}
\begin{align}
v_{A}^{-,0}\epsilon_1 B^{N_1-2} Q^{N_1 -2}&, 
\nonumber\\
v_{A}^{-,0}\epsilon_1\epsilon_2 B^{N_1-2} Q^{N_2 -N_1+2}&, 
\nonumber\\
v_{A}^{-,+}\epsilon_1 B^{N_1-2} Q^{N_1 -2}&, 
\nonumber\\
v_{A}^{0,-}\epsilon_2 Q^{N_2 -2}&, 
\nonumber\\
v_{A}^{0,-}\epsilon_1 \epsilon_2 B^{N_1} Q^{N_2-N_1 -2}&,
\nonumber\\
v_{A}^{+,-}\epsilon_2 Q^{N_2 -2}&,
\nonumber\\
v_{A}^{-,-}\epsilon_1 \epsilon_2 B^{N_1-2} Q^{N_2 -N_1}&
\end{align}
where in any cases where the exponent would indicate an negative power, that type of operator is not present. In the case of zero exponent that component is
not present and this includes the special cases where $N_1 = 2$ and/or
$N_2 = 2$ where some of the combinations above reduce to gauge invariant bare
monopoles.

Theory B has the dualized $SO(\tilde{N}_1 = N_2 - N_1 + 2))$ gauge node and introduces a rank-2 symmetric chiral for $SO(N_2)$. 
Dualizing the $SO(N_2)$ gauge node gives theory C which has $N_f$ flavors for both gauge nodes and rank-2 symmetric chirals for $SO(N_1)$ and the global flavor symmetry $SU(N_f)$.

In general, under Seiberg-like duality of a gauge node $SO(N_I)$ with discrete
fugacities $\chi_I$ and $\zeta_I$, the dual
gauge node $SO(\tilde{N}_I)$ has discrete fugacities
$\tilde{\chi}_I = \zeta_I \chi_I$ and $\tilde{\zeta}_I = \zeta_I$.
For each gauge node $SO(N_J)$ connected to $SO(N_I)$ in the quiver diagram we
also have the mapping $\zeta_J \rightarrow \zeta_I \zeta_J$ under duality of
gauge node $SO(N_I)$.

\begin{figure}
\centering
\scalebox{0.7}{
\begin{tikzpicture}
\path (0,0) node[circle, minimum size=64, fill=green!10, draw](AG1) {$SO(N_1)$}
(4,0) node[circle, minimum size=64, fill=green!10, draw](AG2) {$SO(N_2)$}
(4,4) node[minimum size=50, fill=cyan!10, draw](AF2) {$N_f$};
\draw (AG1) -- (AG2) node [midway, above] {$B$};
\draw (AG2) -- (AF2) node [midway, right] {$Q$};
\path (-6,-8) node[circle, minimum size=64, fill=green!10, draw](BG1) {$SO(\tilde{N_1})$}
(-2,-8) node[circle, minimum size=64, fill=green!10, draw](BG2) {$SO(N_2)$}
(-2,-4) node[minimum size=50, fill=cyan!10, draw](BF2) {$N_f$};
\draw (BG1) -- node[above]{$b$} (BG2);
\draw (BG2) -- node[right]{$Q$}(BF2);
\draw (BG2.south west) .. controls ++(-1, -2) and ++(1, -2) .. node[below]{$\phi_b$} (BG2.south east);
%
\path (6,-8) node[circle, minimum size=64, fill=green!10, draw](CG1) {$SO(N_1)$}
(10,-8) node[circle, minimum size=64, fill=green!10, draw](CG2) {$SO(\tilde{N_2})$}
(10,-4) node[minimum size=50, fill=cyan!10, draw](CF2) {$N_f$};
\draw (CG1) -- node[above]{$c$} (CG2);
\draw (CG1) -- node[right]{$p$}(CF2);
\draw (CG2) -- node[right]{$q$}(CF2);
\draw (CG1.south west) .. controls ++(-1, -2) and ++(1, -2) .. node[below]{$\phi_c$} (CG1.south east);
\draw (CF2.north east) .. controls ++(1, 2) and ++(-1, 2) .. node[below]{$M$} (CF2.north west);
\end{tikzpicture}
}
\caption{Triality of $SO \times SO$ quivers where $\tilde{N_1}=N_2-N_1+2$ and $\tilde{N_2}=N_1+N_f-N_2+2$.
All chirals $\phi_b$, $\phi_c$ and $M$ are in symmetric rank-$2$
representations.
} \label{fig:SO_SO}
\end{figure}
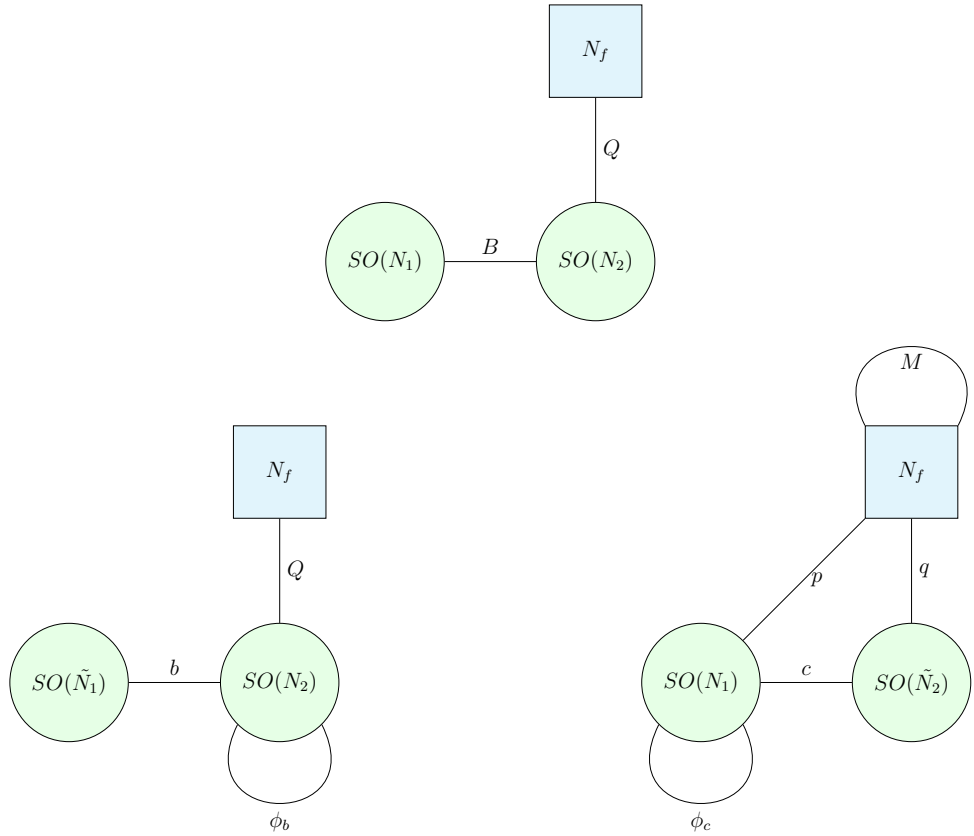

The triality of $SO \times SO$ quiver gauge theories is depicted in Figure \ref{fig:SO_SO}. 

The mapping of monopole operators is quite involved for orthogonal gauge groups so
we refer to \cite{Benvenuti:2020wpc} for details.
The mapping of the other operators (and their R-charge assignment) is similar to the case with symplectic gauge groups so we do not list that again.

\subsection{Supersymmetric indices for orthogonal gauge theories}
\label{sec_SOSOindices}
Writing $N_I = 2n_I + \epsilon_I$ for $n_I \in \Zb$ and
$\epsilon_I \in \{0, 1\}$,
the supersymmetric full-indices \cite{Bhattacharya:2008zy,Bhattacharya:2008bja,Kim:2009wb,Imamura:2011su, Kapustin:2011jm, Dimofte:2011py} for theories A, B and C with orthogonal gauge groups
can be constructed by generalising the expressions for a single gauge node
\cite{Hwang:2011ht, Hwang:2011qt, Aharony:2013kma} and are given by
\begin{align}
I^A = & Z^{SO_1}_{gauge} Z^{SO_2}_{gauge} Z_{matter \; A} \\
I^B = & Z^{\widetilde{SO}_1}_{gauge} Z^{SO_2}_{gauge} Z_{matter \; B} \\
I^C = & Z^{SO_1}_{gauge} Z^{\widetilde{SO}_2}_{gauge} Z_{matter \; C}
\end{align}
where $Z_{gauge}$ and $Z_{matter *}$ are the contributions from the vector and chiral multiplets with $SO_I$ referring to gauge node $SO(N_I)$ and
$\widetilde{SO}_I$ referring to gauge node $SO(\tilde{N}_I)$.
The gauge contributions are given by
\begin{align}
Z^{SO_I}_{gauge} = & \sum_{m_i^{(I)} \in \Zb} \frac{\zeta_I^{\sum_i m_i^{(I)}}}{2^{n_I + \epsilon_I - 1} n_I!} 
\oint \left( \prod_{i=1}^{n_I} \frac{ds_i^{(I)}}{2\pi i s_i^{(I)}} (-s_i^{(I)})^{k_I m_i^{(I)}} \right) \nonumber \\
 & \times
q^{-\sum_{i=1}^{n_I} \epsilon_I |m_i^{(I)}|/2 - \sum_{i < j}^{n_I} |m_i^{(I)} \pm m_j^{(I)}|/2}
 \left( \prod_{i=1}^{n_I} (1 - \chi_I q^{|m_i^{(I)}|/2}s_i^{(I)\pm} \right)^{\epsilon_I} \nonumber \\
 & \times \prod_{i < j}^{n_I}
 ( 1 - q^{|m_i^{(I)} \pm m_j^{(I)}|/2} s_{j}^{(I)} s_j^{(I)\pm}) ( 1 - q^{|-m_i^{(I)} \pm m_j^{(I)}|/2} s_i^{(I)-1} s_j^{(I)\pm})
 \end{align}
except for the case where $\chi_I = -1$ and $\epsilon_I = 0$ where instead
\begin{align}
Z^{SO_I}_{gauge} = & \sum_{m_i^{(I)} \in \Zb} \frac{\zeta_I^{\sum_i m_i^{(I)}}}{2^{n_I - 1} (n_I - 1)!}
\oint \left( \prod_{i=1}^{n_I - 1} \frac{ds_i^{(I)}}{2\pi i s_i^{(I)}} (-s_i^{(I)})^{k_I m_i^{(I)}} \right) \nonumber \\
 & \times
q^{-\sum_{i=1}^{n_I - 1} |m_i^{(I)}| - \sum_{i < j}^{n_I} |m_i^{(I)} \pm m_j^{(I)}|/2}
 \left( \prod_{i=1}^{n_I - 1} (1 - q^{|m_i^{(I)}|}s_i^{(I)\pm 2} \right) \nonumber \\
 & \times \prod_{i < j}^{n_I - 1}
 ( 1 - q^{|m_i^{(I)} \pm m_j^{(I)}|/2} s_{j}^{(I)} s_j^{(I)\pm}) ( 1 - q^{|-m_i^{(I)} \pm m_j^{(I)}|/2} s_i^{(I)-1} s_j^{(I)\pm}) \; .
 \end{align}
Above we have included Chern-Simons levels $k_I$ for $SO(N_I)$ but for now
we take these levels to all vanish.
 
As for the symplectic case the contributions from the chiral multiplets are 
\begin{align}
Z_{matter \; A} = & Z_B Z_Q^{N_f} \\
Z_{matter \; B} = & Z_b Z_{\phi_b} Z_Q^{N_f} Z_{\sigma_B} \\
Z_{matter \; C} = & Z_c Z_{\phi_c} Z_p^{N_f} Z_M^{N_f} Z_q^{N_f} Z_{\sigma_C} \; .
\end{align}
where 
 \begingroup
\allowdisplaybreaks
\begin{align}
&Z_{BiFund \; SO_I-SO_J}(r, a) =  (q^{\frac{1-r}{2}} a^{-1})^{\sum_{i=1}^{N_I} \sum_{j=1}^{N_J} |m_i^{(I)} \pm m_j^{(J)}|} \nonumber \\
 & \times \prod_{i = 1}^{n_I} \prod_{j=1}^{n_J}
 \frac{(q^{1-\frac{r}{2}+\frac{|m_i^{(I)}-m_j^{(J)}|}{2}} a^{-1} s_i^{(I)\mp} s_j^{(J)\pm}; q)_{\infty} 
 (q^{1-\frac{r}{2}+\frac{|m_i^{(I)}+m_j^{(J)}|}{2}} a^{-1} s_i^{(I)\pm} s_j^{(J)\pm}; q)_{\infty}}
 {(q^{\frac{r}{2}+\frac{|m_i^{(I)} - m_j^{(J)}|}{2} } a s_i^{(I)\pm} s_j^{(J)\mp}; q)_{\infty} 
(q^{\frac{r}{2}+\frac{|m_i^{(I)} + m_j^{(J)}|}{2} } a s_i^{(I)\pm} s_j^{(J)\pm}; q)_{\infty}} \nonumber \\
 & \times \left( \prod_{I=1}^{n_I}
 \frac{(q^{1-\frac{r}{2}+\frac{|m_i^{(I)}|}{2}} a^{-1} s_i^{(I)\pm} \chi_J; q)_{\infty}
}
 {(q^{\frac{r}{2}+\frac{|m_i^{(I)}|}{2} } a s_i^{(I)\mp} \chi_J; q)_{\infty}} \right)^{\epsilon_J}
 \left( \prod_{j=1}^{n_J}
 \frac{(q^{1-\frac{r}{2}+\frac{|m_j^{(J)}|}{2}} a^{-1} \chi_I s_j^{(J)\pm}; q)_{\infty}}
 {(q^{\frac{r}{2}+\frac{|m_j^{(J)}|}{2} } a \chi_I s_j^{(J)\mp}; q)_{\infty}} \right)^{\epsilon_I} \nonumber \\
 & \times \left( \prod_{I=1}^{n_I}
 \frac{(q^{1-\frac{r}{2}} a^{-1} \chi_I \chi_J; q)_{\infty}
}
 {(q^{\frac{r}{2}} a \chi_I \chi_J; q)_{\infty}} \right)^{\epsilon_I \epsilon_J} \\
&Z_{N_F Fund \; SO_I}(r, a) =  (q^{\frac{1-r}{2}} a^{-1})^{N_F \sum_{i=1}^{N_I} |m_i^{(I)}|} \prod_{\alpha = 1}^{N_F} \left( \prod_{i=1}^{n_I}
 \frac{(q^{1-\frac{r}{2}+ \frac{|{m_i^{(I)}}|}{2} } a^{-1} {s_j^{(I)}}^{\mp} x_{\alpha}^{-1}; q)_{\infty}}
 {(q^{\frac{r}{2}+\frac{|{m_i^{(I)}}|}{2}} a {s_i^{(I)}}^{\pm} x_{\alpha}; q)_{\infty}} \right) \nonumber \\
 & \times \left( \frac{(q^{1-\frac{r}{2}} a^{-1} \chi_I x_{\alpha}^{-1}; q)_{\infty}}
 {(q^{\frac{r}{2}} a \chi_I x_{\alpha}; q)_{\infty}} \right)^{\epsilon_I} \\
&Z_{Sym \; SO_I}(r, a) =  (q^{\frac{1 - 2r}{2}} a^{-1})^{\sum_{i < j}^{N_I} |{m_i^{(I)}} \pm {m_j^{(I)}}|}  \prod_{i \le j}^{n_I} \nonumber \\
&\times 
\left(
 \frac{(q^{1-r +\frac{|{m_i^{(I)}} - {m_j^{(I)}}|}{2}} a^{-1} {s_i^{(I)}}^{\pm} {s_j^{(I)}}^{\mp}; q)_{\infty}}
 {(q^{r+ \frac{|{m_i^{(I)}} - {m_j^{(I)}}|}{2}} a {s_i^{(I)}}^{\pm} {s_j^{(I)}}^{\mp}; q)_{\infty}}  
 \frac{(q^{1-r + \frac{|{m_i^{(I)}} + {m_j^{(I)}}|}{2}} a^{-1} {s_i^{(I)}}^{\mp} {s_j^{(I)}}^{\mp}; q)_{\infty}}
 {(q^{r+\frac{|{m_i^{(I)}} + {m_j^{(I)}}|}{2}} a {s_i^{(I)}}^{\pm} {s_j^{(I)}}^{\pm}; q)_{\infty}} \right) \nonumber \\
 & \times \left( \left( \prod_{i=1}^{n_I}
 \frac{(q^{1-r +\frac{|m_i^{(I)}|}{2}} a^{-1} \chi_I {s_i^{(I)}}^{\pm}; q)_{\infty}}
 {(q^{r+ \frac{|{m_i^{(I)}}|}{2}} a \chi_I {s_i^{(I)}}^{\pm}; q)_{\infty}} \right)
 \frac{(q^{1-r} a^{-1}; q)_{\infty}}
 {(q^{r} a; q)_{\infty}} \right)^{\epsilon_I} \\
&Z_{Sym \; SU(N_F)}(r, a)  = \prod_{\alpha \le \beta}^{N_F} \frac{(q^{1 - r} a^{-1} x_{\alpha}^{-1} x_{\beta}^{-1}; q)_{\infty}}{(q^{r} a x_{\alpha} x_{\beta}; q)_{\infty}}
\end{align}
\endgroup
with the prescription that for the case of $\chi_I = -1$ and $\epsilon_I = 0$
we must replace every occurrence of $s_{n_I}^{(I) \pm}$ with simply $\pm 1$ and
set $m_{n_I}^{(I)} = 0$~\footnote{See \cite{Aharony:2013kma}
for full details and discussion of how to describe various orthogonal groups with different global structure.}.
So in particular we have
\begingroup
\begin{align}
Z_B = & Z_{BiFund \; SO_I-SO_J}(r_B, a_1) \\
Z^{N_f}_Q = & Z_{N_f Fund \; I}(r_Q, a_2) \\
Z_b = & Z_{BiFund \; SO_I-SO_J}(1 - r_B, a_1^{-1}) \\
Z_{\phi_b} = & Z_{Sym \; SO_I}(2r_B, a_1^2) \\
Z_{\sigma_B} = & Z_{Singlet}(r_{\sigma_B}, a_1^{-N_2}) \\
Z_c = & Z_{BiFund \; SO_I-SO_J}(1 - r_B, a_1^{-1}) \\
Z_{\phi_c} = & Z_{Sym \; SO_I}(2r_B, a_1^2) \\
Z^{N_f}_p = & Z_{BiFund \; SO_I-SO_J}(r_B + r_Q, a_1 a_2) \\
Z^{N_f}_M = & Z_{Sym \; SU(N_F)}(2r_Q, a_2^2) \\
Z^{N_f}_q = & Z_{BiFund \; SO_I-SO_J}(1 - r_Q, a_2^{-1}) \\
Z_{\sigma_C} = & Z_{Singlet}(r_{\sigma_C}, a_1^{-N_1} a_2^{-N_f})
\end{align}
\endgroup
where now $r_{\sigma_B} \equiv (1-r_B)N_2 - (N_1 - 2)$ and
$r_{\sigma_C} \equiv (1-r_B)N_1 + (1-r_Q)N_f - (N_2 - 2)$.

\subsection{Boundary 't Hooft anomalies}
The gauge anomaly cancellation and anomaly matching works in a similar way to the symplectic gauge theories.
For the multiplets we have discussed the contributions to the anomaly polynomial
are given by the following expressions \cite{Dimofte:2017tpi, Okazaki:2021pnc}
if we have Dirichlet boundary conditions.
For Neumann boundary conditions we just take the opposite sign.
Taking the multiplets to have R-charge $q_R$ and a vector of $U(1)_{a_i}$
charges $\underline{q}$ we have
\begin{align}
\Acal_{VM \; SO_I} = & -(N_I - 2) \Tr({\bf s_I}^2) - \frac{N_I(N_I - 1)}{4} {\bf r}^2 \\
\Acal_{N_f \; Fund \; SO_I}(q_R, \underline{q}) = & \frac{N_I}{2} \Tr({\bf x}^2) + N_f \Tr({\bf s_I}^2) + \frac{N_I N_f}{2} \left( \underline{q} \cdot \underline{{\bf a}} + (q_R - 1){\bf r} \right)^2 \\
\Acal_{BiFund \; SO_I-SO_J}(q_R, \underline{q}) = & N_I \Tr({\bf s_J}^2) + N_J \Tr({\bf s_I}^2) + \frac{N_I N_J}{2} \left( \underline{q} \cdot \underline{{\bf a}} + (q_R - 1){\bf r} \right)^2 \\
\Acal_{Sym \; SU(N_f)}(q_R, \underline{q}) = & \frac{N_f + 2}{2} \Tr({\bf x}^2) + \frac{N_f(N_f + 1)}{4} \left( \underline{q} \cdot \underline{{\bf a}} + (q_R - 1){\bf r} \right)^2 \\
\Acal_{Sym \; SO_I}(q_R, \underline{q}) = & (N_I + 2) \Tr({\bf s_I}^2) + \frac{N_I(N_I + 1)}{4} \left( \underline{q} \cdot \underline{{\bf a}} + (q_R - 1){\bf r} \right)^2
\end{align}
where here ${\bf s}_I$ is the $SO(N_I)$ field strength.

We propose a triality of the following sets of boundary conditions:
\begin{itemize}
\item $(\Ncal, \Ncal, N, N)$ for
$(\mathrm{VM}_1, \mathrm{VM}_2, B, Q)$ in theory A.
\item $(\Dcal, \Ncal, D, N, N, D)$ for 
$(\widetilde{\textrm{VM}}_1, \mathrm{VM}_2, b, \phi_b, Q, \sigma_B)$ in theory B.
\item $(\Ncal, \Dcal, D, N, N, N, D, D)$ for 
$(\mathrm{VM}_1, \widetilde{\textrm{VM}}_2, c, \phi_c, p, M, q, \sigma_C)$ in theory C.
\end{itemize}
We will see that with suitable additional 2d boundary multiplets, we can
cancel the gauge anomalies for the gauge group factors with Neumann boundary
conditions, and match the anomalies. We can then check the matching of the
half-indices.

Explicitly, we find the following anomaly polynomials for the bulk fields.
\begin{align}
\Acal^{A \; Bulk}_{\Ncal, \Ncal, N, N} = &
  - \frac{1}{2}N_1 N_2 {\bf a}_1^2 - \frac{1}{2}N_2 N_f {\bf a}_2^2 + N_1 N_2 (1 - r_B){\bf a}_1 {\bf r
} + N_2 N_f (1 - r_Q){\bf a}_2 {\bf r} \nonumber \\
 & + \frac{1}{4}\left( (N_1 - N_2)^2 + 2N_1 N_2 r_B(2 - r_B) - 2N_2 N_f (1 - r_Q)^2 - (N_1 + N_2) \right) {\bf r}^2 \nonumber \\ 
 & - (N_2 - N_1 + 2) \Tr({\bf s}_1^2) - (N_1 + N_f - N_2 + 2)\Tr({\bf s}_2^2) - \frac{1}{2}N_2 \Tr({\bf x}^2) \\
\Acal^{B \; Bulk}_{\Dcal, \Ncal, D, N, N, D} = &
  - \frac{1}{2}N_1 N_2 {\bf a}_1^2 - \frac{1}{2}N_2 N_f {\bf a}_2^2 + N_1 N_2 (1 - r_B){\bf a}_1 {\bf r
} + N_2 N_f (1 - r_Q){\bf a}_2 {\bf r} \nonumber \\
 & + \frac{1}{4} \left( (N_1 - N_2)^2 + 2N_1 N_2 r_B(2 - r_B) - 2N_2 N_f (1 - r_Q)^2 - (N_1 + N_2) \right) {\bf r}^2 \nonumber \\
 & - (N_1 + N_f - N_2 + 2) \Tr({\bf s}_2^2) + N_1 \Tr(\tilde{\bf s}_1^2) - \frac{1}{2} N_2 \Tr({\bf x}^2) \\
\Acal^{C \; Bulk}_{\Ncal, \Dcal, D, N, N, N, D, D} = & 
  - \frac{1}{2} N_1 N_2 {\bf a}_1^2 - \frac{1}{2} N_2 N_f {\bf a}_2^2 + N_1 N_2 (1 - r_B){\bf a}_1 {\bf r
} + N_2 N_f (1 - r_Q){\bf a}_2 {\bf r} \nonumber \\
 & + \frac{1}{4} \left( (N_1 - N_2)^2 + 2N_1 N_2 r_B(2 - r_B) - 2N_2 N_f (1 - r_Q)^2 - (N_1 + N_2) \right) {\bf r}^2 \nonumber \\
 & - (N_2 - N_1 + 2) \Tr({\bf s}_1^2) + N_2 \Tr(\tilde{\bf s}_2^2) - \frac{1}{2} N_2 \Tr({\bf x}^2)
\end{align}

Recalling that $\tilde{N_1} = N_2 - N_1 + 2$ and $\tilde{N_2} = N_1 - N_2 + N_f + 2$, and that due to
the $\Dcal$ boundary conditions $SO(\tilde{N}_1)$ and $SO(\tilde{N}_2)$ are global symmetries on
the boundary, whereas $SO(N_1)$ and $SO(N_2)$ are gauge symmetries due to the
$\Ncal$ boundary conditions, we can cancel all gauge anomalies with the
following 2d multiplets.
\begin{align}
\label{SO_charges}
\begin{array}{c|c|c|c|c|c}
 & & SO(N_1) & SO(N_2) & SO(\tilde{N}_1) & SO(\tilde{N}_2) \\ \hline
\mathrm{Fermi} & \Gamma_{1} & {\bf N_1} & {\bf 1} & {\bf \tilde{N}_1} & {\bf 1} \\
\mathrm{Fermi} & \Gamma_{2} & {\bf 1} & {\bf N_2} & {\bf 1} & {\bf \tilde{N}_2} \\
\end{array}
\end{align}
In particular, we need to include
\begin{itemize}
\item $\Gamma_{1}$, $\Gamma_{2}$ in theory A.
\item $\Gamma_{2}$ in theory B.
\item $\Gamma_{1}$ in theory C.
\end{itemize}

Including the contribution of those 2d multiplets in each theory all gauge anomalies are cancelled and the resulting anomaly polynomials match
\begin{align}
\Acal^{Total} = &
 - \frac{1}{2} N_1 N_2 {\bf a}_1^2 - \frac{1}{2} N_2 N_f {\bf a}_2^2 + N_1 N_2 (1 - r_B){\bf a}_1 {\bf r
} + N_2 N_f (1 - r_Q){\bf a}_2 {\bf r} \nonumber \\
 & + \frac{1}{4} \left( (N_1 - N_2)^2 + 2N_1 N_2 r_B(2 - r_B) - 2N_2 N_f (1 - r_Q)^2 - (N_1 + N_2) \right) {\bf r}^2 \nonumber \\
 & + N_1 \Tr(\tilde{\bf s}_1^2) + N_2 \Tr(\tilde{\bf s}_2^2) - \frac{1}{2} N_2 \Tr({\bf x}^2)
\end{align}

It is possible to consider other boundary conditions. The obvious case is to switch all boundary conditions so we have all Dirichlet in theory A. As noted for
the symplectic case, this will
simply change the sign of the bulk contribution to the anomaly polynomial and
it is easy to see that all gauge anomalies will be cancelled and the anomalies
will match if we include 2d Fermi multiplets
\begin{itemize}
\item None in theory A.
\item $\Gamma_{1}$ in theory B.
\item $\Gamma_{2}$ in theory C.
\end{itemize}
Again, we do not present any examples of these boundary conditions but the dualities should hold for these boundary conditions as was found for theories with a single gauge node with orthogonal group \cite{Okazaki:2021pnc}.

\subsection{Half-indices}
We now give the expressions for the half-indices with the previously chosen boundary conditions following from all Neumann in theory A. For the three theories we have
\begin{align}
\II^A_{\Ncal, \Ncal, N, N} = & \II^{VM \; SO_1}_{\Ncal} \II^{VM \; SO_2}_{\Ncal} \II^B_N \II^Q_N
 \\
\II^B_{\Dcal, \Ncal, D, N, N, D} = & \II^{VM \; SO_{\tilde{1}}}_{\Dcal} \II^{VM \; SO_2}_{\Ncal} \II^b_D \II^{\phi_b}_N \II^Q_N \II^{\sigma_B}_D
 \\
\II^C_{\Ncal, \Dcal, D, N, N, N, D, D} = & \II^{VM \; SO_1}_{\Ncal} \II^{VM \; SO_{\tilde{2}}}_{\Dcal} \II^c_D \II^{\phi_c}_N \II^p_N \II^M_N \II^q_D \II^{\sigma_C}_D
\end{align}
where
\begin{align}
\II^{VM \; SO_I}_{\Ncal} = & \frac{(q)_{\infty}^{n_I}}{2^{n_I + \epsilon_I - 1} n_I!} \left( \prod_{i=1}^{n_I} \oint \frac{ds_i^{(I)}}{2\pi i s_i^{(I)}} \right)
 \left( \prod_{i \ne j}^{n_I} (s_i^{(I)} s_j^{(I)-1}; q)_{\infty} \right) \nonumber \\
 & \times \left( \prod_{i < j}^{n_I} (s_i^{(I)\pm} s_j^{(I)\pm}; q)_{\infty} \right)
 \left( \prod_{i = 1}^{n_I} (\chi_I s_i^{(I)\pm}; q)_{\infty} \right)^{\epsilon_I} \\
\II^{VM \; SO_I}_{\Dcal} = & \frac{1}{(q)_{\infty}^{n_I}} \sum_{m_i^{(I)} \in \Zb^{n_I}} 
\frac{(-1)^{k_{eff \; I} \sum_{i=1}^{n_I} m_i^{(I)}} (\zeta_I \chi_I)^{\sum_{i=1}^{n_I} m_i^{(I)}} q^{\frac{k_{eff \; I}}{2} \sum_{i=1}^{n_I} {m_i^{(I)}}^2} (\prod_i^{n_I} {u_i^{(I)}}^{k_{eff \; I} m_i^{(I)}})}
{\left( \prod_{i \ne j}^{n_1} (q^{1 + m_i^{(I)} - m_j^{(I)}} u_i^{(I)} {u_j^{(I)}}^{-1}; q)_{\infty} \right) \left( \prod_{i < j} (q^{1 \pm (m_i^{(I)} + m_j^{(I)})} u_i^{(I) \pm} u_j^{(I) \pm}; q)_{\infty} \right)} \nonumber \\
 & \times \frac{1}{\left( \prod_{i = 1}^{n_1} (q^{1 \pm m_i^{(I)}} \chi_I {u_i^{(I)}}^{\pm}; q)_{\infty} \right)^{\epsilon_I}} \\
k_{eff \; I} = & \tilde{N}_I
\end{align}
except for the case of $\chi_I = -1$ and $\epsilon_I = 0$ when instead
\begin{align}
\II^{VM \; SO_I}_{\Ncal} = & \frac{(q)_{\infty}^{n_I - 1} (-q; q)_{\infty}}{2^{n_I - 1} (n_I - 1)!} \left( \prod_{i=1}^{n_I - 1} \oint \frac{ds_i^{(I)}}{2\pi i s_i^{(I)}} \right)
 \left( \prod_{i \ne j}^{n_I - 1} (s_i^{(I)} s_j^{(I)-1}; q)_{\infty} \right) \nonumber \\
 & \times \left( \prod_{i < j}^{n_I - 1} (s_i^{(I)\pm} s_j^{(I)\pm}; q)_{\infty} \right)
 \left( \prod_{i = 1}^{n_I - 1} (s_i^{(I)\pm}; q)_{\infty} (-s_i^{(I)\pm}; q)_{\infty} \right) \\
\II^{VM \; SO_I}_{\Dcal} = & \frac{1}{(q)_{\infty}^{n_I - 1} (-q; q)_{\infty}} \sum_{m_i^{(I)} \in \Zb^{n_I - 1}}
\frac{1}{\left( \prod_{i = 1}^{n_1} (q^{1 \pm m_i^{(I)}} {u_i^{(I)}}^{\pm}; q)_{\infty} 
 (-q^{1 \pm m_i^{(I)}} {u_i^{(I)}}^{\pm}; q)_{\infty} \right)} \nonumber \\
 & \times \frac{(-1)^{(k_{eff \; I} + 1) \sum_{i=1}^{n_I - 1} m_i^{(I)}} \zeta_I^{\sum_{i=1}^{n_I - 1} m_i^{(I)}} q^{\frac{k_{eff \; I}}{2} \sum_{i=1}^{n_I} {m_i^{(I)}}^2} (\prod_i^{n_I} {u_i^{(I)}}^{k_{eff \; I} m_i^{(I)}})}
{\left( \prod_{i \ne j}^{n_1} (q^{1 + m_i^{(I)} - m_j^{(I)}} u_i^{(I)} {u_j^{(I)}}^{-1}; q)_{\infty} \right) \left( \prod_{i < j} (q^{1 \pm (m_i^{(I)} + m_j^{(I)})} u_i^{(I) \pm} u_j^{(I) \pm}; q)_{\infty} \right)}
\end{align}
Note that $N_{\tilde{I}} = \tilde{N}_I$ so $\tilde{N}_{\tilde{I}} = N_I$.
We also have the prescription that for Dirichlet boundary conditions for the
vector multiplet we make the replacement
$s_i^{(I)} \rightarrow q^{m_i^{(I)}} u_i^{(I)}$ in the following matter
contributions.

The 3d matter contributions are given, for either Dirichlet or Neumann boundary conditions by
\begingroup
\begin{align}
\II^B = & \II^{BiFund \; SO_I-SO_J}(r_B, a_1) \\
\II^Q = & \II^{N_f Fund \; SO_I}(r_Q, a_2) \\
\II^b = & \II^{BiFund \; SO_I-SO_J}(1 - r_B, a_1^{-1}) \\
\II^{\phi_b} = & \II^{Sym \; SO_I}(2r_B, a_1^2) \\
\II^{\sigma_B} = & \II^{Singlet}(r_{\sigma_B}, a_1^{-N_2}) \\
\II^c = & \II^{BiFund \; SO_I-SO_J}(1 - r_B, a_1^{-1}) \\
\II^{\phi_c} = & \II^{Sym \; SO_I}(2r_B, a_1^2) \\
\II^p = & \II^{BiFund \; SO_I-SO_J}(r_B + r_Q, a_1 a_2) \\
\II^M = & \II^{Sym \; SU(N_F)}(2r_Q, a_2^2) \\
\II^q = & \II^{BiFund \; SO_I-SO_J}(1 - r_Q, a_2^{-1}) \\
\II^{\sigma_C} = & \II^{Singlet}(r_{\sigma_C}, a_1^{-N_1} a_2^{-N_f})
\end{align}
\endgroup

For Neumann boundary conditions the 3d matter contributions are
\begingroup
\allowdisplaybreaks
\begin{align}
\II^{BiFund \; SO_I-SO_J}_N(r,a) = & \left( \prod_{i=1}^{n_I} \prod_{j=1}^{n_J} 
\frac{1}{(q^{\frac{r}{2}} a s_i^{(I)\pm} {s_j^{(J)}}^{\mp} ; q)_{\infty} 
(q^{\frac{r}{2}} a s_i^{(I)\pm} {s_j^{(J)}}^{\pm} ; q)_{\infty}} \right) \nonumber \\
 & \times \left( \prod_{j=1}^{n_J} \frac{1}{(q^{\frac{r}{2}} a \chi_I {s_j^{(J)}}^{\pm} ; q)_{\infty}} \right)^{\epsilon_I}
 \left( \prod_{i=1}^{n_I} \frac{1}{(q^{\frac{r}{2}} a \chi_J {s_i^{(I)}}^{\pm} ; q)_{\infty}} \right)^{\epsilon_J} \nonumber \\
 & \times \left( \frac{1}{(q^{\frac{r}{2}} a \chi_I \chi_J; q)_{\infty}} \right)^{\epsilon_I \epsilon_J} \\
\II^{N_f Fund \; SO_I}_{N}(r,a) = & \prod_{\alpha=1}^{N_f} \left( \prod_{i=1}^{n_I}
\frac{1}{(q^{\frac{r}{2}} a {s_i^{(I)}}^{\pm} x_{\alpha} ; q)_{\infty}} \right)
 \left( \frac{1}{(q^{\frac{r}{2}} a \chi_I x_{\alpha}; q)_{\infty}} \right)^{\epsilon_I} \\
\II^{Sym \; SO_I}_{N}(r,a) = & \left( \prod_{i,j=1}^{n_I} \frac{1}{(q^{\frac{r}{2}} a {s_i^{(I)}} {s_j^{(I)}}^{-1} ; q)_{\infty}} \right) 
 \left( \prod_{i \le j}^{n_I} \frac{1}{(q^{\frac{r}{2}} a {s_i^{(I)}}^{\pm} {s_j^{(I)}}^{\pm} ; q)_{\infty}} \right) \nonumber \\
 & \times \left( \frac{1}{(q^{\frac{r}{2}} a; q)_{\infty} \prod_{i=1}^{n_I} (q^{\frac{r}{2}} a \chi_I {s_i^{(I)}}^{\pm}; q)_{\infty}} \right)^{\epsilon_I} \\
\II^{Sym \; SU(N_f)}_{N}(r,a) = & \prod_{\alpha \le \beta}^{N_f} 
\frac{1}{(q^{\frac{r}{2}} a x_{\alpha} x_{\beta} ; q)_{\infty}}
\end{align}
\endgroup
while for Dirichlet boundary conditions the 3d matter contributions are
\begingroup
\allowdisplaybreaks
\begin{align}
\II^{BiFund \; SO_I-SO_J}_D(r,a) = & \left( \prod_{i=1}^{n_I} \prod_{j=1}^{n_J} 
(q^{1 - \frac{r}{2}} a^{-1} s_i^{(I)\pm} {s_j^{(J)}}^{\mp} ; q)_{\infty} 
(q^{1 - \frac{r}{2}} a^{-1} s_i^{(I)\pm} {s_j^{(J)}}^{\pm} ; q)_{\infty} \right) \nonumber \\
 & \times \left( \prod_{j=1}^{n_J} (q^{1 - \frac{r}{2}} a^{-1} \chi_I {s_j^{(J)}}^{\pm} ; q)_{\infty} \right)^{\epsilon_I}
 \left( \prod_{i=1}^{n_I} (q^{1 - \frac{r}{2}} a^{-1} \chi_J {s_i^{(I)}}^{\pm} ; q)_{\infty} \right)^{\epsilon_J} \nonumber \\
 & \times (q^{1 - \frac{r}{2}} a^{-1} \chi_I \chi_J; q)_{\infty}^{\epsilon_I \epsilon_J} \\
\II^{N_f Fund \; SO_I}_{D}(r,a) = & \prod_{\alpha=1}^{N_f} \left( \prod_{i=1}^{n_I}
(q^{1 - \frac{r}{2}} a^{-1} {s_i^{(I)}}^{\pm} x_{\alpha} ; q)_{\infty} \right)
 \left( (q^{1 - \frac{r}{2}} a^{-1} \chi_I x_{\alpha}; q)_{\infty} \right)^{\epsilon_I} \\
\II^{Sym \; SO_I}_{D}(r,a) = & \left( \prod_{i,j=1}^{n_I} (q^{1 - \frac{r}{2}} a^{-1} {s_i^{(I)}} {s_j^{(I)}}^{-1} ; q)_{\infty} \right) 
 \left( \prod_{i \le j}^{n_I} (q^{1 - \frac{r}{2}} a^{-1} {s_i^{(I)}}^{\pm} {s_j^{(I)}}^{\pm} ; q)_{\infty} \right) \nonumber \\
 & \times \left( (q^{1 - \frac{r}{2}} a^{-1}; q)_{\infty} \prod_{i=1}^{n_I} (q^{1 - \frac{r}{2}} a^{-1} \chi_I {s_i^{(I)}}^{\pm}; q)_{\infty} \right)^{\epsilon_I} \\
\II^{Sym \; SU(N_f)}_{D}(r,a) = & \prod_{\alpha \le \beta}^{N_f} 
(q^{1 - \frac{r}{2}} a^{-1} x_{\alpha} x_{\beta} ; q)_{\infty}
\end{align}
\endgroup

Finally, the contributions from the 2d matter multiplets are
\begin{align}
I^{\Gamma_{I}} = & \left( \prod_{i=1}^{n_I} \prod_{j=1}^{\tilde{n}_I} 
(q^{\frac12} s_i^{(I) \pm} u_j^{(I) \pm}; q)_{\infty} (q^{\frac12} s_i^{(I) \pm} u_j^{(I) \mp}; q)_{\infty} \right) \nonumber \\
 & \times \left( \prod_{j=1}^{\tilde{n}_I} (q^{\frac12} \tilde{\chi}_I u_j^{(I) \pm}; q)_{\infty} \right)^{\epsilon_{\tilde{I}}}
 \left( \prod_{i=1}^{n_I} (q^{\frac12} \chi_I s_i^{(I) \pm}; q)_{\infty} \right)^{\epsilon_I}
 (q^{\frac12} \chi_I \tilde{\chi}_I; q)_{\infty}^{\epsilon_I \epsilon_{\tilde{I}}}
\end{align}
and we recall that $\tilde{\chi}_I = \chi_I \zeta_I$.

As for the full indices, all the above 3d and 2d matter contributions are
modified in the case of $\chi_I = -1$ and $\epsilon_I = 0$ by
replacing every occurrence of $s_{n_I}^{(I) \pm}$ with simply $\pm 1$, and
similarly for the dual group in the case of the 2d Fermi.

\subsection{$SO(3)\times SO(4)-[3]$ $(N_1 = 3$, $N_2 = 4$, $N_f = 3)$}
Here $\tilde{N}_1 = 3$ and $\tilde{N}_2 = 4$ so both dual theories also have gauge group $SO(3) \times SO(4)$.

The indices for the dual pair actually coincide. 
For $r_B=1/2$ and $r_Q=1/3$ we have the
full-indices~\footnote{This example is also presented in \cite{Benvenuti:2020wpc} but we
include a complete identification of the operators counted by the indices.}
\begin{align}
&I^A = I^B = I^C
\nonumber\\
=&1+\underbrace{6 a_2^2 q^{1/3}}_{\Tr (QQ)}
+(\underbrace{a_1^2}_{\Tr (BB)}+\underbrace{\frac{1}{a_1^4}}_{v_A^{+,0}}) q^{1/2}
+\underbrace{21 a_2^4 q^{2/3}}_{\Tr(QQ)^2}
+( \underbrace{\frac{1}{a_1^3 a_2^3}}_{v_A^{0,+}} 
+\underbrace{\frac{1}{a_1^5 a_2^3}}_{v_A^{+,+}} ) q^{3/4}
 +6 (
 \underbrace{2a_1^2}_{\begin{smallmatrix} \Tr(QQ) \Tr(BB), \\ \Tr(Q(BB)Q) \end{smallmatrix}}
 +\underbrace{\frac{1}{a_1^4}}_{v_A^{+,0}\Tr(QQ)} ) a_2^2 q^{5/6}
   \nonumber\\
   &
   +3 (\underbrace{a_1^3}_{\epsilon_1 \epsilon_2 B^{3} Q}
   +\underbrace{\frac{1}{a_1^3}}_{v_A^{-,0}\epsilon_1 B Q}) a_2 q^{11/12} 
   +q (\underbrace{2a_1^4}_{\begin{smallmatrix} \Tr(BB)^2, \\ \Tr(BBBB) \end{smallmatrix}}
   +\underbrace{\frac{1}{a_1^2}}_{v_{A}^{+,0}\Tr(BB)}
   +\underbrace{\frac{1}{a_1^8}}_{v_A^{\pm2,0}}
   +\underbrace{56 a_2^6}_{\Tr(QQ)^3}
   -\underbrace{10}_{\begin{smallmatrix}\Tr(Q\psi_Q) \\ \Tr(B\psi_B)
   \end{smallmatrix}} )
   \nonumber\\
   &
   + 9 ( \underbrace{\frac{1}{a_1^3a_2}}_{\begin{smallmatrix} v_A^{0,+}\Tr(QQ), \\ v_A^{0,-}\epsilon_2Q^{2} \end{smallmatrix}}
   +\underbrace{\frac{1}{a_1^5 a_2}}_{\begin{smallmatrix} v_A^{+,+}\Tr(QQ), \\v_A^{+,-}\epsilon_2Q^{2} \end{smallmatrix}} ) q^{13/12} 
   + q^{7/6} (3 ( 
   \underbrace{20a_1^2 a_2^4}_{\begin{smallmatrix} \Tr(QQ)^2\Tr(BB), \\ \Tr(QQ) \Tr(Q(BB)Q), \\ (\epsilon_2 BQQQ) \Tr(BQ) \end{smallmatrix}}
   +\underbrace{\frac{7a_2^4}{a_1^4}}_{v_A^{+,0}\Tr(QQ)^2} )
   +\underbrace{\frac{6}{a_1^4 a_2^2}}_{\begin{smallmatrix} v_A^{-,+}\epsilon_1B Q,\\ 
   v_A^{-,-}\epsilon_1 \epsilon_2 BQ \end{smallmatrix}}) 
   \nonumber\\
  & + q^{5/4} (\underbrace{19 a_1^{3}a_2^3}_{\begin{smallmatrix} (\epsilon_1\epsilon_2 BBBQ) \Tr(QQ), \\ \epsilon_1 \Tr(BQ)^3 \end{smallmatrix}}
 + \frac{1}{a_1^3 a_2^3} 
  (\underbrace{19a_2^6}_{\begin{smallmatrix} v_A^{-,0} \epsilon_1 B Q \Tr(QQ), \\ v_A^{-,0} \epsilon_1 \epsilon_2 B Q^{3} \end{smallmatrix}}
 + \underbrace{2}_{v_A^{+,+} \Tr(BB)} )
  +\underbrace{\frac{1}{a_1a_2^3}}_{v_A^{0,+} \Tr(BB)}
  +\underbrace{\frac{1}{a_1^9 a_2^3}}_{v_A^{+2,+}})
+\cdots
\end{align}
We note the coefficient $2$ of the $q^{5/4}/(a_1^3 a_2^3)$ term. This arises
from the contributions of two different dressed monopole operators
schematically written $v_A^{+,+} \Tr(BB)$. The main point is that for each
$SO(N)$ group the monopole flux breaks the gauge symmetry so that (loosely speaking) a fundamental splits into a ``massive'' fundamental of $SO(2)$ due to the flux and a ``massless'' $SO(N-2)$ fundamental following the description in \cite{Cremonesi:2015dja}. For a bifundamental of
$SO(N_1) \times SO(N_2)$ we will get a massless bifundamental of
$SO(N_1 - 2) \times SO(N_2 - 2)$. However, in the case where the monopole has
equal magnitude fluxes for both groups, we can also get a massless
bifundamental of $SO(2) \times SO(2)$ due to cancellation of the flux
contributions to the mass. Hence, in the $v_A^{+,+}$ background there are
two distinct gauge invariant operators by dressing the monopole with the
traces of the squares of these two massless bifundamentals arising from B
after symmetry breaking.~\footnote{We thank Stefano Cremonesi for explaining
this point.}

Above we have taken the discrete fugacities $\zeta_I = 1$ and $\chi_I = 1$.
The duality holds for other values and the indices have a similar expansion
and identification of operators. E.g.\ if we take $\zeta_1 = -1$ the effect
is simply to include a factor $-1$ with each of the monopole operators with
flux $\pm$ for the $SO(3)$ gauge node. For simplicity here and in the following
examples we list only the results with all discrete fugacities set to $1$.

The half-indices are given by
\begin{align}
&\II^A_{\mathcal{N},\mathcal{N}} 
=\II^B_{\mathcal{D},\mathcal{N}}  
=\II^C_{\mathcal{N},\mathcal{D}}
\nonumber\\
=&
1+\underbrace{6 a_2^2 q^{1/3}}_{\Tr(QQ)}
+\underbrace{a_1^2 q^{1/2}}_{\Tr(BB)}
+q^{2/3} (
\underbrace{21 a_2^4}_{\Tr(QQ)^2} - \underbrace{3 a_2 \left( w_1 + \frac{1}{w_1} + w_2 + \frac{1}{w_2} \right) }_{Q \Gamma_2})
   +\underbrace{12 a_1^2 a_2^2 q^{5/6}}_{\begin{smallmatrix} \Tr(QQ) \Tr(BB), \\ \Tr(Q(BB)Q) \end{smallmatrix}}
   \nonumber\\
   &+q^{11/12} (\underbrace{3 a_1^3 a_2}_{\epsilon_1 \epsilon_2 B^3 Q}
 - \underbrace{3 a_1 a_2 \left( v + \frac{1}{v} + 1 \right)}_{QB\Gamma_1})
  + q (\underbrace{2 a_1^4}_{\Tr(BB)^2}
 - \underbrace{19 a_2^3 \left( w_1 + \frac{1}{w_1} + w_2 - \frac{1}{w_2} \right)}_{\Tr(QQ) Q \Gamma_2, \; \epsilon_2 QQQ \Gamma_2}
   \nonumber\\
   &
   +\underbrace{56a_2^6}_{\Tr(QQ)^3}+ \underbrace{v+\frac{1}{v}+w_1 w_2+\frac{w_2}{w_1}+\frac{w_1}{w_2}+\frac{1}{w_1 w_2}+3}_{\Gamma_1 \Gamma_1, \; \Gamma_2 \Gamma_2})+\cdots
\end{align}

\subsection{$SO(N_1) \times SO(N_2)-[N_f] \times SO(N_3)$}
\label{sec_SOSOSO}
Now consider the case with three gauge nodes. 
Dualizing either left or right gauge node leads to the ones discussed in section \ref{sec_SOSO}
The Seiberg-like dual of the $SO(N_2)$ gauge node gives a $SO(\tilde{N}_2 = N_1 + N_3 + N_f - N_2 + 2)$ gauge node, 
all three gauge nodes have $N_f$ flavors, there are bifundamental chirals for each pair of gauge nodes, 
there are symmetric rank-$2$ chirals for $SO(N_1)$ and $SO(N_3)$, 
and a singlet chiral in the symmetric rank-$2$ representation of the flavor symmetry $SU(N_f)$. 

\begin{figure}
\centering
\scalebox{0.7}{
\begin{tikzpicture}
\path (-4,0) node[circle, minimum size=64, fill=green!10, draw](AG1) {$SO(N_1)$}
(0,0) node[circle, minimum size=64, fill=green!10, draw](AG2) {$SO(N_2)$}
(4,0) node[circle, minimum size=64, fill=green!10, draw](AG3) {$SO(N_3)$}
(0,4) node[minimum size=50, fill=cyan!10, draw](AF2) {$N_f$};
\draw (AG1) -- (AG2) node [midway, above] {$B_1$};
\draw (AG3) -- (AG2) node [midway, above] {$B_2$};
\draw (AG2) -- (AF2) node [midway, right] {$Q$};
\path (-10,-8) node[circle, minimum size=64, fill=green!10, draw](BG1) {$SO(\tilde{N_1})$}
(-6,-8) node[circle, minimum size=64, fill=green!10, draw](BG2) {$SO(N_2)$}
(-2,-8) node[circle, minimum size=64, fill=green!10, draw](BG3) {$SO(N_3)$}
(-6,-4) node[minimum size=50, fill=cyan!10, draw](BF2) {$N_f$};
\draw (BG1) -- node[above]{$b_1$} (BG2);
\draw (BG3) -- node[above]{$B_2$} (BG2);
\draw (BG2) -- node[right]{$Q$}(BF2);
\draw (BG2.south west) .. controls ++(-1, -2) and ++(1, -2) .. node[below]{$\phi_b$} (BG2.south east);
%
%
\path (2,-8) node[circle, minimum size=64, fill=green!10, draw](DG1) {$SO(N_1)$}
(6,-8) node[circle, minimum size=64, fill=green!10, draw](DG2) {$SO(N_2)$}
(10,-8) node[circle, minimum size=64, fill=green!10, draw](DG3) {$SO(\tilde{N_3})$}
(6,-4) node[minimum size=50, fill=cyan!10, draw](DF2) {$N_f$};
\draw (DG1) -- node[above]{$B_1$} (DG2);
\draw (DG3) -- node[above]{$b_2$} (DG2);
\draw (DG2) -- node[right]{$Q$}(DF2);
\draw (DG2.south west) .. controls ++(-1, -2) and ++(1, -2) .. node[below]{$\phi_d$} (DG2.south east);
%
\path (-4,-18) node[circle, minimum size=64, fill=green!10, draw](CG1) {$SO(N_1)$}
(0,-18) node[circle, minimum size=64, fill=green!10, draw](CG2) {$SO(\tilde{N_2})$}
(4,-18) node[circle, minimum size=64, fill=green!10, draw](CG3) {$SO(N_3)$}
(0,-14) node[minimum size=50, fill=cyan!10, draw](CF2) {$N_f$};
\draw (CG1) -- node[above]{$c_1$} (CG2);
\draw (CG3) -- node[above]{$c_2$} (CG2);
\draw (CG1) -- node[right]{$p_1$}(CF2);
\draw (CG3) -- node[right]{$p_2$}(CF2);
\draw (CG2) -- node[right]{$q$}(CF2);
\draw (CG1) to [out=-45, in=-135] node[below]{$c_3$}(CG3);
\draw (CG1.south west) .. controls ++(-1, -2) and ++(1, -2) .. node[below]{$\phi_c^1$} (CG1.south east);
\draw (CG3.south west) .. controls ++(-1, -2) and ++(1, -2) .. node[below]{$\phi_c^2$} (CG3.south east);
\draw (CF2.north east) .. controls ++(1, 2) and ++(-1, 2) .. node[below]{$M$} (CF2.north west);
\end{tikzpicture}
}
\caption{Quadrality of $SO(N_1) \times SO(N_2)-[N_f] \times SO(N_3)$ quiver 
where $\tilde{N_1}=N_2-N_1+2$, $2\tilde{N_2}=N_1+N_3+N_f-2N_2+2$ and $\tilde{N_3}=N_2-N_3+2$.
All chirals $\phi_b$, $\phi_c^1$, $\phi_c^2$, $\phi_d$ and $M$ are in symmetric rank-$2$
representations.} \label{fig:SO_SO_SO}
\end{figure}
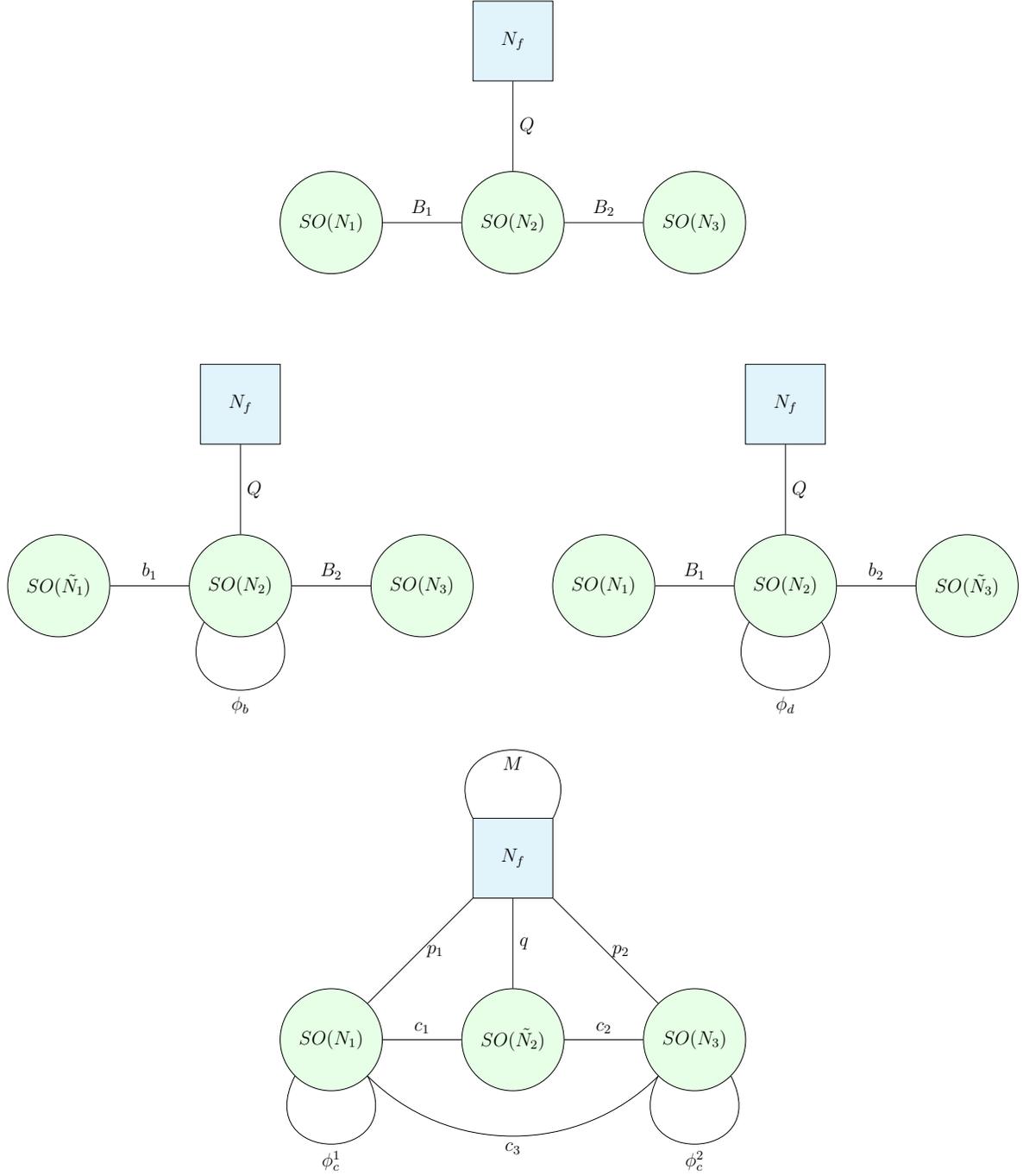

We propose the quadrality of the $SO\times SO\times SO$ quivers in Figure \ref{fig:SO_SO_SO}. 
The monopole operators can be understood by generalizing the results for the case of two gauge nodes. We now have possible fluxes for each of the three gauge nodes. Monopole operators without any negative fluxes are gauge invariant so will contribute to the index and can be dressed with any gauge invariant operator. A monopole with flux $-1$ for gauge node $I$ must be combined with a baryonic combination of operators, specifically it will have a contraction with the antisymmetric tensor $\epsilon_I$ and dressed with an operator with $N_I - 2$ indices. In addition there are monopoles with more than one flux.
E.g.\ for any gauge node $SO(N)$ with $N \ge 4$ there are monopole operators with two non-zero fluxes for that node which need to be contracted with the antisymmetric tensor and dressed with an operator with $N-4$ free indices \cite{Benvenuti:2020wpc}. The mapping of monopole operators also generalizes naturally from the two gauge nodes case. As there are many options we do not list everything explicitly here.

\subsection{$SO(2)\times SO(4)-[1]-SO(2)$ $(N_1 = 2$, $N_2 = 4$, $N_3 = 2$, $N_f = 1)$}
\label{sec_SOSOSO_example}

Here $\tilde{N}_2 = 3$ so the dual theory has gauge group $SO(2) \times SO(3) \times SO(2)$.

As we conjecture, the supersymmetric indices precisely agree with each other. 
For $r_B=r_C=2/5$ and $r_Q=2/7$ we get the following full indices: 
\begingroup
\allowdisplaybreaks
\begin{align}
&I^A = I^B= I^C = I^D
\nonumber\\
&=1+\underbrace{a_2^2 q^{2/7}}_{\Tr(QQ)}
+q^{2/5} 
(\underbrace{a_1^2}_{\Tr(B_1B_1)}+\underbrace{a_3^2}_{\Tr(B_2B_2)})
+\underbrace{\frac{q^{39/70}}{a_1^2 a_2 a_3^2}}_{v_A^{0,+,0}}
+\underbrace{a_2^4 q^{4/7}}_{\Tr(QQ)^2}
+2 a_2^2 q^{24/35}
   (\underbrace{a_1^2}_{\begin{smallmatrix} \Tr(QQ) \Tr(B_1 B_1) \\ \Tr(Q B_1 B_1 Q) \end{smallmatrix}}
   +\underbrace{a_3^2}_{\begin{smallmatrix} \Tr(QQ) \Tr(B_2 B_2) \\ \Tr(Q B_2 B_2 Q) \end{smallmatrix}})
   \nonumber\\
   & + 2 q^{4/5} (
   \underbrace{a_1^4}_{\begin{smallmatrix} \Tr(B_1B_1)^2 \\ \Tr(B_1 B_1 B_1 B_1) \end{smallmatrix}}
   +\underbrace{2a_1^2 a_3^2}_{\begin{smallmatrix}\Tr(B_1B_1)\Tr(B_2B_2)\\ \Tr(B_1B_2B_2B_1) \\ \epsilon_1 \epsilon_3 (B_1 B_2) (B_1 B_2) \\ \epsilon_1 \epsilon_3 (B_1 B_1) (B_2 B_2) \end{smallmatrix}}
   +\underbrace{a_3^4}_{\begin{smallmatrix} \Tr(B_2B_2)^2 \\ \Tr(B_2 B_2 B_2 B_2)^2 \end{smallmatrix}})
   +\underbrace{\frac{a_2 q^{59/70}}{a_1^2 a_3^2}}_{v_A^{0,+,0}\Tr(QQ)}
   +\underbrace{a_2^6 q^{6/7}}_{\Tr(QQ)^3}
      \nonumber\\
   &
   + 2 q^{67/70}
   (
   \underbrace{\frac{1}{a_2 a_3^2}}_{\begin{smallmatrix} v_A^{0,+,0}B_1B_1, \\ v_A^{0,-,0}\epsilon_1B_1B_1 \end{smallmatrix}}
   +\underbrace{\frac{1}{a_2 a_1^2}}_{\begin{smallmatrix} v_A^{0,+,0}B_2B_2, \\ v_A^{0,-,0}\epsilon_3B_2B_2 \end{smallmatrix}} )
   +2 a_2^4 q^{34/35} (\underbrace{a_1^2}_{\begin{smallmatrix} \Tr(QQ)^2 \Tr(B_1B_1) \\ \Tr(QQ) \Tr(Q B_1 B_1 Q) \end{smallmatrix}}
   +\underbrace{a_3^2}_{\begin{smallmatrix} \Tr(QQ)^2 \Tr(B_2B_2) \\ \Tr(QQ) \Tr(Q B_2 B_2 Q) \end{smallmatrix}}) \nonumber \\
 & - \underbrace{5 q}_{\begin{smallmatrix}\Tr(Q\psi_Q)\\ \Tr(B_1 \psi_{B_1}), \; \epsilon_1 B_1 \psi_{B_1} \\ \Tr(B_2 \psi_{B_2}), \; \epsilon_1 B_2 \psi_{B_2} \end{smallmatrix}}
+a_2^2 q^{38/35} (\underbrace{13 a_1^2 a_3^2}_{\begin{smallmatrix} \Tr(QQ)\Tr(B_1B_1) \Tr(B_2B_2) , \\ \cdots \end{smallmatrix}}
 + \underbrace{5 a_1^4}_{\begin{smallmatrix} \Tr(QQ)\Tr(B_1B_1)^2 , \\ \Tr(QQ) \Tr(B_1 B_1 B_1B_1) , \\ \Tr(QB_1B_1Q)\Tr(B_1B_1) , \\ \Tr(QB_1B_1B_1B_1Q) , \\ (QB_1)\epsilon_1(B_1 B_1)\epsilon_1(B_1 Q) \end{smallmatrix}}
 + \underbrace{5 a_3^4}_{\begin{smallmatrix} \Tr(QQ)\Tr(B_2B_2)^2 , \\ \Tr(QQ) \Tr(B_2 B_2 B_2B_2) , \\ \Tr(QB_2B_2Q)\Tr(B_2B_2) , \\ \Tr(QB_2B_2B_2B_2Q) , \\ (QB_2)\epsilon_3(B_2 B_2)\epsilon_3(B_2 Q) \end{smallmatrix}} ) \nonumber \\
 & + \underbrace{\frac{q^{39/35}}{a_1^4 a_2^2 a_3^4}}_{v_A^{0,+2,0}}
+\underbrace{\frac{a_2^3 q^{79/70}}{a_1^2 a_3^2}}_{v_A^{0,+,0}\Tr(QQ)^2}
+\underbrace{a_2^8 q^{8/7}}_{\Tr(QQ)^4}
   +\cdots
\end{align}
\endgroup
where the counting can get a bit tedious, e.g.\ $(\epsilon_1 B_1 B_1)^2$ is
a linear combination of $\Tr(B_1B_1)^2$ and $\Tr(B_1B_1B_1B_1)$, and we have
not explicitly listed the $13$ different gauge invariant combinations of the
$QQB_1B_1B_2B_2$ operators.
Note that we only see monopole operators of the form $v_A^{0,*,0}$ to this order but others will contribute at higher order.

The half-indices are given by
\begin{align}
&\II^A_{\mathcal{N},\mathcal{N},\mathcal{N}} 
= \II^B_{\mathcal{D},\mathcal{N},\mathcal{N}} 
= \II^C_{\mathcal{N},\mathcal{D},\mathcal{N}} 
= \II^D_{\mathcal{N},\mathcal{N},\mathcal{D}} 
\nonumber\\
&=1
+\underbrace{a_2^2 q^{2/7}}_{\Tr(QQ)}
+q^{2/5} (\underbrace{a_1^2}_{\Tr(B_1B_1)}
+\underbrace{a_3^2}_{\Tr(B_2B_2)})
+\underbrace{a_2^4 q^{4/7}}_{\Tr(QQ)^2}
- \underbrace{\frac{a_2 q^{9/14} \left(w^2+w+1\right)}{w}}_{Q \Gamma_2}
   \nonumber\\
   &
   +2 a_2^2 q^{24/35}
   (\underbrace{a_1^2}_{\begin{smallmatrix} \Tr(QQ) \Tr(B_1 B_1) \\ \Tr(Q B_1 B_1 Q) \end{smallmatrix}}
   +\underbrace{a_3^2}_{\begin{smallmatrix} \Tr(QQ) \Tr(B_2 B_2) \\ \Tr(Q B_2 B_2 Q) \end{smallmatrix}})
 + 2 q^{4/5} (
   \underbrace{a_1^4}_{\begin{smallmatrix} \Tr(B_1B_1)^2 \\ (\epsilon_1 B_1 B_1)^2 \end{smallmatrix}}
   +\underbrace{2a_1^2 a_3^2}_{\begin{smallmatrix}\Tr(B_1B_1)\Tr(B_2B_2)\\ \Tr(B_1B_2B_2B_1) \\ \epsilon_1 \epsilon_2 (B_1 B_2) (B_1 B_2) \\ \epsilon_1 \epsilon_2 (B_1 B_1) (B_2 B_2) \end{smallmatrix}}
   +\underbrace{a_3^4}_{\begin{smallmatrix} \Tr(B_2B_2)^2 \\ (\epsilon_2 B_2 B_2)^2 \end{smallmatrix}})
      \nonumber\\
   &
- 2 q^{59/70} \left( \underbrace{\frac{\left(a_2 a_1 \left(v_1^2 v_2+v_1
   v_2^2+v_1+v_2\right)\right)}{v_1 v_2}}_{Q B_1 \Gamma_1}
  + \underbrace{\frac{\left(a_2 a_3 \left(u_1^2 u_2+u_1
   u_2^2+u_1+u_2\right)\right)}{u_1 u_2}}_{Q B_2 \Gamma_3} \right)
      \nonumber\\
   &
   +\underbrace{a_2^6 q^{6/7}}_{\Tr(QQ)^3}
- \underbrace{\frac{a_2^3 q^{13/14} \left(w^2+w+1\right)}{w}}_{\Tr(QQ) Q \Gamma_2}
+2 a_2^4 q^{34/35} (\underbrace{a_1^2}_{\begin{smallmatrix} \Tr(QQ)^2 \Tr(B_1B_1) \\ \Tr(QQ) \Tr(Q B_1 B_1 Q) \end{smallmatrix}}
   +\underbrace{a_3^2}_{\begin{smallmatrix} \Tr(QQ)^2 \Tr(B_2B_2) \\ \Tr(QQ) \Tr(Q B_2 B_2 Q) \end{smallmatrix}}) + \cdots
   \end{align}

\section{Ortho-symplectic linear quivers}
\label{sec_orthosym}
In this section we study the quiver gauge theories with both orthogonal and symplectic gauge groups.
\subsection{$SO(N_1) \times USp(2N_2)$}
\label{sec_SOUSp}
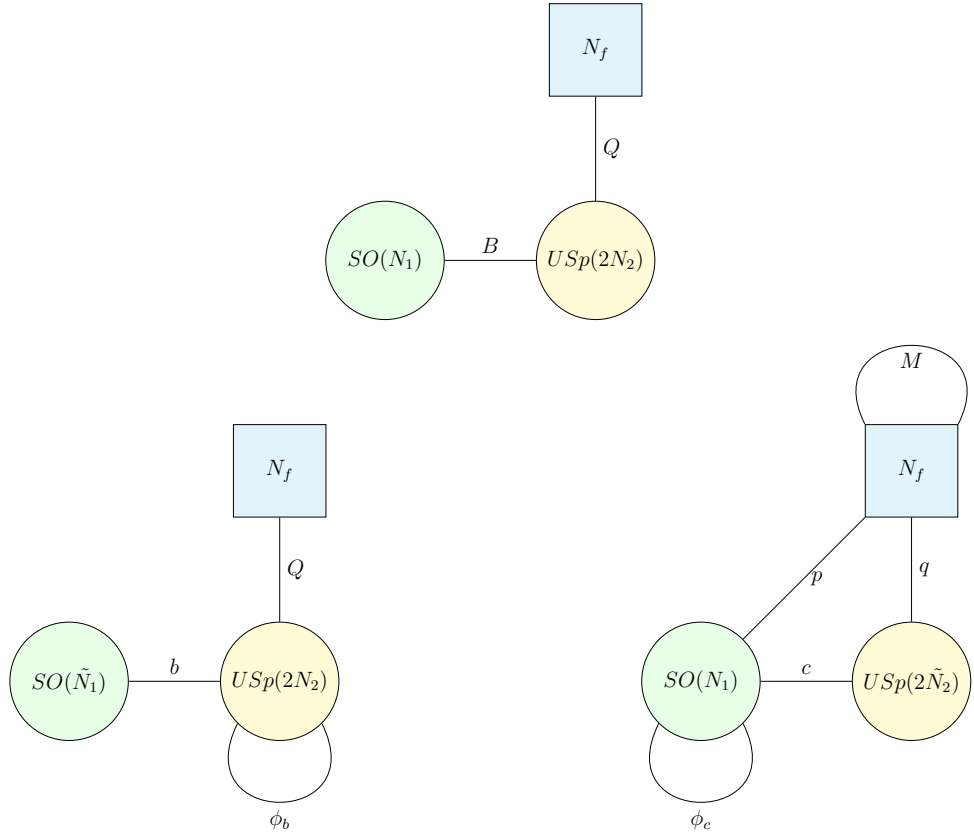
\begin{figure}
\centering
\scalebox{0.7}{
\begin{tikzpicture}
\path (0,0) node[circle, minimum size=64, fill=green!10, draw](AG1) {$SO(N_1)$}
(4,0) node[circle, minimum size=64, fill=yellow!20, draw](AG2) {$USp(2N_2)$}
(4,4) node[minimum size=50, fill=cyan!10, draw](AF2) {$N_f$};
\draw (AG1) -- (AG2) node [midway, above] {$B$};
\draw (AG2) -- (AF2) node [midway, right] {$Q$};
\path (-6,-8) node[circle, minimum size=64, fill=green!10, draw](BG1) {$SO(\tilde{N_1})$}
(-2,-8) node[circle, minimum size=64, fill=yellow!20, draw](BG2) {$USp(2N_2)$}
(-2,-4) node[minimum size=50, fill=cyan!10, draw](BF2) {$N_f$};
\draw (BG1) -- node[above]{$b$} (BG2);
\draw (BG2) -- node[right]{$Q$}(BF2);
\draw (BG2.south west) .. controls ++(-1, -2) and ++(1, -2) .. node[below]{$\phi_b$} (BG2.south east);
%
\path (6,-8) node[circle, minimum size=64, fill=green!10, draw](CG1) {$SO(N_1)$}
(10,-8) node[circle, minimum size=64, fill=yellow!20, draw](CG2) {$USp(2\tilde{N_2})$}
(10,-4) node[minimum size=50, fill=cyan!10, draw](CF2) {$N_f$};
\draw (CG1) -- node[above]{$c$} (CG2);
\draw (CG1) -- node[right]{$p$}(CF2);
\draw (CG2) -- node[right]{$q$}(CF2);
\draw (CG1.south west) .. controls ++(-1, -2) and ++(1, -2) .. node[below]{$\phi_c$} (CG1.south east);
\draw (CF2.north east) .. controls ++(1, 2) and ++(-1, 2) .. node[below]{$M$} (CF2.north west);
\end{tikzpicture}
}
\caption{Triality of $SO(N_1) \times USp(2N_2)$ quiver where $\tilde{N_1}=2N_2-N_1+2$ and $2\tilde{N_2}=N_1+N_f-2N_2-2$.
Here $\phi_b$ is in the rank-$2$
symmetric representation while $\phi_c$ and $M$ are in antisymmetric
representations. Note also that for consistency we must have
$N_1 + N_f$ even.}
\label{fig:SO_USp}
\end{figure}

The triality of $SO\times USp$ quiver is depicted in Figure \ref{fig:SO_USp}. 

The required details for the full indices, half-indices and anomalies are
mostly covered in the previous section for symplectic and orthogonal groups.
The extra ingredients required here are
\begin{align}
&Z_{BiFund \; SO_I-USp_J}(r, a) =  (q^{\frac{1-r}{2}} a^{-1})^{\sum_{i=1}^{n_I} \sum_{j=1}^{N_J} |m_i^{(I)} \pm m_j^{(J)}|} \nonumber \\
 & \times \prod_{i = 1}^{n_I} \prod_{j=1}^{N_J}
 \frac{(q^{1-\frac{r}{2}+\frac{|m_i^{(I)}-m_j^{(J)}|}{2}} a^{-1} s_i^{(I)\mp} s_j^{(J)\pm}; q)_{\infty} 
 (q^{1-\frac{r}{2}+\frac{|m_i^{(I)}+m_j^{(J)}|}{2}} a^{-1} s_i^{(I)\pm} s_j^{(J)\pm}; q)_{\infty}}
 {(q^{\frac{r}{2}+\frac{|m_i^{(I)} - m_j^{(J)}|}{2} } a s_i^{(I)\pm} s_j^{(J)\mp}; q)_{\infty} 
(q^{\frac{r}{2}+\frac{|m_i^{(I)} + m_j^{(J)}|}{2} } a s_i^{(I)\pm} s_j^{(J)\pm}; q)_{\infty}} \nonumber \\
 & \times \left(
 \prod_{j=1}^{N_J}
 \frac{(q^{1-\frac{r}{2}+\frac{|m_j^{(J)}|}{2}} a^{-1} \chi_I s_j^{(J)\pm}; q)_{\infty}}
 {(q^{\frac{r}{2}+\frac{|m_j^{(J)}|}{2} } a \chi_I s_j^{(J)\mp}; q)_{\infty}} \right)^{\epsilon_I}
 \\
&Z_{Sym \; USp_I}(r, a) =  (q^{\frac{1 - 2r}{2}} a^{-1})^{\sum_{i \le j}^{N_I} |{m_i^{(I)}} \pm {m_j^{(I)}}|} 
\nonumber \\
 & \times
\left( \prod_{i, j = 1}^{N_I}
 \frac{(q^{1-r +\frac{|{m_i^{(I)}} - {m_j^{(I)}}|}{2}} a^{-1} {s_i^{(I)}}^{-1} {s_j^{(I)}}; q)_{\infty}}
 {(q^{r+ \frac{|{m_i^{(I)}} - {m_j^{(I)}}|}{2}} a {s_i^{(I)}} {s_j^{(I)}}^{-1}; q)_{\infty}} \right)  \left( \prod_{i \le j}^{N_I}
 \frac{(q^{1-r + \frac{|{m_i^{(I)}} + {m_j^{(I)}}|}{2}} a^{-1} {s_i^{(I)}}^{\mp} {s_j^{(I)}}^{\mp}; q)_{\infty}}
 {(q^{r+\frac{|{m_i^{(I)}} + {m_j^{(I)}}|}{2}} a {s_i^{(I)}}^{\pm} {s_j^{(I)}}^{\pm}; q)_{\infty}} \right)
 \\
&Z_{Antisym \; SO_I}(r, a) =  (q^{\frac{1 - 2r}{2}} a^{-1})^{\sum_{i < j}^{N_I} |{m_i^{(I)}} \pm {m_j^{(I)}}|} \left( \prod_{i,j = 1}^{n_I}
 \frac{(q^{1-r +\frac{|{m_i^{(I)}} - {m_j^{(I)}}|}{2}} a^{-1} {s_i^{(I)}}^{-1} s_j^{(I)}; q)_{\infty}}
 {(q^{r+ \frac{|{m_i^{(I)}} - {m_j^{(I)}}|}{2}} a s_i^{(I)} {s_j^{(I)}}^{-1}; q)_{\infty}} \right) \nonumber \\
 & \times \left( \prod_{i < j}^{n_I}
 \frac{(q^{1-r + \frac{|{m_i^{(I)}} + {m_j^{(I)}}|}{2}} a^{-1} {s_i^{(I)}}^{\mp} {s_j^{(I)}}^{\mp}; q)_{\infty}}
 {(q^{r+\frac{|{m_i^{(I)}} + {m_j^{(I)}}|}{2}} a {s_i^{(I)}}^{\pm} {s_j^{(I)}}^{\pm}; q)_{\infty}} \right) \left( \prod_{i=1}^{n_I}
 \frac{(q^{1-r +\frac{|m_i^{(I)}|}{2}} a^{-1} \chi_I {s_i^{(I)}}^{\pm}; q)_{\infty}}
 {(q^{r+ \frac{|{m_i^{(I)}}|}{2}} a \chi_I {s_i^{(I)}}^{\pm}; q)_{\infty}}
 \right)^{\epsilon_I}
\end{align}
for the full indices and
\begin{align}
\II^{BiFund \; SO_I-USp_J}_N(r,a) = & \left( \prod_{i=1}^{n_I} \prod_{j=1}^{N_J} 
\frac{1}{(q^{\frac{r}{2}} a s_i^{(I)\pm} {s_j^{(J)}}^{\mp} ; q)_{\infty} 
(q^{\frac{r}{2}} a s_i^{(I)\pm} {s_j^{(J)}}^{\pm} ; q)_{\infty}} \right) \nonumber \\
 & \times
 \left( \prod_{j=1}^{n_J} \frac{1}{(q^{\frac{r}{2}} a \chi_I {s_j^{(J)}}^{\pm} ; q)_{\infty}} \right)^{\epsilon_I}
 \\
\II^{Antisym \; SO_I}_{N}(r,a) = & \left( \prod_{i,j=1}^{n_I} \frac{1}{(q^{\frac{r}{2}} a {s_i^{(I)}} {s_j^{(I)}}^{-1} ; q)_{\infty}} \right) 
 \left( \prod_{i < j}^{n_I} \frac{1}{(q^{\frac{r}{2}} a {s_i^{(I)}}^{\pm} {s_j^{(I)}}^{\pm} ; q)_{\infty}} \right) \nonumber \\
 & \times \left( \frac{1}{\prod_{i=1}^{n_I} (q^{\frac{r}{2}} a \chi_I {s_i^{(I)}}^{\pm}; q)_{\infty}} \right)^{\epsilon_I} \\
\II^{Sym \; USp_I}_{N}(r,a) = & \left( \prod_{i,j=1}^{N_I} \frac{1}{(q^{\frac{r}{2}} a {s_i^{(I)}} {s_j^{(I)}}^{-1} ; q)_{\infty}} \right)
 \left( \prod_{j \le l}^{N_2} \frac{1}{(q^{\frac{r}{2}} a {s_i^{(I)}}^{\pm} {s_j^{(I)}}^{\pm} ; q)_{\infty}} \right) \\
\II^{BiFund \; SO_I-USp_J}_D(r,a) = & \left( \prod_{i=1}^{n_I} \prod_{j=1}^{n_J} 
(q^{1 - \frac{r}{2}} a^{-1} s_i^{(I)\pm} {s_j^{(J)}}^{\mp} ; q)_{\infty} 
(q^{1 - \frac{r}{2}} a^{-1} s_i^{(I)\pm} {s_j^{(J)}}^{\pm} ; q)_{\infty} \right) \nonumber \\
 & \times \left( \prod_{j=1}^{n_J} (q^{1 - \frac{r}{2}} a^{-1} \chi_I {s_j^{(J)}}^{\pm} ; q)_{\infty} \right)^{\epsilon_I}
 \\
\II^{Antisym \; SO_I}_{D}(r,a) = & \left( \prod_{i,j=1}^{n_I} (q^{1 - \frac{r}{2}} a^{-1} {s_i^{(I)}} {s_j^{(I)}}^{-1} ; q)_{\infty} \right) 
 \left( \prod_{i < j}^{n_I} (q^{1 - \frac{r}{2}} a^{-1} {s_i^{(I)}}^{\pm} {s_j^{(I)}}^{\pm} ; q)_{\infty} \right) \nonumber \\
 & \times \left( \prod_{i=1}^{n_I} (q^{1 - \frac{r}{2}} a^{-1} \chi_I {s_i^{(I)}}^{\pm}; q)_{\infty} \right)^{\epsilon_I} \\
\II^{Sym \; USp_I}_{D}(r,a) = & \left( \prod_{i,j=1}^{N_I}
(q^{1 - \frac{r}{2}} a^{-1} {s_i^{(I)}} {s_j^{(I)}}^{-1} ; q)_{\infty} \right)
\left( \prod_{i \le j}^{N_I} (q^{1 - \frac{r}{2}} a^{-1} {s_i^{(I)}}^{\pm} {s_j^{(I)}}^{\pm} ; q)_{\infty} \right)
\end{align}
for the half-indices. Again, for the orthogonal groups if $\chi_I = -1$ and
$\epsilon_I = 0$ we must replace $s_i^{(I) \pm}$ with $\pm 1$.

The new contributions to the anomaly polynomials are given by
\begin{align}
\Acal_{BiFund \; SO_I-USp_J}(q_R, \underline{q}) = & \frac{N_I}{2} \Tr({\bf s_J}^2) + 2N_J \Tr({\bf s_I}^2) + N_I N_J \left( \underline{q} \cdot \underline{{\bf a}} + (q_R - 1){\bf r} \right)^2 \\
\Acal_{Antisym \; SO_I}(q_R, \underline{q}) = & (N_I - 2) \Tr({\bf s_I}^2) + \frac{N_I(N_I - 1)}{4} \left( \underline{q} \cdot \underline{{\bf a}} + (q_R - 1){\bf r} \right)^2 \\
\Acal_{Sym \; USp_I}(q_R, \underline{q}) = & (N_I + 1) \Tr({\bf s_I}^2) + \frac{N_I(2N_I + 1)}{2} \left( \underline{q} \cdot \underline{{\bf a}} + (q_R - 1){\bf r} \right)^2 \; .
\end{align}

\subsection{$SO(3)\times USp(4)-[7]$ $(N_1 = 3$, $N_2 = 2$, $N_f = 7)$}
Here we have $\tilde{N_1} = 3$ and $\tilde{N_2} = 2$ so the gauge group for theories A, B and
C is $SO(3) \times USp(4)$.

The supersymmetric indices agree with each other. 
For $r_B=2/5$ and $r_Q=1/2$ the expansions of full-indices
are~\footnote{See also \cite{Benvenuti:2020wpc}. }
\begin{align}
&I^A = I^B =I^C
\nonumber\\
=&1+\underbrace{21 a_2^2 q^{1/2}}_{\Tr (QQ)}
+\underbrace{\frac{q^{13/20}}{a_1^3 a_2^7}}_{v_A^{0, \pm}}
+\underbrace{\frac{q^{7/10}}{a_1^4}}_{v_A^{+,0}}
+\underbrace{\frac{q^{3/4}}{a_1^5 a_2^7}}_{v_A^{+, \pm}}
+\underbrace{a_1^4 q^{4/5}}_{\Tr(BB)^2}
+\underbrace{7 a_1^3 a_2 q^{17/20}}_{\epsilon BBBQ}
+\underbrace{28 a_1^2 a_2^2 q^{9/10}}_{\Tr (Q(BB)Q)}
   \nonumber\\
   &+(\underbrace{231a_2^4}_{\Tr(QQ)^2}
   -\underbrace{50}_{\begin{smallmatrix}\Tr(Q\Psi_Q) \\ \Tr(B\psi_B) \end{smallmatrix}}) q
   +(\underbrace{7 \frac{a_2}{a_1^3}}_{v_A^{-,0}B^{N_1-2}Q}
   +\underbrace{21 \frac{1}{a_1^3 a_2^5}}_{v_A^{0, \pm}\Tr(QQ)}
   +\underbrace{\frac{1}{a_1^3 a_2^7 }}_{v_A^{+, \pm}\Tr(BB)})
   q^{23/20} +(
   \underbrace{21 \frac{a_2^2}{a_1^4}}_{v_A^{+,0}\Tr(QQ)}
   +\underbrace{\frac{7}{a_1^4a_2^6}}_{v_A^{-, \pm}B^{N_1-2}Q} ) q^{6/5}
   \nonumber\\
   &
   +\underbrace{\frac{21 q^{5/4}}{a_1^5 a_2^5}}_{v_A^{+, \pm}\Tr(QQ)}
   +q^{13/10} (
   \underbrace{42 a_1^4 a_2^2}_{\begin{smallmatrix} \Tr(QQ)\Tr(BB)^2, \\ \Tr(BB) \Tr(Q(BB)Q) \end{smallmatrix}}
   + \underbrace{\frac{1}{a_1^6 a_2^{14}}}_{v_A^{0, \pm 2}})
 + \cdots
  \end{align}

The half-indices are given by
\begin{align}
&\II^A_{\mathcal{N},\mathcal{N}} = \II^B_{\mathcal{D},\mathcal{N}} =\II^C_{\mathcal{N},\mathcal{D}}
\nonumber\\
=&1+\underbrace{21  a_2^2 q^{1/2}}_{\Tr (QQ)}
 - \underbrace{\frac{7 q^{3/4} \left( a_2 \left( w_1^2  w_2+ w_1  w_2^2+ w_1+ w_2\right)\right)}{ w_1 w_2}}_{Q \Gamma_2}
  + \underbrace{a_1^4 q^{4/5}}_{\Tr (BB)^2}+\underbrace{7  a_1^3  a_2 q^{17/20}}_{\epsilon BBBQ}
    \nonumber\\
 & + a_1^2 q^{9/10} \left( - \underbrace{28  a_2^2}_{\Tr(Q(BB)Q)} + \underbrace{v + \frac{1}{v} + 1}_{\epsilon BB\Gamma_1} \right)
 - \underbrace{\frac{7 q^{19/20}
   \left( a_1  a_2 \left(v^2+v+1\right)\right)}{v}}_{Q B \Gamma_1}
\nonumber\\
&+q (\underbrace{231  a_2^4}_{\Tr(QQ)^2}
 + \underbrace{v+\frac{1}{v}+ w_1^2+\frac{1}{ w_1^2}+ w_1
    w_2+\frac{ w_2}{ w_1}+\frac{ w_1}{ w_2}+\frac{1}{ w_1  w_2}+ w_2^2+\frac{1}{ w_2^2}+3}_{\Gamma_1 \Gamma_1, \; \Gamma_2 \Gamma_2})+\cdots
\end{align}

\subsection{$USp(2N_1) \times USp(2N_2)-[N_f] \times SO(N_3)$}
\label{sec_USpUSpSO}
For consistency we need $N_3 + N_f$ to be even.
The dualization of the $USp(2N_2)$ gauge node gives a $USp(2\tilde{N}_2 = 2N_1 + N_3 + N_f - 2N_2 - 2)$ gauge node, all three gauge nodes have $N_f$ flavors, there are antisymmetric rank-$2$ chirals for $USp(2N_1)$ and $SO(N_3)$, and a singlet chiral in the antisymmetric rank-$2$ representation of the flavor symmetry $SU(N_f)$.

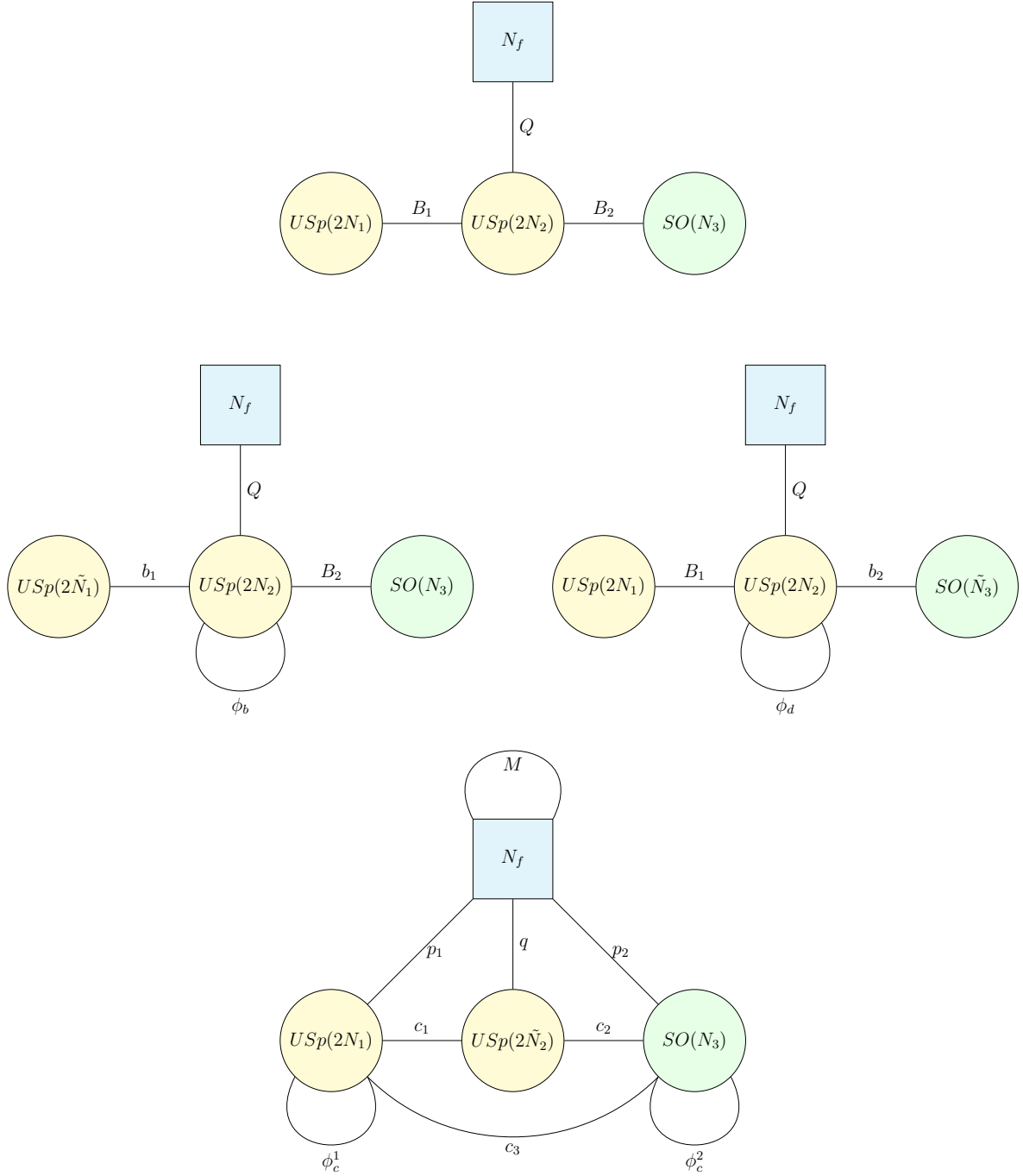
\begin{figure}
\centering
\scalebox{0.7}{
\begin{tikzpicture}
\path (-4,0) node[circle, minimum size=64, fill=yellow!20, draw](AG1) {$USp(2N_1)$}
(0,0) node[circle, minimum size=64, fill=yellow!20, draw](AG2) {$USp(2N_2)$}
(4,0) node[circle, minimum size=64, fill=green!10, draw](AG3) {$SO(N_3)$}
(0,4) node[minimum size=50, fill=cyan!10, draw](AF2) {$N_f$};
\draw (AG1) -- (AG2) node [midway, above] {$B_1$};
\draw (AG3) -- (AG2) node [midway, above] {$B_2$};
\draw (AG2) -- (AF2) node [midway, right] {$Q$};
\path (-10,-8) node[circle, minimum size=64, fill=yellow!20, draw](BG1) {$USp(2\tilde{N_1})$}
(-6,-8) node[circle, minimum size=64, fill=yellow!20, draw](BG2) {$USp(2N_2)$}
(-2,-8) node[circle, minimum size=64, fill=green!10, draw](BG3) {$SO(N_3)$}
(-6,-4) node[minimum size=50, fill=cyan!10, draw](BF2) {$N_f$};
\draw (BG1) -- node[above]{$b_1$} (BG2);
\draw (BG3) -- node[above]{$B_2$} (BG2);
\draw (BG2) -- node[right]{$Q$}(BF2);
\draw (BG2.south west) .. controls ++(-1, -2) and ++(1, -2) .. node[below]{$\phi_b$} (BG2.south east);
%
%
\path (2,-8) node[circle, minimum size=64, fill=yellow!20, draw](DG1) {$USp(2N_1)$}
(6,-8) node[circle, minimum size=64, fill=yellow!20, draw](DG2) {$USp(2N_2)$}
(10,-8) node[circle, minimum size=64, fill=green!10, draw](DG3) {$SO(\tilde{N_3})$}
(6,-4) node[minimum size=50, fill=cyan!10, draw](DF2) {$N_f$};
\draw (DG1) -- node[above]{$B_1$} (DG2);
\draw (DG3) -- node[above]{$b_2$} (DG2);
\draw (DG2) -- node[right]{$Q$}(DF2);
\draw (DG2.south west) .. controls ++(-1, -2) and ++(1, -2) .. node[below]{$\phi_d$} (DG2.south east);
%
\path (-4,-18) node[circle, minimum size=64, fill=yellow!20, draw](CG1) {$USp(2N_1)$}
(0,-18) node[circle, minimum size=64, fill=yellow!20, draw](CG2) {$USp(2\tilde{N_2})$}
(4,-18) node[circle, minimum size=64, fill=green!10, draw](CG3) {$SO(N_3)$}
(0,-14) node[minimum size=50, fill=cyan!10, draw](CF2) {$N_f$};
\draw (CG1) -- node[above]{$c_1$} (CG2);
\draw (CG3) -- node[above]{$c_2$} (CG2);
\draw (CG1) -- node[right]{$p_1$}(CF2);
\draw (CG3) -- node[right]{$p_2$}(CF2);
\draw (CG2) -- node[right]{$q$}(CF2);
\draw (CG1) to [out=-45, in=-135] node[below]{$c_3$}(CG3);
\draw (CG1.south west) .. controls ++(-1, -2) and ++(1, -2) .. node[below]{$\phi_c^1$} (CG1.south east);
\draw (CG3.south west) .. controls ++(-1, -2) and ++(1, -2) .. node[below]{$\phi_c^2$} (CG3.south east);
\draw (CF2.north east) .. controls ++(1, 2) and ++(-1, 2) .. node[below]{$M$} (CF2.north west);
\end{tikzpicture}
}
\caption{Quadrality of $USp(2N_1) \times USp(2N_2)-[N_f] \times SO(N_3)$ quiver 
where $\tilde{N_1}=N_2-N_1-1$, $2\tilde{N_2}=2N_1+N_3+N_f-2N_2-2$ and $\tilde{N_3}=2N_2-N_3+2$, and we need $N_3 + N_f$ to be even. Note that $\phi_b$, $\phi_c^1$, $\phi_c^2$ and $M$ are in antisymmetric rank-$2$ representations while
$\phi_d$ is in the symmetric representation.} \label{fig:USp_USp_SO}
\end{figure}

We propose the quadrality in Figure \ref{fig:USp_USp_SO}. 

\subsection{$USp(2)\times USp(6)-[5]\times SO(3)$ $(N_1 = 1$, $N_2 = 3$, $N_3 = 3$, $N_f = 5)$}
Here $\tilde{N}_2 = 1$ so the dual theory has gauge group $USp(2) \times USp(2) \times SO(3)$.

The full indices are given by
\begin{align}
&I^A = I^B= I^C = I^D
\nonumber\\
=&1+q^{2/7} \left(\frac{1}{a_1^2 a_2^5 a_3^3}+10 a_2^2\right)+a_1^2 q^{2/5}+\frac{q^{17/35}}{a_1^6 a_2^5 a_3^3}+q^{4/7}
   \left(\frac{10}{a_1^2 a_2^3 a_3^3}+\frac{1}{a_1^4 a_2^{10} a_3^6}+55 a_2^4\right)
   \nonumber\\
   &+q^{24/35} \left(20 a_1^2 a_2^2+15 a_2^2
   a_3^2+\frac{1}{a_2^5 a_3^3}\right)+5 a_2 a_3^3 q^{26/35}+\frac{q^{27/35} \left(10 a_1^2 a_2^7 a_3^3+1\right)}{a_1^8 a_2^{10}
   a_3^6}
   \nonumber\\
   &+q^{4/5} \left(a_1^4+\frac{1}{a_1^6}+a_3^4\right)+q^{6/7} \left(\frac{10}{a_1^4 a_2^8 a_3^6}+\frac{1}{a_1^6 a_2^{15}
   a_3^9}+\frac{55}{a_1^2 a_2 a_3^3}+220 a_2^6\right)+\cdots
\end{align}

The half-indices are given by
\begin{align}
&\II^A_{\mathcal{N},\mathcal{N},\mathcal{N}} 
= \II^B_{\mathcal{D},\mathcal{N},\mathcal{N}}
= \II^C_{\mathcal{N},\mathcal{D},\mathcal{N}}
= \II^D_{\mathcal{N},\mathcal{N},\mathcal{D}}
\nonumber\\
&= 1+10  a_2^2 q^{2/7}+ a_1^2 q^{2/5}+55  a_2^4 q^{4/7}-\frac{5 q^{9/14} \left( a_2 \left(w^2+1\right)\right)}{w}
+5  a_2^2 q^{24/35} \left(4a_1^2+3  a_3^2\right) 
\nonumber\\
&+5  a_2  a_3^3 q^{26/35}+q^{4/5} \left( a_1^4+ a_3^4\right)-\frac{5 q^{59/70} \left( a_2  a_3
   \left( v_1^2  v_2+ v_1 \left( v_2^2+ v_2+1\right)+ v_2\right)\right)}{ v_1  v_2} 
\nonumber\\
&+220  a_2^6 q^{6/7}-\frac{ a_3^2
   q^{9/10} \left( v_1^2  v_2+ v_1 \left( v_2^2+ v_2+1\right)+ v_2\right)}{ v_1  v_2}-\frac{50 q^{13/14} \left( a_2^3
   \left(w^2+1\right)\right)}{w}
 + \cdots
\end{align}

For these indices and half-indices we do not write the explicit operators contribution to each term as these can easily be understood from the previous discussions.

\subsection{$USp(2N_1) \times SO(N_2)-[N_f] \times SO(N_3)$}
\label{sec_USpSOSO}
Here we need $N_2$ to be even.
Dualizing the $SO(N_2)$ gauge node leads to an $SO(\tilde{N}_2 = 2N_1 + N_3 + N_f - N_2 + 2)$ gauge node 
and all three gauge nodes have $N_f$ flavors, there are bifundamental chirals for each pair of gauge nodes, there are symmetric rank-$2$ chirals for $USp(2N_1)$ and $SO(N_3)$, and a singlet chiral in the symmetric rank-$2$ representation of the flavor symmetry $SU(N_f)$.

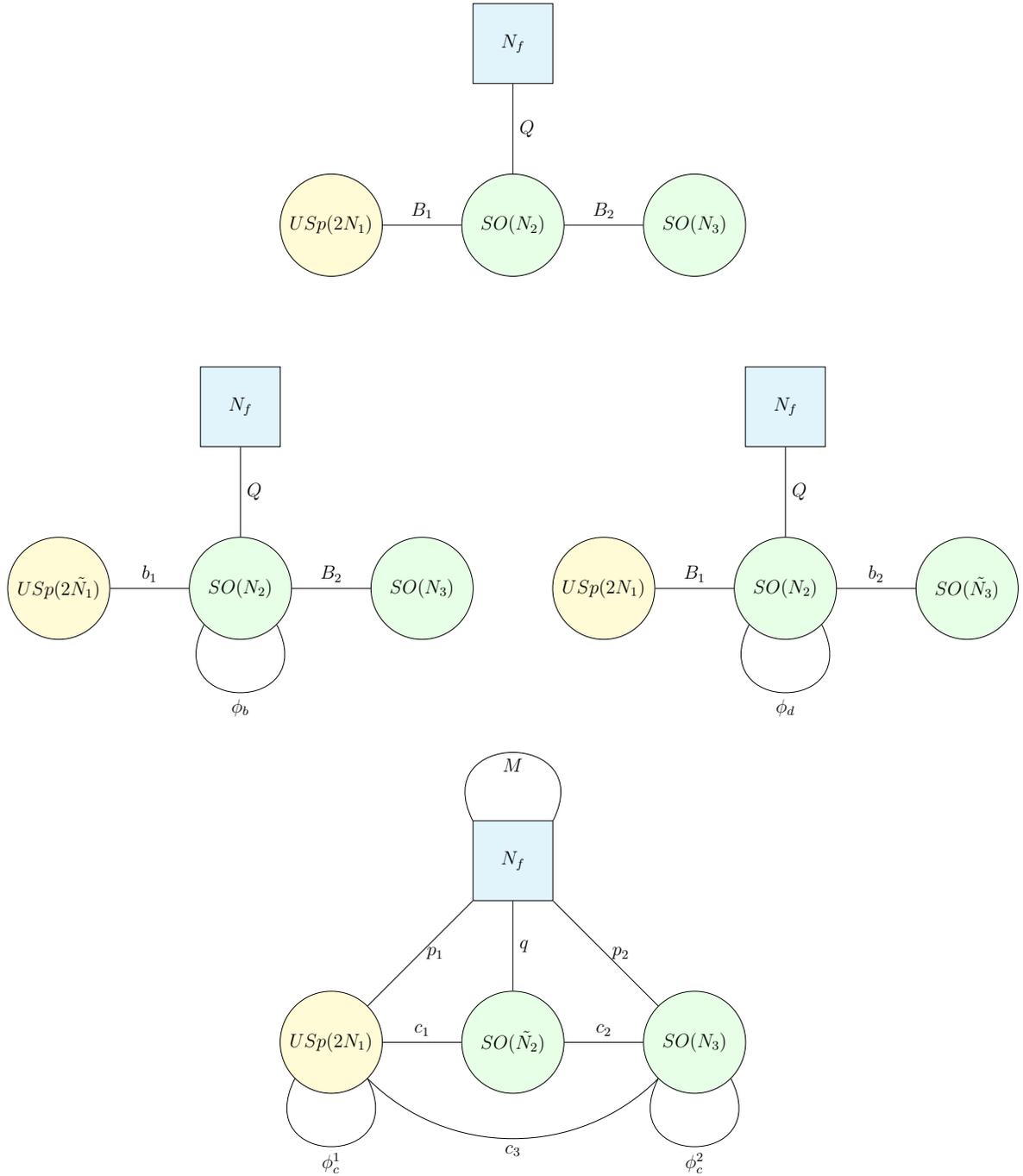
\begin{figure}
\centering
\scalebox{0.7}{
\begin{tikzpicture}
\path (-4,0) node[circle, minimum size=64, fill=yellow!20, draw](AG1) {$USp(2N_1)$}
(0,0) node[circle, minimum size=64, fill=green!10, draw](AG2) {$SO(N_2)$}
(4,0) node[circle, minimum size=64, fill=green!10, draw](AG3) {$SO(N_3)$}
(0,4) node[minimum size=50, fill=cyan!10, draw](AF2) {$N_f$};
\draw (AG1) -- (AG2) node [midway, above] {$B_1$};
\draw (AG3) -- (AG2) node [midway, above] {$B_2$};
\draw (AG2) -- (AF2) node [midway, right] {$Q$};
\path (-10,-8) node[circle, minimum size=64, fill=yellow!20, draw](BG1) {$USp(2\tilde{N_1})$}
(-6,-8) node[circle, minimum size=64, fill=green!10, draw](BG2) {$SO(N_2)$}
(-2,-8) node[circle, minimum size=64, fill=green!10, draw](BG3) {$SO(N_3)$}
(-6,-4) node[minimum size=50, fill=cyan!10, draw](BF2) {$N_f$};
\draw (BG1) -- node[above]{$b_1$} (BG2);
\draw (BG3) -- node[above]{$B_2$} (BG2);
\draw (BG2) -- node[right]{$Q$}(BF2);
\draw (BG2.south west) .. controls ++(-1, -2) and ++(1, -2) .. node[below]{$\phi_b$} (BG2.south east);
%
%
\path (2,-8) node[circle, minimum size=64, fill=yellow!20, draw](DG1) {$USp(2N_1)$}
(6,-8) node[circle, minimum size=64, fill=green!10, draw](DG2) {$SO(N_2)$}
(10,-8) node[circle, minimum size=64, fill=green!10, draw](DG3) {$SO(\tilde{N_3})$}
(6,-4) node[minimum size=50, fill=cyan!10, draw](DF2) {$N_f$};
\draw (DG1) -- node[above]{$B_1$} (DG2);
\draw (DG3) -- node[above]{$b_2$} (DG2);
\draw (DG2) -- node[right]{$Q$}(DF2);
\draw (DG2.south west) .. controls ++(-1, -2) and ++(1, -2) .. node[below]{$\phi_d$} (DG2.south east);
%
\path (-4,-18) node[circle, minimum size=64, fill=yellow!20, draw](CG1) {$USp(2N_1)$}
(0,-18) node[circle, minimum size=64, fill=green!10, draw](CG2) {$SO(\tilde{N_2})$}
(4,-18) node[circle, minimum size=64, fill=green!10, draw](CG3) {$SO(N_3)$}
(0,-14) node[minimum size=50, fill=cyan!10, draw](CF2) {$N_f$};
\draw (CG1) -- node[above]{$c_1$} (CG2);
\draw (CG3) -- node[above]{$c_2$} (CG2);
\draw (CG1) -- node[right]{$p_1$}(CF2);
\draw (CG3) -- node[right]{$p_2$}(CF2);
\draw (CG2) -- node[right]{$q$}(CF2);
\draw (CG1) to [out=-45, in=-135] node[below]{$c_3$}(CG3);
\draw (CG1.south west) .. controls ++(-1, -2) and ++(1, -2) .. node[below]{$\phi_c^1$} (CG1.south east);
\draw (CG3.south west) .. controls ++(-1, -2) and ++(1, -2) .. node[below]{$\phi_c^2$} (CG3.south east);
\draw (CF2.north east) .. controls ++(1, 2) and ++(-1, 2) .. node[below]{$M$} (CF2.north west);
\end{tikzpicture}
}
\caption{Quadrality of $USp(2N_1) \times SO(N_2)-[N_f] \times SO(N_3)$ quiver 
with even $N_2$ where $2\tilde{N_1}=N_2-2N_1-2$, $\tilde{N_2}=2N_1 + N_3 + N_f - N_2 + 2$ and $\tilde{N_3}=N_2-N_3+2$.
Note that $\phi_d$, $\phi_c^1$, $\phi_c^2$ and $M$ are in symmetric rank-$2$ representations while
$\phi_b$ is in the antisymmetric representation.} \label{fig:USp_SO_SO}
\end{figure}

The proposed quadrality of the $USp\times SO\times SO$ quiver is illustrated in Figure \ref{fig:USp_SO_SO}. 

\subsection{$USp(2)\times SO(4)-[1]-SO(2)$ $(N_1 = 1$, $N_2 = 4$, $N_3 = 2$, $N_f = 1)$}
Here $\tilde{N}_2 = 3$ so the dual theory has gauge group $USp(2) \times SO(3) \times SO(2)$.

The full indices of theory A and theory C perfectly match. 
They are 
\begin{align}
&I^A = I^B = I^C = I^D 
\nonumber\\
=&
1+\frac{q^{11/70}}{a_1^4 a_2 a_3^2}+\frac{q^{1/5}}{a_1^4}+a_2^2 q^{2/7}+\frac{q^{11/35}}{a_1^8 a_2^2 a_3^4}+\frac{q^{5/14}}{a_1^8
   a_2 a_3^2}+q^{2/5} \left(\frac{1}{a_1^8}+a_3^2\right)+\frac{a_2 q^{31/70}}{a_1^4 a_3^2}+\frac{q^{33/70}}{a_1^{12} a_2^3
   a_3^6}
   \nonumber\\
   &
   +\frac{a_2^2 q^{17/35}}{a_1^4}+\frac{q^{18/35}}{a_1^{12} a_2^2 a_3^4}+\frac{q^{39/70} \left(2 a_1^8 a_3^2+2
   a_1^{10}+1\right)}{a_1^{12} a_2 a_3^2}+a_2^4 q^{4/7}+q^{3/5} \left(\frac{a_3^2}{a_1^4}+\frac{1}{a_1^8
   a_3^4}+\frac{1}{a_1^{12}}\right)
   \nonumber\\
   &
   +\frac{q^{22/35}}{a_1^{16} a_2^4 a_3^8}+\frac{a_2 q^{9/14}}{a_1^8 a_3^2}+\frac{q^{47/70}}{a_1^{16}
   a_2^3 a_3^6}+a_2^2 q^{24/35} \left(\frac{1}{a_1^8}+2 a_3^2\right)
 + \cdots
\end{align}

The half-indices are given by
\begin{align}
&\II^A_{\mathcal{N},\mathcal{N},\mathcal{N}} 
= \II^B_{\mathcal{D},\mathcal{N},\mathcal{N}}
= \II^C_{\mathcal{N},\mathcal{D},\mathcal{N}}
= \II^D_{\mathcal{N},\mathcal{N},\mathcal{D}}
\nonumber\\
=&
1+a_2^2 q^{2/7}+a_3^2 q^{2/5}+a_2^4 q^{4/7}-\frac{a_2 q^{9/14} \left(w^2+w+1\right)}{w}+2 a_2^2 a_3^2 q^{24/35}
\nonumber\\
&+q^{4/5} \left(2
   a_1^2 a_3^2+a_1^4+2 a_3^4\right)-\frac{2 q^{59/70} \left(a_2 a_3 \left(v_1^2 v_2+v_1
   v_2^2+v_1+v_2\right)\right)}{v_1 v_2}+a_2^6 q^{6/7}
   \nonumber\\
   &-\frac{a_2^3 q^{13/14} \left(w^2+w+1\right)}{w}+2 a_2^4 a_3^2
   q^{34/35}+\cdots
\end{align}

Again, for these indices and half-indices we do not write the explicit operators
contribution to each term as these can easily be understood from the previous discussions.

\section{Unitary linear quivers}
\label{sec_unitary}
We now come to the case of unitary gauge groups. Again, for two gauge nodes
the general description of these bulk dualities was given in
\cite{Benvenuti:2020wpc}. We will generalize this to include Chern-Simons
levels (see also \cite{Benvenuti:2020gvy}) and to cases with additional gauge
nodes. As we will discuss, we expect the duality extends to the case with
boundary in a similar way to the symplectic and orthogonal cases, but it appears
we need to include 2d charged chiral multiplets (not just Fermis) to cancel
gauge anomalies and match 't Hooft anomalies. This complicates the evaluation
of the half-indices by requiring use of the JK residue prescription which we
leave to future work. Here we discuss the anomaly cancellation and matching,
and simply conjecture the dualities with boundaries.

\subsection{$U(N_1) \times U(N_2)-[N_f]$}
\label{sec_UNUN}
We summarize the case with two gauge nodes which was presented in \cite{Benvenuti:2020wpc}. 
We start with theory A, and $U(N_1) \times U(N_2)$ gauge theory with
bifundamental chirals $B$ and $\tilde{B}$ in the $(N_1, \bar{N}_2)$ and
$(\bar{N}_1, N_2)$ representations, as well as $N_f$ chirals $Q$ in the
fundamental representation of $U(N_2)$, and $N_a$ chirals $\tilde{Q}$ in the
anti-fundamental representation of $U(N_2)$. For now we fix $N_a = N_f$ and
consider cases with $N_a \ne N_f$ to arise from integrating out some
fundamental or anti-fundamental chirals.
The Chern-Simons levels vanish (but can be generated by integrating out
fundamental and/or anti-fundamental chirals)
but in comparison to orthogonal and symplectic gauge groups we have
$U(1)$ topological symmetries and background FI terms.
Theory A has vanishing superpotential.

Provided $\tilde{N}_1 \equiv N_2 - N_1 \ge 0$ there is a dual theory B given by
Seiberg-like duality on the $U(N_1)$ gauge factor. The result is a
$U(\tilde{N}_1) \times U(N_2)$ gauge theory with
bifundamental chirals $b$ and $\tilde{b}$ in the $(\tilde{N}_1, \bar{N}_2)$ and
$(\overline{\tilde{N}}_1, N_2)$ representations, an adjoint chiral $\phi_b$ of $U(N_2)$, as well as
$N_f$ chirals $Q$ in the
fundamental representation of $U(N_2)$, and $N_a$ chirals $\tilde{Q}$ in the
anti-fundamental representation of $U(N_2)$.
We also have gauge singlets $\sigma_B^{\pm}$ which are dual to monopoles in
theory A.
The Chern-Simons levels vanish
but we again have background FI terms.
The superpotential is given by
\begin{align}
\Wcal = & \sigma_B^{\pm} v_B^{\pm,0} + \Tr(\tilde{b} \phi_b b)
\end{align}

Provided $\tilde{N}_2 \equiv N_1 - N_2 + N_f \ge 0$ there is a dual theory C given by 
Seiberg-like duality on the $U(N_2)$ gauge factor. The result is a
$U(N_1) \times U(\tilde{N}_2)$ gauge theory with
bifundamental chirals $c$ and $\tilde{c}$ in the $(N_1, \overline{\tilde{N}}_2)$ and
$(\bar{N}_1, \tilde{N}_2)$ representations, an adjoint chiral $\phi_c$ of $U(N_1)$, as well as 
$N_f$ chirals $p$ and $q$ in the
fundamental representations of $U(N_1)$ and $U(\tilde{N}_2)$, and $N_a$ chirals
$\tilde{p}$ and $\tilde{q}$ in the
anti-fundamental representations of $U(N_1)$ and $U(\tilde{N}_2)$.
Again we have gauge singlets $\sigma_B^{\pm}$ which are dual to monopoles in 
theory A, but now also an additional $N_f \times N_a$ matrix of singlets.
The Chern-Simons levels vanish
but we again have background FI terms.
The superpotential is given by
\begin{align}
\Wcal = & \sigma_C^{\pm} v_C^{0, \pm} + \Tr(\tilde{c} \phi_c c) +
  \Tr(qcp) + \Tr(\tilde{q} \tilde{c} \tilde{p}) + \Tr(qM\tilde{q})
\end{align}

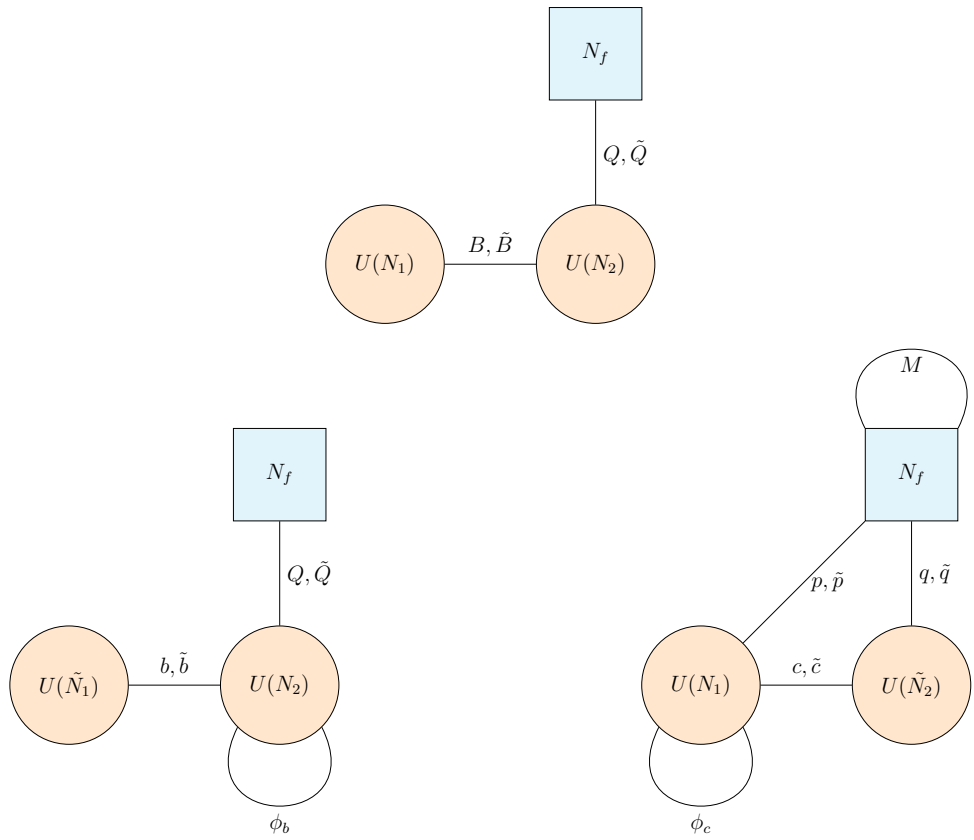
\begin{figure}
\centering
\scalebox{0.7}{
\begin{tikzpicture}
\path (0,0) node[circle, minimum size=64, fill=orange!20, draw](AG1) {$U(N_1)$}
(4,0) node[circle, minimum size=64, fill=orange!20, draw](AG2) {$U(N_2)$}
(4,4) node[minimum size=50, fill=cyan!10, draw](AF2) {$N_f$};
\draw (AG1) -- (AG2) node [midway, above] {$B,\tilde{B}$};
\draw (AG2) -- (AF2) node [midway, right] {$Q, \tilde{Q}$};
\path (-6,-8) node[circle, minimum size=64, fill=orange!20, draw](BG1) {$U(\tilde{N_1})$}
(-2,-8) node[circle, minimum size=64, fill=orange!20, draw](BG2) {$U(N_2)$}
(-2,-4) node[minimum size=50, fill=cyan!10, draw](BF2) {$N_f$};
\draw (BG1) -- node[above]{$b,\tilde{b}$} (BG2);
\draw (BG2) -- node[right]{$Q, \tilde{Q}$}(BF2);
\draw (BG2.south west) .. controls ++(-1, -2) and ++(1, -2) .. node[below]{$\phi_b$} (BG2.south east);
%
\path (6,-8) node[circle, minimum size=64, fill=orange!20, draw](CG1) {$U(N_1)$}
(10,-8) node[circle, minimum size=64, fill=orange!20, draw](CG2) {$U(\tilde{N_2})$}
(10,-4) node[minimum size=50, fill=cyan!10, draw](CF2) {$N_f$};
\draw (CG1) -- node[above]{$c, \tilde{c}$} (CG2);
\draw (CG1) -- node[right]{$p, \tilde{p}$}(CF2);
\draw (CG2) -- node[right]{$q, \tilde{q}$}(CF2);
\draw (CG1.south west) .. controls ++(-1, -2) and ++(1, -2) .. node[below]{$\phi_c$} (CG1.south east);
\draw (CF2.north east) .. controls ++(1, 2) and ++(-1, 2) .. node[below]{$M$} (CF2.north west);
\end{tikzpicture}
}
\caption{Triality of $U \times U$ quivers where $\tilde{N_1}=N_2-N_1$ and $\tilde{N_2}=N_1+N_f-N_2$.} \label{fig:U_U}
\end{figure}

The triality of the $U\times U$ quivers is shown in Figure \ref{fig:U_U}. 

The global symmetry group is given by a $SU(N_f) \times SU(N_a)$ flavor
symmetry, two axial symmetries $U(1)_{a_1} \times U(1)_{a_2}$, two topological
symmetries $U(1)_{y_1} \times U(1)_{y_2}$ and the $U(1)_R$ R-symmetry. The
following table lists the charges and representations of the fields in each
theory (for simplicity still restricted to the case $N_a = N_f$).
\begin{align}
\label{UN_charges}
\begin{array}{c|c|c|c|c|c|c|c}
& SU(N_f) & SU(N_a) & U(1)_{y_1} & U(1)_{y_2} & U(1)_{a_1} & U(1)_{a_2} & U(1)_R \\ \hline
B, \tilde{B} & {\bf 1} & {\bf 1} & 0 & 0 & 1 & 0 & r_B \\
Q & {\bf \bar{N}_f} & {\bf 1} & 0 & 0 & 0 & 1 & r_Q \\
\tilde{Q} & {\bf 1} & {\bf N_a} & 0 & 0 & 0 & 1 & r_Q \\ \hline
b, \tilde{b} & {\bf 1} & {\bf 1} & 0 & 0 & -1 & 0 & 1 - r_B \\
\phi_b & {\bf 1} & {\bf 1} & 0 & 0 & 2 & 0 & 2r_B \\
Q & {\bf \bar{N}_f} & {\bf 1} & 0 & 0 & 0 & 1 & r_Q \\
\tilde{Q} & {\bf 1} & {\bf N_a} & 0 & 0 & 0 & 1 & r_Q \\
\sigma_B^{\pm} & {\bf 1} & {\bf 1} & \pm 1 & 0 & -N_2 & 0 & N_2(1 - r_B) - (N_1 - 1) \\ \hline
c, \tilde{c} & {\bf 1} & {\bf 1} & 0 & 0 & -1 & 0 & 1 - r_B \\
\phi_c & {\bf 1} & {\bf 1} & 0 & 0 & 2 & 0 & 2r_B \\
p & {\bf \bar{N}_f} & {\bf 1} & 0 & 0 & 1 & 1 & r_B + r_Q \\
\tilde{p} & {\bf 1} & {\bf N_a} & 0 & 0 & 1 & 1 & r_B + r_Q \\
M & {\bf N_f} & {\bf \bar{N}_a} & 0 & 0 & 0 & 2 & 2r_Q \\
q & {\bf \bar{N}_f} & {\bf 1} & 0 & 0 & 0 & -1 & 1 - r_Q \\ 
\tilde{q} & {\bf 1} & {\bf N_a} & 0 & 0 & 0 & -1 & 1 - r_Q \\
\sigma_C^{\pm} & {\bf 1} & {\bf 1} & 0 & \pm 1 & -N_1 & -N_f & N_1(1 - r_B) + N_f(1 - r_Q) - (N_2 - 1)
\end{array}
\end{align}

\subsection{Supersymmetric indices for unitary gauge theories}
\label{sec_UNUNindices}
Our notation for the indices \cite{Bhattacharya:2008zy,Bhattacharya:2008bja,Kim:2009wb,Imamura:2011su, Kapustin:2011jm, Dimofte:2011py} for the unitary cases is similar to that for other gauge groups. 
However, we now have separate fugacities $x$ and $\tilde{x}$ for
the $SU(N_f)$ and $SU(N_a)$ flavor symmetries and we introduce fugacities
$z_1, z_2$ for the topological symmetries.

The superconformal indices for theories A, B and C are
\begin{align}
I^A = & Z^{U_1}_{gauge} Z^{U_2}_{gauge} Z_{matter \; A} \\
I^B = & Z^{\widetilde{U}_1}_{gauge} Z^{U_2}_{gauge} Z_{matter \; B} \\
I^C = & Z^{U_1}_{gauge} Z^{\widetilde{U}_2}_{gauge} Z_{matter \; C}
\end{align}
where
\begin{align}
Z^{U_I}_{gauge} & = \sum_{m^{(I)}_i \in \Zb} \frac{1}{N_I!} \oint \left( \prod_{i=1}^{N_I} \frac{ds^{(I)}_i}{2\pi i s^{(I)}_i} (-s^{(I)}_i)^{k_I m^{(I)}_i} z_I^{m^{(I)}_i} \right) \nonumber \\
& \times q^{-\sum_{i < j}^{N_I} |m^{(I)}_i - m^{(I)}_j|/2}
 \prod_{i \ne j}^{N_I} ( 1 - q^{|m^{(I)}_i - m^{(I)}_j|/2} s_i^{(I)} s_j^{(I) -1})
\end{align}
and
\begin{align}
Z_{matter \; A} = & Z_{B, \tilde{B}} Z_Q^{N_f} Z_{\tilde{Q}}^{N_a} \\
Z_{matter \; B} = & Z_{b, \tilde{b}} Z_{\phi_b} Z_Q^{N_f} Z_{\tilde{Q}}^{N_a} Z_{\sigma_B^+} Z_{\sigma_B^-} \\
Z_{matter \; C} = & Z_{c, \tilde{c}} Z_{\phi_c} Z_p^{N_f} Z_{\tilde{p}}^{N_a} Z_M^{N_fN_a} Z_q^{N_f} Z_{\tilde{q}}^{N_a} Z_{\sigma_C^+} Z_{\sigma_C^-} \; .
\end{align}
Above we have included Chern-Simons levels $k_1$ for $U(N_1)$ etc.\ but for now
we take these levels to all vanish. Note also that for
$\tilde{U}_I = U(\tilde{N}_I)$ the topological symmetry is the same as for
$U_I = U(N_I)$ so in both cases we label the fugacity $z_I$. 
There can be a non-trivial
mapping of topological fugacities under Seiberg-like duality. Specifically, while the
fugacities in theory B are the same as those in theory A, in theory C the
matching requires the replacement of $z_2 \rightarrow 1/z_2$ and
$z_1 \rightarrow z_1 z_2$.
Indeed, this mapping for theory C is to be expected from the mapping of FI
terms as seen from analysis of contact terms in similar contexts of Seiberg
duality \cite{Nishinaka:2013pua, Benini:2014mia, Hwang:2017kmk, Pasquetti:2019uop} and it was observed explicitly in this
context in \cite{Benvenuti:2020wpc}.~\footnote{The mapping given in
\cite{Benvenuti:2020wpc} differs by exchanging the indices
$1 \leftrightarrow 2$ on the topological fugacities but this appears to be a
typo.}
The reason the mapping is trivial when the gauge node being dualized is only
attached to other gauge nodes, i.e.\ has no flavors, is not entirely clear. However,
the explanation seems to be that the same mapping holds~\footnote{I.e.\ specifically $z_1 \rightarrow 1/z_1$ and
$z_2 \rightarrow z_1 z_2$ for the map to theory B.}
but is a symmetry exchanging some monopole operators. This can be seen explicitly in the examples later in this section.
Note also that for a single gauge node the index is symmetric under the inversion of the topological fugacity $z \rightarrow 1/z$ so this mapping of topological fugacities also cannot be observed in those indices.

The matter contributions are given by
\begingroup
\allowdisplaybreaks
\begin{align}
Z_{B, \tilde{B}} = & Z_{BiFund \; U_1-U_2}(r_B, a_1) \\
Z_Q^{N_f} = & Z_{N_f \; Fund \; U_2}(r_Q, a_2) \\
Z_{\tilde{Q}}^{N_a} = & Z_{N_a \; AntiFund \; U_2}(r_Q, a_2) \\
Z_{b, \tilde{b}} = & Z_{BiFund \; U_1-U_2}(1 - r_B, a_1^{-1}) \\
Z_{\phi_b} = & Z_{Adj \; U_1}(2r_B, a_2^2) \\
Z_{\sigma_B^{\pm}} = & Z_{Singlet}(r_{\sigma_B}, z_1^{\pm} a_1^{-2N_2}) \\
Z_c = & Z_{BiFund \; U_I-U_J}(1 - r_B, a_1^{-1}) \\
Z_{\phi_c} = & Z_{Adj \; U_I}(2r_B, a_1^2) \\
Z^{N_f}_p = & Z_{Fund \; U_1}(r_B + r_Q, a_1 a_2) \\
Z^{N_f}_{\tilde{p}} = & Z_{AntiFund \; U_1}(r_B + r_Q, a_1 a_2) \\
Z^{N_f}_q = & Z_{Fund \; U_2}(1 - r_Q, a_2^{-1}) \\
Z^{N_f}_{\tilde{q}} = & Z_{Fund \; U_2}(1 - r_Q, a_2^{-1}) \\
Z_{\sigma_C^{\pm}} = & Z_{Singlet}(r_{\sigma_C}, z_2^{\pm}a _1^{-2N_1} a_2^{-2N_f})
\end{align}
\endgroup
where $r_{\sigma_B} \equiv (1-r_B)N_2 - (N_1 - 1)$ and
$r_{\sigma_C} \equiv (1-r_B)N_1 + (1-r_Q)N_f - (N_2 - 1)$ and
\begingroup
\allowdisplaybreaks
\begin{align}
Z_{BiFund \; U_I-U_J}(r, a) & = \prod_{i = 1}^{N_I} \prod_{j=1}^{N_J} (q^{1-r} a^{-2})^{(|m^{(I)}_i - m^{(J)}_j|)/2}
 \frac{(q^{1 + (|m^{(I)}_i - m^{(J)}_j| - r)/2} a^{-1} s_i^{(I) \mp} s_j^{(J) \pm}; q)_{\infty}}{(q^{(|m^{(I)}_i - m^{(J)}_j| + r)/2} a s_i^{(I) \pm} s_j^{(J) \mp}; q)_{\infty}} \\
Z^{N_f}_{Fund \; U_I} & = \prod_{\alpha = 1}^{N_f} \prod_{i=1}^{N_I} (q^{1-r} a^{-2} (s^{(I)}_i)^{-2})^{|m^{(I)}_i|/4}
 \frac{(q^{1 + (|m^{(I)}_i| - r)/2} a^{-1} (s^{(I)}_i)^{-1} x_{\alpha}^{-1}; q)_{\infty}}{(q^{(|m^{(I)}_i| + r)/2} a s^{(I)}_i x_{\alpha}; q)_{\infty}} \\
Z^{N_a}_{AntiFund \; U_I} & = \prod_{\alpha = 1}^{N_a} \prod_{i=1}^{N_I} (q^{1-r} a^{-2} (s^{(I)}_i)^{2})^{|m^{(I)}_i|/4}
 \frac{(q^{1 + (|m^{(I)}_i| - r)/2} a^{-1} s^{(I)}_i \tilde{x}_{\alpha}^{-1}; q)_{\infty}}{(q^{(|m^{(I)}_i| + r)/2} a (s^{(I)}_i)^{-1} \tilde{x}_{\alpha}; q)_{\infty}} \\
Z_{Adj \; U_I} & = \prod_{i, j = 1}^{N_I} (q^{1 - r} a^{-2})^{|m^{(I)}_i - m^{(I)}_j|/4}
 \frac{(q^{1 + (|m^{(I)}_i - m^{(I)}_j| - r)/2} a^{-1} (s^{(I)}_i)^{-1} s^{(I)}_j; q)_{\infty}}{(q^{(|m^{(I)}_i - m^{(I)}_j| + r)/2} a  s^{(I)}_i (s^{(I)}_j)^{-1}; q)_{\infty}} \\
Z^{N_fN_a}_M & = \prod_{\alpha = 1}^{N_f} \prod_{\beta = 1}^{N_a} \frac{(q^{1 - r_Q} a_2^{-2} x_{\alpha}^{-1} \tilde{x}_{\beta}; q)_{\infty}}{(q^{r_Q} a_2^2 x_{\alpha} \tilde{x}_{\beta}^{-1}; q)_{\infty}} \; .
\end{align}
\endgroup

\subsection{Boundary 't Hooft anomalies}
Initially, we focus on the following sets of boundary conditions:
\begin{itemize}
\item $(\Ncal, \Ncal, N, N, N, N)$ for
$(\mathrm{VM}_1, \mathrm{VM}_2, B, \tilde{B}, Q, \tilde{Q})$ in theory A.
\item $(\Dcal, \Ncal, D, D, N, N, N, D)$ for 
$(\mathrm{VM}_3, \mathrm{VM}_2, b, \tilde{b}, \phi_b, Q, \tilde{Q}, \sigma_B^{\pm})$ in theory B.
\item $(\Ncal, \Dcal, D, D, N, N, N, N, D, D, D)$ for 
$(\mathrm{VM}_1, \mathrm{VM}_4, c, \tilde{c}, \phi_c, p, \tilde{p}, M, q, \tilde{q}, \sigma_C^{\pm})$ in theory C.
\end{itemize}
We will see that with suitable additional 2d boundary multiplets, we can
cancel the gauge anomalies for the gauge group factors with Neumann boundary
conditions, and match the anomalies.

For the multiplets we have discussed the contributions to the anomaly polynomial
are given by the following expressions \cite{Dimofte:2017tpi, Okazaki:2021pnc}
if we have Dirichlet boundary conditions.
For Neumann boundary conditions we just take the opposite sign.
Taking the multiplets to have R-charge $q_R$, a vector of $U(1)_{a_i}$
charges $\underline{q}$ we have
\begin{align}
\Acal_{VM \; U_I} = & -N_I \Tr({\bf s_I}^2) + \Tr({\bf s_I})^2 - \frac{N_I^2}{2} {\bf r}^2 \\
\Acal_{N_f \; Fund \; U_I}(q_R, \underline{q}) = & \frac{N_I}{2} \Tr({\bf x}^2) + \frac{N_f}{2} \Tr({\bf s_I}^2) + 2N_f \Tr({\bf s_I}) \left( \underline{q} \cdot \underline{{\bf a}} + (q_R - 1){\bf r} \right) \nonumber \\
 & + N_I N_f \left( \underline{q} \cdot \underline{{\bf a}} + (q_R - 1){\bf r} \right)^2 \nonumber \\
\Acal_{N_a \; AntiFund \; U_I}(q_R, \underline{q}) = & \frac{N_I}{2} \Tr({\bf \tilde{x}}^2) + \frac{N_a}{2} \Tr({\bf s_I}^2) - 2N_a \Tr({\bf s_I}) \left( \underline{q} \cdot \underline{{\bf a}} + (q_R - 1){\bf r} \right) \nonumber \\ 
 & + N_I N_a \left( \underline{q} \cdot \underline{{\bf a}} + (q_R - 1){\bf r} \right)^2 \nonumber \\
\Acal_{BiFund \; U_I-U_J}(q_R, \underline{q}) = & \frac{N_I}{2} \Tr({\bf s_J}^2) + \frac{N_J}{2} \Tr({\bf s_I}^2) + N_I N_J \left( \underline{q} \cdot \underline{{\bf a}} + (q_R - 1){\bf r} \right)^2 \nonumber \\
 & + \left( N_J \Tr({\bf s_I}) - N_I \Tr({\bf s_J}) \right) \left( \underline{q} \cdot \underline{{\bf a}} + (q_R - 1){\bf r} \right) \\
\Acal_{Adj \; U_I}(q_R, \underline{q}) = & N_I \Tr({\bf s_I}^2) - \Tr({\bf s_I})^2 + \frac{N_I^2}{2} \left( \underline{q} \cdot \underline{{\bf a}} + (q_R - 1){\bf r} \right)^2 \; .
\end{align}
In this notation ${\bf s}_I$ is the $U(N_I)$ field strength,
$\underline{\bf a}$ is a vector of $U(1)_{a_i}$ field strengths.

We then find for $N_a = N_f$ the following anomaly polynomials for the bulk fields, including FI
contributions $2 {\bf s}_I {\bf y}_I$ or $2 \tilde{\bf s}_I {\bf y}_I$ for each
$U(N_I)$ or $U(\tilde{N}_I)$ gauge group factor
where ${\bf y}_I$ are the topological $U(1)_{y_I}$ field strengths
\begin{align}
\Acal^{A \; Bulk}_{\Ncal, \Ncal, N, N, N, N} = &
  N_1 N_2 {\bf a}_1^2 - N_2 N_f {\bf a}_2^2 + 2N_1 N_2 (1 - r_B){\bf a}_1 {\bf r
} + 2N_2 N_f (1 - r_B){\bf a}_2 {\bf r} \nonumber \\
 & + \left( \frac{1}{2}(N_1 - N_2)^2 + N_1 N_2 r_B(2 - r_B) - N_2 N_f (1 - r_Q)^2 \right) {\bf r}^2 \nonumber \\ 
 & + \Tr({\bf s}_1)^2 + (N_1 - N_2) \Tr({\bf s}_1^2) - \Tr({\bf s}_2)^2 + (-N_1 + N_2 - N_f)\Tr({\bf s}_2^2) \nonumber \\
 & + 2 {\bf y}_1 \Tr({\bf s}_1) + 2 {\bf y}_2 \Tr({\bf s}_2) - \frac{1}{2}N_2 \left( \Tr({\bf x}_f^2) + \Tr({\bf x}_a^2) \right) \\
\Acal^{B \; Bulk}_{\Dcal, \Ncal, D, D, N, N, N, D} = &
  N_1 N_2 {\bf a}_1^2 - N_2 N_f {\bf a}_2^2 + 2N_1 N_2 (1 - r_B){\bf a}_1 {\bf r
} + 2N_2 N_f (1 - r_B){\bf a}_2 {\bf r} \nonumber \\
 & + \left( \frac{1}{2}(N_1 - N_2)^2 + N_1 N_2 r_B(2 - r_B) - N_2 N_f (1 - r_Q)^2 \right) {\bf r}^2 \nonumber \\ 
 & + (-N_1 + N_2 - N_f) \Tr({\bf s}_2^2) + \Tr(\tilde{\bf s}_1)^2 + N_1\Tr(\tilde{\bf s}_1^2) \nonumber \\
 & + 2 {\bf y}_1 \Tr(\tilde{\bf s}_1) + {\bf y}_1^2 + 2 {\bf y}_2 \Tr({\bf s}_2) - \frac{1}{2}N_2 \left( \Tr({\bf x}_f^2) + \Tr({\bf x}_a^2) \right) \\
\Acal^{C \; Bulk}_{\Ncal, \Dcal, D, D, N, N, N, N, D, D, D} = & 
  N_1 N_2 {\bf a}_1^2 - N_2 N_f {\bf a}_2^2 + 2N_1 N_2 (1 - r_B){\bf a}_1 {\bf r} + 2N_2 N_f (1 - r_B){\bf a}_2 {\bf r} \nonumber \\
 & + \left( \frac{1}{2}(N_1 - N_2)^2 + N_1 N_2 r_B(2 - r_B) - N_2 N_f (1 - r_Q)^2 \right) {\bf r}^2 \nonumber \\
 & + (N_1 - N_2) \Tr({\bf s}_1^2) + \Tr(\tilde{\bf s}_2)^2 + N_2\Tr(\tilde{\bf s}_2^2) \nonumber \\
 & + 2 {\bf y}_1 \Tr({\bf s}_1) + 2 {\bf y}_2 \Tr(\tilde{\bf s}_2) + {\bf y}_2^2 - \frac{1}{2}N_2 \left( \Tr({\bf x}_f^2) + \Tr({\bf x}_a^2) \right)
\end{align}

Recalling that $\tilde{N}_1 = N_2 - N_1$ and $\tilde{N}_2 = N_1 - N_2 + N_f$, and that due to
the $\Dcal$ boundary conditions $U(\tilde{N}_1)$ and $U(\tilde{N}_2)$ are global symmetries on
the boundary, whereas $U(N_1)$ and $U(N_2)$ are gauge symmetries due to the
$\Ncal$ boundary conditions, we can cancel all gauge anomalies with the
following 2d multiplets.
\begin{align}
\label{UN_charges_2d}
\begin{array}{c|c|c|c|c|c|c|c}
 & & U(N_1) & U(N_2) & U(\tilde{N}_1) & U(\tilde{N}_2) & U(1)_{y_1} & U(1)_{y_2} \\ \hline
\mathrm{Fermi} & \eta_{1} & {\bf det} & {\bf 1} & {\bf det^{-1}} & {\bf 1} & -1 & 0 \\
\mathrm{Fermi} & \eta_{2} & {\bf 1} & {\bf det} & {\bf 1} & {\bf det^{-1}} & 0 & -1 \\
\mathrm{Fermi} & \Gamma_{1} & {\bf N_1} & {\bf 1} & {\bf \tilde{N}_1} & {\bf 1} & 0 & 0 \\
\mathrm{Fermi} & \Gamma_{2} & {\bf 1} & {\bf N_2} & {\bf 1} & {\bf \tilde{N}_2} & 0 & 0 \\
\mathrm{Chiral} & \chi_1 & {\bf det} & {\bf 1} & {\bf 1} & {\bf 1} & 0 & 0 \\
\mathrm{Chiral} & \chi_2 & {\bf 1} & {\bf det} & {\bf 1} & {\bf 1} & 0 & 0
\end{array}
\end{align}
In particular, subject the the ambiguity described below, we need to include
\begin{itemize}
\item $\eta_{1}$, $\eta_{2}$, $\Gamma_{1}$, $\Gamma_{2}$ in theory A.
\item $\eta_{2}$, $\Gamma_{2}$, $\chi_2$ in theory B.
\item $\eta_{1}$, $\Gamma_{1}$, $\chi_1$ in theory C.
\end{itemize}

Including the contribution of those 2d multiplets in each theory all gauge anomalies are cancelled and the resulting anomaly polynomials match
\begin{align}
\Acal^{Total} = &
  N_1 N_2 {\bf a}_1^2 - N_2 N_f {\bf a}_2^2 + 2N_1 N_2 (1 - r_B){\bf a}_1 {\bf r
} + 2N_2 N_f (1 - r_B){\bf a}_2 {\bf r} \nonumber \\
 & + \left( \frac{1}{2}(N_1 - N_2)^2 + N_1 N_2 r_B(2 - r_B) - N_2 N_f (1 - r_Q)^2 \right) {\bf r}^2 \nonumber \\ 
 & + \Tr(\tilde{\bf s}_1)^2 + N_1 \Tr(\tilde{\bf s}_1^2) + \Tr(\tilde{\bf s}_2)^2 + N_2\Tr(\tilde{\bf s}_2^2) \nonumber \\ 
 & + 2 {\bf y}_1 \Tr(\tilde{\bf s}_1) + {\bf y}_1^2 + 2 {\bf y}_2 \Tr(\tilde{\bf s}_2) + {\bf y}_2^2 - \frac{1}{2}N_2 \left( \Tr({\bf x}_f^2) + \Tr({\bf x}_a^2) \right)
\end{align}

It should be noted that conjugate representations give the same contribution to
the anomaly polynomial, so we cannot distinguish between them here. For Fermi
multiplets this is not important as conjugate representations also give the
same contribution to the half-index. However, for 2d chirals the contributions
are not quite the same, so we should also consider the possibility that the
chirals $\chi_I$ are in the ${\bf det^{-1}}$ representations.

While the theory A half-indices can be straightforwardly evaluated, due to the presence of 2d chiral multiplets the theory B and C half-indices cannot be
evaluated simply by taking a contour around the origin. Instead it is necessary to use the JK residue prescription
\cite{MR1318878, Benini:2013nda, Benini:2013xpa}. 
This involves some computational complexity so we
do not evaluate these in this article,
leaving the issue of checking the matching of half-indices to future work.

We could also consider the case with all boundary conditions reversed. As for
the symplectic and orthogonal groups we can cancel the gauge anomalies and match the global anomalies by rearranging the bifundamental Fermis, but now we also
need to change the sign of the topological charges of the bi-determinant Fermis.
As there are fewer constraints from cancelling gauge anomalies in this case,
there are two obvious ways to include the determinant representation 2d matter
to match anomalies. In particular we find either
\begin{itemize}
\item $\chi_{1}$, $\chi_{2}$ in theory A.
\item $\eta'_{1}$, $\Gamma_{1}$, $\chi_1$ in theory B.
\item $\eta'_{2}$, $\Gamma_{2}$, $\chi_2$ in theory C.
\end{itemize}
or
\begin{itemize}
\item None in theory A.
\item $\eta'_{1}$, $\Gamma_{1}$, $\tilde{\chi}_2$ in theory B.
\item $\eta'_{2}$, $\Gamma_{2}$, $\tilde{\chi}_1$ in theory C.
\end{itemize}
where we have defined the following additional 2d matter multiplets
\begin{align}
\label{UN_charges2}
\begin{array}{c|c|c|c|c|c|c|c}
 & & U(N_1) & U(N_2) & U(\tilde{N}_1) & U(\tilde{N}_2) & U(1)_{y_1} & U(1)_{y_2} \\ \hline
\mathrm{Fermi} & \eta'_{1} & {\bf det} & {\bf 1} & {\bf det^{-1}} & {\bf 1} & 1 & 0 \\
\mathrm{Fermi} & \eta'_{2} & {\bf 1} & {\bf det} & {\bf 1} & {\bf det^{-1}} & 0 & 1 \\
\mathrm{Fermi} & \tilde{\chi}_1 & {\bf det} & {\bf 1} & {\bf 1} & {\bf 1} & 0 & 0 \\
\mathrm{Fermi} & \tilde{\chi}_2 & {\bf 1} & {\bf det} & {\bf 1} & {\bf 1} & 0 & 0
\end{array}
\end{align}

\subsection{$U(1)\times U(2)-[2]$ $(N_1 = 1, N_2 = 2, N_f = N_a = 2)$}
\label{sec:U1U2_2}
We have checked that the three supersymmetric indices coincide. 
For $r_B=1/5$, $r_Q=3/8$ we find the full indices
\begingroup
\allowdisplaybreaks
\begin{align}
\label{eq:U1U2_2}
&I^A = I^B = I^C
\nonumber\\
=&1+\underbrace{a_1^2 q^{1/5}}_{\Tr(B\tilde{B})}
+\underbrace{4 a_2^2 q^{3/8}}_{\Tr(\tilde{Q}Q)}
+\underbrace{a_1^4 q^{2/5}}_{\Tr(B\tilde{B})^2}
+\frac{q^{21/40}}{a_1 a_2^2}
   ( \underbrace{z_1 z_2 + z_1^{-1}z_2^{-1}}_{v_A^{\pm,\pm}} +
    \underbrace{z_2 + z_2^{-1}}_{v_A^{0,\pm}} )
   +\underbrace{8 a_1^2 a_2^2 q^{23/40}}_{\begin{smallmatrix} \Tr(\tilde{Q}Q) \Tr(B \tilde{B}) , \\ \Tr(\tilde{Q}B\tilde{B}Q) \end{smallmatrix}}
   +\underbrace{a_1^6 q^{3/5}}_{\Tr(B\tilde{B})^3}
   \nonumber\\
   & + \frac{a_1 q^{29/40}}{a_2^2}
   ( \underbrace{z_1 z_2 + z_1^{-1}z_2^{-1}}_{v_A^{\pm,\pm} \Tr(B \tilde{B})} +
    \underbrace{z_2 + z_2^{-1}}_{v_A^{0,\pm} \Tr(B \tilde{B})} )
   +\underbrace{10 a_2^4 q^{3/4}}_{\Tr(\tilde{Q}Q)^2}
   +\underbrace{8 a_1^4 a_2^2 q^{31/40}}_{\begin{smallmatrix} \Tr(\tilde{Q}Q) \Tr(B \tilde{B})^2 , \\ \Tr(\tilde{Q}B\tilde{B}Q) \Tr(B \Tilde{B}) \end{smallmatrix}}
   + q^{4/5}
   \Big( \underbrace{a_1^{8}}_{\Tr(B \tilde{B})^4} + \underbrace{\frac{1}{a_1^2} (z_1 + z_1^{-1})}_{v_A^{\pm, 0}} \Big)
   \nonumber\\
   & + \frac{4 q^{9/10}}{a_1}
   ( \underbrace{z_1 z_2 + z_1^{-1}z_2^{-1}}_{v_A^{\pm,\pm} \Tr(\tilde{Q} Q)} +
    \underbrace{z_2 + z_2^{-1}}_{v_A^{0,\pm} \Tr(\tilde{Q} Q)} )
   + \frac{a_1^3 q^{37/40}}{a_2^2}
   ( \underbrace{z_1 z_2 + z_1^{-1}z_2^{-1}}_{v_A^{\pm,\pm} \Tr(B \tilde{B})^2} +
    \underbrace{z_2 + z_2^{-1}}_{v_A^{0,\pm} \Tr(B \tilde{B})^2} ) \nonumber \\
 & + \underbrace{25 a_1^2 a_2^4
   q^{19/20}}_{\begin{smallmatrix} \Tr(\tilde{Q}Q)^2 \Tr(B \tilde{B}) , \\ \Tr(\tilde{Q}Q) \Tr(\tilde{Q}B\tilde{B}Q) \end{smallmatrix}}
 + \underbrace{8 a_1^6 a_2^2 q^{39/40}}_{\begin{smallmatrix} \Tr(\tilde{Q}Q) \Tr(B \tilde{B})^3 , \\ \Tr(\tilde{Q}B\tilde{B}Q) \Tr(B \Tilde{B})^3 \end{smallmatrix}}
 + \Big(\underbrace{a_1^{10}}_{\Tr(B \Tilde{B})^5}
  - \underbrace{10}_{B \psi_{\tilde{B}} , \tilde{B} \psi_B , Q \psi_{\tilde{Q}} , \tilde{Q} \psi_Q} \Big) q + \cdots \; .
\end{align}
\endgroup
To this order we have identified all the operators in theory A which contribute.
Note that the
contribution
at order $q^{19/20}$ would naively have a coefficient $26$ from
$\Tr(\tilde{Q}Q)^2$ $\Tr(B \tilde{B})$ giving a
contribution $10$ to the coefficient and
$\Tr(\tilde{Q}Q)$ $\Tr(\tilde{Q}B\tilde{B}Q)$ giving a contribution $16$.
However, noting the particular gauge groups involved, it turns out that there
is one linear relation between the $6$ operators which would generically be
formed in these ways from $Q^1, Q^2, \tilde{Q}^1, \tilde{Q}^2, B, \tilde{B}$.
Therefore the coefficient is in fact expected to be $25$, in agreement with
the index calculations.

While the identification of operators above is given for theory A, we can
make the identification in theories B and C too. The details of the operator
map are given in \cite{Benvenuti:2020wpc} and are similar to the maps for the
symplectic and orthogonal gauge groups so we do not repeat them here.

\subsection{Chern-Simons levels}
As is well known, we can also generate non-zero Chern-Simons levels by giving large positive or
negative masses to the chirals $Q$ as first derived for unitary groups
by taking a limit of the partition function in \cite{Willett:2011gp}, following the original proposal using brane constructions in \cite{Giveon:2008zn}.
We can also generate non-zero Chern-Simons levels by giving large positive or
negative masses to the chirals $Q$ and $\tilde{Q}$. In the simplest case we
start with $N_f = N_a = \hat{N}_f + |k|$ fundamental and anti-fundamental
chirals and give masses (of the same sign) to $|k|$ of them to leave
$N_f = N_a = \hat{N}_f$ fundamental and anti-fundamental chirals. This generates
a Chern-Simons level $k$ for the $U(N_2)$ gauge node in theories A and B with
$k>0$ by sending the masses to $+\infty$ and $k<0$ by sending the masses to
$-\infty$. There are no other changes to theories A and B so the result is
that the duality holds for arbitrary Chern-Simons level $k$ for $U(N_2)$.
However, the effect on theory C is more involved. In the case where we send the
masses of $|k|$ multiplets $Q$ and $\tilde{Q}$ to $\pm \infty$, in theory C we
must also send the masses of $|k|$ multiplets $q$ and $\tilde{q}$ and of
$\sigma_C^{+}$ and $\sigma_C^{-}$ to $\mp \infty$ while also sending the masses
of $|k|$ multiplets $p$ and $\tilde{p}$ to $\pm \infty$. This modifies theory
C by having $\tilde{N}_2 = N_1 - N_2 + \hat{N}_f + |k|$ while having $\hat{N}_f$
flavors, removing the $\sigma_C^{\pm}$ multiplets, and giving gauge group with
Chern-Simons levels $U(N_1)_k \times U(\tilde{N}_2)_{-k}$.

\subsection{$U(1)\times U(2)_k-[2-|k|]$ $(N_1 = 1, N_2 = 2, \hat{N}_f = 2 - |k|$)}
By sending some or all chiral flavor masses to infinity for the example
given in section~\ref{sec:U1U2_2}, we can produce
examples with $k = \pm 1$ or $k = \pm 2$ where theory A has gauge
group $U(1) \times U(2)_k$ with $2 - |k|$ flavors, theory B has gauge group
$U(1) \times U(2)_k$ with $2 - |k|$ flavors, and theory C has gauge group
$U(1)_k \times U(1)_{-k}$ with $2 - |k|$ flavors and no $\sigma_C^{\pm}$
multiplets.

For $k=2$ we have removed all flavors and we find agreement of the full
indices given by
\begingroup
\allowdisplaybreaks
\begin{align}
I^A = & I^B = I^C = \nonumber \\
 & 1+a_1^2 q^{1/5}+a_1^4 q^{2/5}+a_1^6 q^{3/5}+\frac{q^{4/5}
   \left(a_1^{10}+z_1+\frac{1}{z_1}\right)}{a_1^2}+\left(a_1^{10}-2\right) q+a_1^{12} q^{6/5}
   \nonumber\\
   &
   +a_1^{14} q^{7/5}+\frac{q^{8/5}
   \left(a_1^{20} z_1^2+z_1^4+1\right)}{a_1^4 z_1^2}+\left(a_1^{18}+\frac{1}{a_1^2}\right)
   q^{9/5}+\left(a_1^{20}-3\right) q^2+a_1^2 \left(a_1^{20}-2\right) q^{11/5}
   \nonumber\\
   &
   +\frac{q^{12/5} \left(a_1^{30}
   z_1^3+z_1^6+1\right)}{a_1^6 z_1^3}+a_1^{26} q^{13/5}+\frac{\left(a_1^{30}+2\right) q^{14/5}}{a_1^2}+q^3
   \left(a_1^{30}+z_1+\frac{1}{z_1}\right)
   \nonumber\\
   &
   +\frac{q^{16/5} \left(a_1^{40}-4
   a_1^{10}+z_1^4+\frac{1}{z_1^4}\right)}{a_1^8}+a_1^4 \left(a_1^{30}-2\right) q^{17/5}+a_1^{36} q^{18/5}
   \nonumber\\
   &
   +\frac{q^{19/5}
   \left(a_1^{40} z_1-2 z_1^2+z_1-2\right)}{a_1^2 z_1}+\cdots
\end{align}
\endgroup
We can see that, at least to order $q$, the indices agree with the results for the $k=0$ case given in
equation~(\ref{eq:U1U2_2}) subject to the removal of all terms (in the theory A interpretation) involving the
fundamental chiral multiplets (containing $Q$ and $\psi_Q$) or any monopole
operators with flux in the $U(2)$ gauge group -- monopoles with flux only in
$U(1)$ survive.

\subsection{$U(N_1) \times U(N_2)-[N_f] \times U(N_3)$}
As for the previous gauge groups, we can easily generalize to a linear
quiver with three gauge nodes, with a theory D corresponding to dualizing
the $U(N_3)$ gauge node. These theories are described by the quiver diagrams
in Figure~\ref{fig:U_U_U}. The structure of theory D is the same as of theory B
while, up to some notational differences, theories A and B are the same as for
the case of two gauge nodes other than the additional $U(N_3)$ gauge node and
bifundamental matter connecting gauge nodes $2$ and $3$. The generalization of
theory C to the case of 3 gauge nodes is a little more involved but still
straightforward.
The theory C superpotential is given by
\begin{align}
\Wcal = & \sigma_C^{\pm} v_C^{0, \pm, 0} + \sum_{I=1}^2 \left(
 \Tr(\tilde{c}_I \phi_c^I c_I) + \Tr(q c_I p_I)
 + \Tr(\tilde{q} \tilde{c}_I \tilde{p}_I) \right)
 \nonumber\\
 &
 + \Tr(c_1 c_2 c_3) + \Tr(\tilde{c}_1 \tilde{c}_2 \tilde{c}_3)
 + \Tr(qM\tilde{q})
\end{align}
The topological fugacities are mapped non-trivially but in a
generalization of the result for the case of two gauge nodes so that in
theory C (assuming $N_f \ne 0$) we need to replace $z_1 \rightarrow z_1z_2$, $z_2 \rightarrow 1/z_2$
and $z_3 \rightarrow z_2z_3$.

\subsection{$U(1) \times U(2)-[2] \times U(1)$}

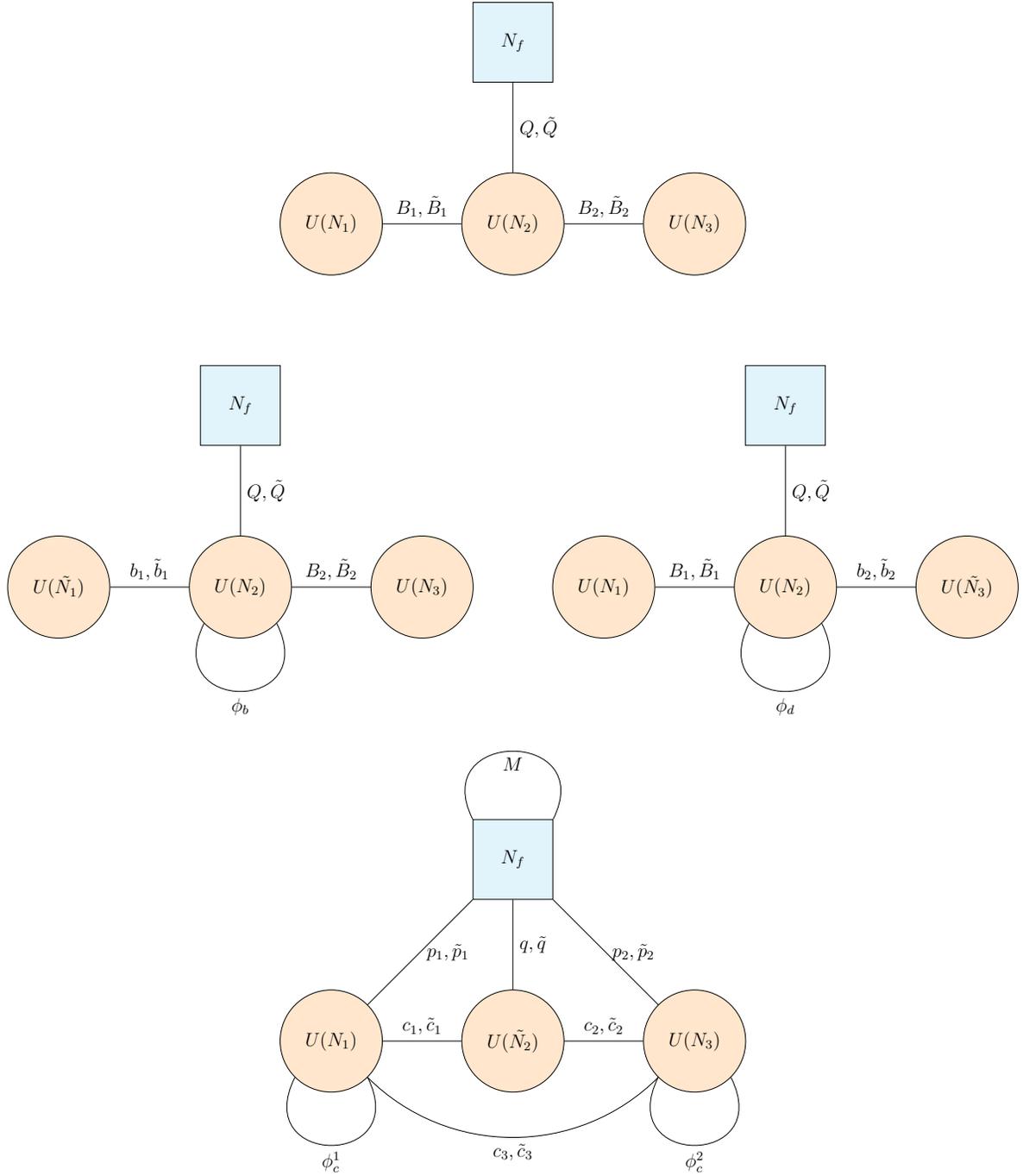
\begin{figure}
\centering
\scalebox{0.7}{
\begin{tikzpicture}
\path (-4,0) node[circle, minimum size=64, fill=orange!20, draw](AG1) {$U(N_1)$}
(0,0) node[circle, minimum size=64, fill=orange!20, draw](AG2) {$U(N_2)$}
(4,0) node[circle, minimum size=64, fill=orange!20, draw](AG3) {$U(N_3)$}
(0,4) node[minimum size=50, fill=cyan!10, draw](AF2) {$N_f$};
\draw (AG1) -- (AG2) node [midway, above] {$B_1, \tilde{B}_1$};
\draw (AG3) -- (AG2) node [midway, above] {$B_2, \tilde{B}_2$};
\draw (AG2) -- (AF2) node [midway, right] {$Q, \tilde{Q}$};
\path (-10,-8) node[circle, minimum size=64, fill=orange!20, draw](BG1) {$U(\tilde{N_1})$}
(-6,-8) node[circle, minimum size=64, fill=orange!20, draw](BG2) {$U(N_2)$}
(-2,-8) node[circle, minimum size=64, fill=orange!20, draw](BG3) {$U(N_3)$}
(-6,-4) node[minimum size=50, fill=cyan!10, draw](BF2) {$N_f$};
\draw (BG1) -- node[above]{$b_1, \tilde{b}_1$} (BG2);
\draw (BG3) -- node[above]{$B_2, \tilde{B}_2$} (BG2);
\draw (BG2) -- node[right]{$Q, \tilde{Q}$}(BF2);
\draw (BG2.south west) .. controls ++(-1, -2) and ++(1, -2) .. node[below]{$\phi_b$} (BG2.south east);
%
%
\path (2,-8) node[circle, minimum size=64, fill=orange!20, draw](DG1) {$U(N_1)$}
(6,-8) node[circle, minimum size=64, fill=orange!20, draw](DG2) {$U(N_2)$}
(10,-8) node[circle, minimum size=64, fill=orange!20, draw](DG3) {$U(\tilde{N_3})$}
(6,-4) node[minimum size=50, fill=cyan!10, draw](DF2) {$N_f$};
\draw (DG1) -- node[above]{$B_1, \tilde{B}_1$} (DG2);
\draw (DG3) -- node[above]{$b_2, \tilde{b}_2$} (DG2);
\draw (DG2) -- node[right]{$Q, \tilde{Q}$}(DF2);
\draw (DG2.south west) .. controls ++(-1, -2) and ++(1, -2) .. node[below]{$\phi_d$} (DG2.south east);
%
\path (-4,-18) node[circle, minimum size=64, fill=orange!20, draw](CG1) {$U(N_1)$}
(0,-18) node[circle, minimum size=64, fill=orange!20, draw](CG2) {$U(\tilde{N_2})$}
(4,-18) node[circle, minimum size=64, fill=orange!20, draw](CG3) {$U(N_3)$}
(0,-14) node[minimum size=50, fill=cyan!10, draw](CF2) {$N_f$};
\draw (CG1) -- node[above]{$c_1, \tilde{c}_1$} (CG2);
\draw (CG3) -- node[above]{$c_2, \tilde{c}_2$} (CG2);
\draw (CG1) -- node[right]{$p_1, \tilde{p}_1$}(CF2);
\draw (CG3) -- node[right]{$p_2, \tilde{p}_2$}(CF2);
\draw (CG2) -- node[right]{$q, \tilde{q}$}(CF2);
\draw (CG1) to [out=-45, in=-135] node[below]{$c_3, \tilde{c}_3$}(CG3);
\draw (CG1.south west) .. controls ++(-1, -2) and ++(1, -2) .. node[below]{$\phi_c^1$} (CG1.south east);
\draw (CG3.south west) .. controls ++(-1, -2) and ++(1, -2) .. node[below]{$\phi_c^2$} (CG3.south east);
\draw (CF2.north east) .. controls ++(1, 2) and ++(-1, 2) .. node[below]{$M$} (CF2.north west);
\end{tikzpicture}
}
\caption{Quadrality of $U(N_1) \times U(N_2)-[N_f] \times U(N_3)$ quiver 
where $\tilde{N_1}=N_2-N_1$, $\tilde{N_2}=N_1 + N_3 + N_f - N_2 $ and $\tilde{N_3}=N_2-N_3$. } \label{fig:U_U_U}
\end{figure}

Theory A and its dual theory C, by dualizing the $U(2)$ gauge node, are given
by the quivers in Figure \ref{fig:U_U_U}. We have confirmed that the full indices match. 
For $r_B=2/5$ and $r_Q=2/7$ they are
given by
\begin{align}
&I^A = I^B = I^C = I^D
= \nonumber \\
 &
1+4 a_2^2 q^{2/7}+q^{2/5} \left(a_1^2+a_3^2\right)+10 a_2^4 q^{4/7}+q^{3/5}
   \left(\frac{z_1+\frac{1}{z_1}}{a_1^2}+\frac{z_3^2+1}{a_3^2
   z_3}\right)+8 a_2^2 q^{24/35} \left(a_1^2+a_3^2\right)
   \nonumber\\
   &
   +q^{4/5}
   \left(a_1^2+a_3^2\right)^2+\frac{q^{57/70} (z_1+1) (z_3+1)
   \left(z_1 z_2^2 z_3+1\right)}{a_1 a_2^2 a_3 z_1
   z_2 z_3}+20 a_2^6 q^{6/7}
   \nonumber\\
   &+4 a_2^2 q^{31/35}
   \left(\frac{z_1+\frac{1}{z_1}}{a_1^2}+\frac{z_3^2+1}{a_3^2
   z_3}\right)+25 a_2^4 q^{34/35}
   \left(a_1^2+a_3^2\right)+\cdots
   \end{align}

\section{Circular linear quivers}
\label{sec_linear}

As we have confirmed Seiberg-like dualities of linear quivers, we generalize
the dualities to circular quiver gauge theories. 
This is very straightforward. We just introduce an extra bifundamental multiplet to connect the two gauge nodes at the ends of the linear quiver. Dualizing
any interior gauge node is the same with the additional bifundamental present in the original and dual theory, and the superpotential is not altered by the presence of this additional bifundamental. dualizing the end nodes is changed, but now they dualize in the same way as all the other nodes. We illustrate with some examples
with three gauge nodes and due to the symmetry we focus only on the dualization of what was the middle node (i.e.\ to theory C).

\subsection{Circular $USp(2)\times USp(6)-[6]\times USp(2)$}
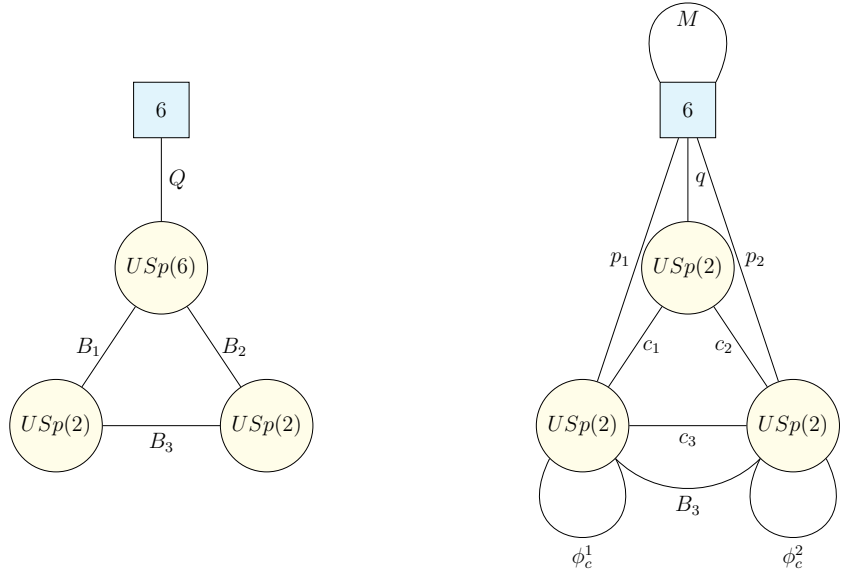
\begin{figure}
\centering
\scalebox{0.7}{
\begin{tikzpicture}
\path 
(0,0) node[circle, minimum size=40, fill=yellow!10, draw](AG1) {$USp(6)$}
(-2,-3) node[circle, minimum size=40, fill=yellow!10, draw](AG2) {$USp(2)$}
(2,-3) node[circle, minimum size=40, fill=yellow!10, draw](AG3) {$USp(2)$}
(0,3) node[minimum size=30, fill=cyan!10, draw](AF2) {$6$};
\draw (AG2) -- (AG1) node [midway, left] {$B_1$};
\draw (AG3) -- (AG1) node [midway, right] {$B_2$};
\draw (AG2) -- (AG3) node [midway, below] {$B_3$};
\draw (AG1) -- (AF2) node [midway, right] {$Q$};
\path 
(10,0) node[circle, minimum size=40, fill=yellow!10, draw](DG1) {$USp(2)$}
(8,-3) node[circle, minimum size=40, fill=yellow!10, draw](DG2) {$USp(2)$}
(12,-3) node[circle, minimum size=40, fill=yellow!10, draw](DG3) {$USp(2)$}
(10,3) node[minimum size=30, fill=cyan!10, draw](DF2) {$6$};
\draw (DG2) -- (DG1) node [midway, right] {$c_1$};
\draw (DG3) -- (DG1) node [midway, left] {$c_2$};
\draw (DG2) -- (DG3) node [midway, below] {$c_3$};
\draw (DG1) -- (DF2) node [midway, right] {$q$};
\draw (DG2) -- (DF2) node [midway, left] {$p_1$};
\draw (DG3) -- (DF2) node [midway, right] {$p_2$};
\draw (DG2) to [out=-45, in=-135] node[below]{$B_3$}(DG3);
\draw (DG2.south west) .. controls ++(-1, -2) and ++(1, -2) .. node[below]{$\phi_c^1$} (DG2.south east);
\draw (DG3.south west) .. controls ++(-1, -2) and ++(1, -2) .. node[below]{$\phi_c^2$} (DG3.south east);
\draw (DF2.north east) .. controls ++(1, 2) and ++(-1, 2) .. node[below]{$M$} (DF2.north west);
\end{tikzpicture}
}
\caption{Circular quiver with gauge group $USp(2)\times USp(6)\times USp(2)$ and its dual circular quiver with gauge group $USp(2)\times USp(2)\times USp(2)$.
All chirals $\phi_c^1$, $\phi_c^2$ and $M$ are in antisymmetric rank-$2$
representations.} \label{fig:cUSpUSpUSp}
\end{figure}

These theories have the same field content as the linear quivers described in
section~\ref{sec_USpUSpUSp}, with the specific example in section~\ref{sec_USpUSpUSp_example}, with the addition of a bifundamental chiral for the two $USp(2)$ gauge nodes with an additional $U(1)_{a_4}$ axial symmetry with fugacity $a_4$. Theories A and C are summarized in the quiver diagram in Figure~\ref{fig:cUSpUSpUSp}.

The full-indices match for theory A and C. 
For $r_B=r_C=\frac25$, $r_Q=\frac27$ we have 
\begin{align}
&
1+15 a_2^2 q^{2/7}+\frac{q^{12/35}}{a_1^2 a_2^6 a_3^2}+q^{2/5} \left(a_1^2+a_3^2+a_4^2\right)+120 a_2^4 q^{4/7}+a_1 a_3
   a_4 q^{3/5}+\frac{15 q^{22/35}}{a_1^2 a_2^4 a_3^2}
   \nonumber\\
   &+q^{24/35} \left(\frac{1}{a_1^4 a_2^{12} a_3^4}+30 a_1^2 a_2^2+15
   a_2^2 \left(2 a_3^2+a_4^2\right)\right)+\frac{q^{26/35} \left(a_1^2+a_3^2+a_4^2\right)}{a_1^2 a_2^6 a_3^2}
   \nonumber\\
   &+q^{4/5}
   \left(a_1^2 \left(2 a_3^2+a_4^2\right)+a_1^4+a_3^2 a_4^2+a_3^4+a_4^4\right)+680 a_2^6 q^{6/7}+51 a_1 a_2^2
   a_3 a_4 q^{31/35}
   \nonumber\\
   &+\frac{120 q^{32/35}}{a_1^2 a_2^2 a_3^2}+\frac{a_4 q^{33/35}}{a_1 a_2^6 a_3}+q^{34/35}
   \left(\frac{15}{a_1^4 a_2^{10} a_3^4}+345 a_1^2 a_2^4+15 a_2^4 \left(23 a_3^2+8 a_4^2\right)\right)
   \nonumber\\
   &+q \left(a_1^3
   a_3 a_4+a_1 a_3 a_4 \left(a_3^2+a_4^2\right)-39\right)
   \nonumber\\
   &+\frac{q^{36/35} \left(\frac{15 a_2^{14} a_3^4 \left(2
   a_3^2+a_4^2\right)}{a_1^2}+\frac{1}{a_1^6}+30 a_2^{14} a_3^4\right)}{a_2^{18} a_3^6}
   \nonumber\\
   &+\frac{q^{38/35} \left(6 a_1^6
   a_2^{14} a_3^4 \left(16 a_3^2+5 a_4^2\right)+15 a_1^4 a_2^{14} a_3^4 \left(2 a_3^2 a_4^2+2
   a_3^4+a_4^4\right)+30 a_1^8 a_2^{14} a_3^4+a_1^2+a_3^2+a_4^2\right)}{a_1^4 a_2^{12} a_3^4}
   \nonumber\\
   &+\frac{q^{8/7}
   \left(a_1^6 a_3^4 a_4^2 \left(\left(3060 a_2^{14}+2\right) a_3^2+a_4^2\right)+a_1^8 a_3^4 a_4^2+a_1^4
   \left(a_3^8 a_4^2+a_3^6 a_4^4+a_3^4 a_4^6+1\right)+a_3^4\right)}{a_1^6 a_2^6 a_3^6 a_4^2}
   \nonumber\\
   &
   +660 a_1
   a_2^4 a_3 a_4 q^{41/35}
   \nonumber\\
   &+q^{6/5} \left(a_1^4 \left(2 a_3^2+a_4^2\right)+a_1^2 \left(3 a_3^2 a_4^2+2
   a_3^4+a_4^4\right)+\frac{679}{a_1^2 a_3^2}
   \right. 
   \nonumber\\
   &\left. 
   +a_1^6-\frac{a_1 \left(a_3^2+a_4^2\right)}{a_3 a_4}-\frac{a_3
   a_4}{a_1}+a_3^2 a_4^4+a_3^4 a_4^2+a_3^6+a_4^6\right)+\frac{51 a_4 q^{43/35}}{a_1 a_2^4 a_3}
   +\cdots
\end{align}

Compared to the non-circular linear quiver with the same gauge group for theories
$A$ and $C$ we see that there are obvious additional terms arising from the new
gauge invariant operators constructed using the new $USp(2) \times USp(2)$
bifundamental fields $B_3$. E.g.\ we have $\Tr(B_3B_3)$ contributing
$q^{2/5} a_4^2$ and $\Tr(B_1B_2B_3)$ contributing $q^{3/5} a_1a_3a_4$.
Also, due to the circular quiver structure, monopole operators with fluxes for
either $USp(2)$ gauge node have larger R-charge. For example the
$v_A^{0, \pm, \pm}$ and
$v_A^{\pm, \pm, 0}$ bare monopole operator contributions now appear at order
$q^{8/7}$ with fugacities $1/(a_1^2a_2^6a_3^6a_4^2)$ and
$1/(a_1^6a_2^6a_3^2a_4^2)$.

The half-indices will similarly match and can be understood in term of the
non-circular quiver with the additional terms involving $B_3$. We do not list
the half-indices here, but we do give an example in the next section with a circular quiver with orthogonal gauge groups.

\subsection{Circular $SO(2)\times SO(4)-[1]\times SO(2)$}

\begin{figure}
\centering
\scalebox{0.7}{
\begin{tikzpicture}
\path 
(0,0) node[circle, minimum size=40, fill=green!10, draw](AG1) {$SO(4)$}
(-2,-3) node[circle, minimum size=40, fill=green!10, draw](AG2) {$SO(2)$}
(2,-3) node[circle, minimum size=40, fill=green!10, draw](AG3) {$SO(2)$}
(0,3) node[minimum size=30, fill=cyan!10, draw](AF2) {$1$};
\draw (AG2) -- (AG1) node [midway, left] {$B_1$};
\draw (AG3) -- (AG1) node [midway, right] {$B_2$};
\draw (AG2) -- (AG3) node [midway, below] {$B_3$};
\draw (AG1) -- (AF2) node [midway, right] {$Q$};
\path 
(10,0) node[circle, minimum size=40, fill=green!10, draw](DG1) {$SO(3)$}
(8,-3) node[circle, minimum size=40, fill=green!10, draw](DG2) {$SO(2)$}
(12,-3) node[circle, minimum size=40, fill=green!10, draw](DG3) {$SO(2)$}
(10,3) node[minimum size=30, fill=cyan!10, draw](DF2) {$1$};
\draw (DG2) -- (DG1) node [midway, right] {$c_1$};
\draw (DG3) -- (DG1) node [midway, left] {$c_2$};
\draw (DG2) -- (DG3) node [midway, below] {$c_3$};
\draw (DG1) -- (DF2) node [midway, right] {$q$};
\draw (DG2) -- (DF2) node [midway, left] {$p_1$};
\draw (DG3) -- (DF2) node [midway, right] {$p_2$};
\draw (DG2) to [out=-45, in=-135] node[below]{$B_3$}(DG3);
\draw (DG2.south west) .. controls ++(-1, -2) and ++(1, -2) .. node[below]{$\phi_c^1$} (DG2.south east);
\draw (DG3.south west) .. controls ++(-1, -2) and ++(1, -2) .. node[below]{$\phi_c^2$} (DG3.south east);
\draw (DF2.north east) .. controls ++(1, 2) and ++(-1, 2) .. node[below]{$M$} (DF2.north west);
\end{tikzpicture}
}
\caption{Circular quiver with gauge group $SO(2)\times SO(4)\times SO(2)$ and its dual circular quiver with gauge group $SO(2)\times SO(3)\times SO(2)$.
All chirals $\phi_b$, $\phi_c^1$, $\phi_c^2$, $\phi_d$ and $M$ are in symmetric rank-$2$
representations.} \label{fig:cSOSOSO}
\end{figure}
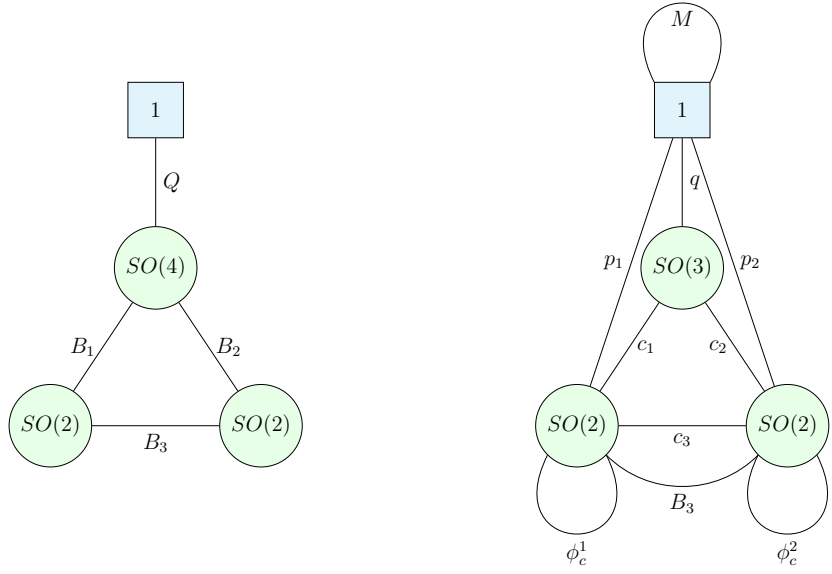

The theories are similar to the linear quivers in section~\ref{sec_SOSOSO}.
In this example we have the dual gauge theory whose gauge group is
$SO(2) \times SO(3) \times SO(2)$. 

The full indices are given by
\begin{align}
&I^A =I^B= I^C=I^D
\nonumber\\
&=
1+2 a_4^2 q^{1/4}+a_2^2 q^{2/7}+q^{2/5} \left(a_1^2+a_3^2\right)+3 a_4^4 q^{1/2}+4 a_1 a_3 a_4 q^{21/40}+2 a_2^2 a_4^2
   q^{15/28}
   \nonumber\\
   &+\frac{q^{39/70}}{a_1^2 a_2 a_3^2}+a_2^4 q^{4/7}+4 a_4^2 q^{13/20} \left(a_1^2+a_3^2\right)+2 a_2^2 q^{24/35}
   \left(a_1^2+a_3^2\right)+4 a_4^6 q^{3/4}
   \nonumber\\
   &+8 a_1 a_3 a_4^3 q^{31/40}+3 a_2^2 a_4^4 q^{11/14}+2 q^{4/5}
   \left(a_1^2+a_3^2\right)^2+\frac{2 a_4^2 q^{113/140}}{a_1^2 a_2 a_3^2}+8 a_1 a_2^2 a_3 a_4 q^{227/280}
   \nonumber\\
   &+2 a_2^4
   a_4^2 q^{23/28}+\frac{a_2 q^{59/70}}{a_1^2 a_3^2}+a_2^6 q^{6/7}+7 a_4^4 q^{9/10} \left(v_2(a_1^2+a_3^2\right)+8 a_1 a_3 a_4
   q^{37/40} \left(a_1^2+a_3^2\right)
   \nonumber\\
   &+8 a_2^2 a_4^2 q^{131/140} \left(a_1^2+a_3^2\right)+\frac{2 q^{67/70}
   \left(a_1^2+a_3^2\right)}{a_1^2 a_2 a_3^2}+2 a_2^4 q^{34/35} \left(a_1^2+a_3^2\right)+\cdots
\end{align}

The half-indices are given by
\begin{align}
&\II^A_{\mathcal{N},\mathcal{N},\mathcal{N}} 
= \II^C_{\mathcal{D},\mathcal{N},\mathcal{N}} 
= \II^C_{\mathcal{N},\mathcal{D},\mathcal{N}} 
= \II^D_{\mathcal{N},\mathcal{N},\mathcal{D}} 
\nonumber\\
&=
1+2 a_4^2 q^{1/4}+a_2^2 q^{2/7}+q^{2/5} \left(a_1^2+a_3^2\right)+3 a_4^4 q^{1/2}+4 a_1 a_3 a_4 q^{21/40}+2
   a_2^2 a_4^2 q^{15/28}
   \nonumber\\
   &
   +a_2^4 q^{4/7}-\frac{a_2 q^{9/14} \left(w^2+w+1\right)}{w}+4 a_4^2 q^{13/20} \left(a_1^2+a_3^2\right)+2
   a_2^2 q^{24/35} \left(a_1^2+a_3^2\right)+4 a_4^6 q^{3/4}
   \nonumber\\
   &+8 a_1 a_3 a_4^3 q^{31/40}+3 a_2^2 a_4^4 q^{11/14}+2
   q^{4/5} \left(a_1^2+a_3^2\right)^2
   +8 a_1 a_2^2 a_3 a_4 q^{227/280}+2 a_2^4 a_4^2 q^{23/28}
      \nonumber\\
   &
  -\frac{2 q^{59/70}
   a_1 a_2 \left(v_1^2 v_2 v_3+v_1 \left(v_2^2 v_3+v_2
   v_3^2+v_2+v_3\right)+v_2 v_3\right)}{v_1 v_2 v_3}
  \nonumber \\
 & 
   -\frac{2 q^{59/70}
   a_2 a_3 \left(u_1^2 u_2 u_3+u_1 \left(u_2^2 u_3+u_2
   u_3^2+u_2+u_3\right)+u_2 u_3\right)}{u_1 u_2 u_3}
   + a_2^6 q^{6/7}+\cdots
   \end{align}

The full index and the half-index can easily be compared to
the linear non-circular
example in section~\ref{sec_SOSOSO_example}. In the circular case here the full
index is the same with the addition of gauge invariant operators which can be
constructed using the bifundamental $B_3$ linking the first and third nodes in
the quiver. Note that there are some specific features arising from the
fact that in this example the first and third gauge groups are $SO(2)$. E.g.\
the $a_4^2 q^{1/4}$ term has coefficient $2$ since we can construct two
independent gauge-invariant operators from $B_3 B_3$. These are $\Tr(B_2 B_3)$
of course, but also $\epsilon_1 \epsilon_3 B_3 B_3$. The
$a_1 a_2 a_3 q^{21/40}$ term illustrates the circular nature of the quiver,
getting a contribution from $\Tr(B_1 B_2 B_3)$. The coefficient $4$ is due to
the fact we can replace the contraction of either or both $SO(2)$ index
contractions with $\epsilon_1$ and/or $\epsilon_3$. The story is similar for the
half-index, with the additional point that the Fermis $\Gamma_1$ and $\Gamma_3$
are different from the non-circular case. Here they are in bifundamental
representations of $SO(2) \times SO(6)$ rather than $SO(2) \times SO(4)$ as
the circular nature of the quiver modified the dual group for the end nodes.

\subsection{Summary}
It is now straightforward to generalize to linear quiver gauge theories with $n$ gauge nodes. 
Let $G_i$, $i=1,2,\cdots,n$ be a classical gauge group of the vector multiplet of $i$-th node which is coupled to 
$F_i$ fundamental chiral multiplets $Q_i$ and bifundamental chiral multiplets $B_{i-1,i}$ and $B_{i,i+1}$ 
which are also coupled to $(i-1)$th node and $(i+1)$th gauge nodes. 

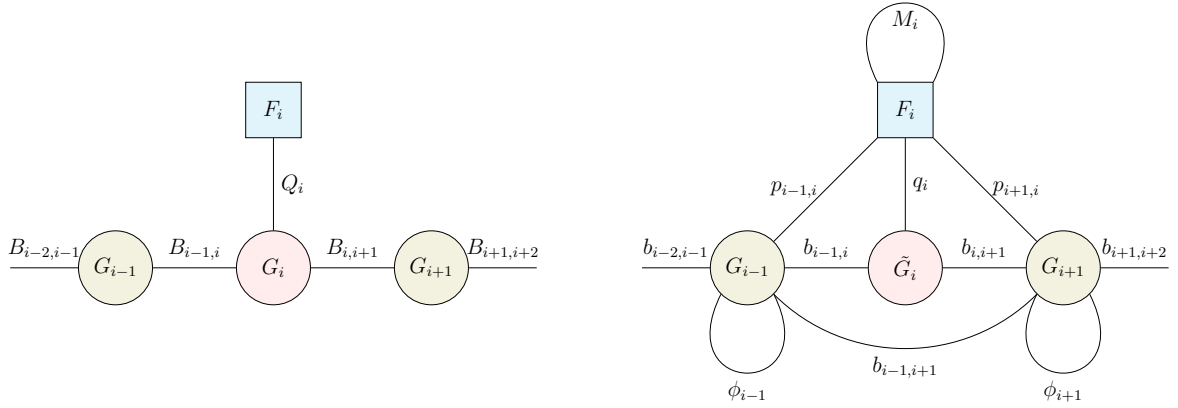
\begin{figure}
\centering
\scalebox{0.7}{
\begin{tikzpicture}
\path 
(0,0) node[circle, minimum size=40, fill=pink!30, draw](AG1) {$G_i$}
(-3,0) node[circle, minimum size=40, fill=olive!10, draw](AG2) {$G_{i-1}$}
(3,0) node[circle, minimum size=40, fill=olive!10, draw](AG3) {$G_{i+1}$}
(0,3) node[minimum size=30, fill=cyan!10, draw](AF2) {$F_i$};
\draw (AG1) -- (AG2) node [midway, above] {$B_{i-1,i}$};
\draw (AG1) -- (AG3) node [midway, above] {$B_{i,i+1}$};
\draw (AG1) -- (AF2) node [midway, right] {$Q_i$};
\draw (AG2) -- (-5,0) node [midway, above] {$B_{i-2,i-1}$}; 
\draw (AG3) -- (5,0) node [midway, above] {$B_{i+1,i+2}$}; 
\path 
(12,0) node[circle, minimum size=40, fill=pink!30, draw](BG1) {$\tilde{G_i}$}
(9,0) node[circle, minimum size=40, fill=olive!10, draw](BG2) {$G_{i-1}$}
(15,0) node[circle, minimum size=40, fill=olive!10, draw](BG3) {$G_{i+1}$}
(12,3) node[minimum size=30, fill=cyan!10, draw](BF2) {$F_i$};
\draw (BG1) -- (BG2) node [midway, above] {$b_{i-1,i}$};
\draw (BG1) -- (BG3) node [midway, above] {$b_{i,i+1}$};
\draw (BG1) -- (BF2) node [midway, right] {$q_i$};
\draw (BG2) -- (BF2) node [midway, left] {$p_{i-1,i}$};
\draw (BG3) -- (BF2) node [midway, right] {$p_{i+1,i}$};
\draw (BG2) -- (7,0) node [midway, above] {$b_{i-2,i-1}$}; 
\draw (BG3) -- (17,0) node [midway, above] {$b_{i+1,i+2}$}; 
\draw (BG2) to [out=-45, in=-135] node[below]{$b_{i-1,i+1}$}(BG3);
\draw (BG2.south west) .. controls ++(-1, -2) and ++(1, -2) .. node[below]{$\phi_{i-1}$} (BG2.south east);
\draw (BG3.south west) .. controls ++(-1, -2) and ++(1, -2) .. node[below]{$\phi_{i+1}$} (BG3.south east);
\draw (BF2.north west) .. controls ++(-1, 2) and ++(1, 2) .. node[below]{$M_i$} (BF2.north east);
\end{tikzpicture}
}
\caption{Dualization on the the $i$th node of gauge group $G_i$ in linear or circular quivers. } \label{fig:Glinear}
\end{figure}

One can find $n$ dual theories by taking the Seiberg-like dual of $n$ distinct gauge nodes. 
When the $i$th gauge node of gauge group $G_i$ is dualized, 
the resulting theory has similar fundamental chirals $q_i$ and bifundamental chirals $b_{i-1,i}$, $b_{i,i+1}$ with modified charges 
as well as additional bifundamental chirals $b_{i-1,i+1}$ between the adjacent gauge nodes $G_{i-1}$ and $G_{i+1}$, 
$F_i$ chirals $p_{i-1,i}$ transforming as the fundamental representation under the $G_{i-1}$, 
$F_i$ chirals $p_{i+1, i}$ transforming as the fundamental representation under the $G_{i}$, 
chirals $\phi_{i-1}$ transforming in a rank-$2$ representation under the $G_{i-1}$, 
chirals $\phi_{i+1}$ transforming in a rank-$2$ representation under the, $G_{i+1}$, 
and gauge singlets $M_i$ in a rank-$2$ representation of the flavor symmetry group, and $\sigma_i$ (see Figure \ref{fig:Glinear}). The
specific rank-$2$ representation depends on the group $G_i$ being dualized. In particular, it is
antisymmetric if $G_i$ is symplectic and symmetric if $G_i$ is orthogonal.

Note that while we can dualize any of the gauge nodes, once we have dualized one of them we cannot then immediately dualize either of the neighboring gauge nodes as they now have a different matter content, specifically a rank-$2$ tensor. It is known how to dualize such 3d theories but we do not pursue this in this article. We also leave for future work the question of how to dualize such cases with a boundary as this is not a straightforward problem.

\section{Star-shaped quivers}
\label{sec_star}

We now state our general conjecture for Seiberg-like duals of quivers with
fundamental and bifundamental matter which contains star-shaped quiver gauge theories. 
We describe this in terms of Seiberg-like duality of a
gauge node $G$ connected with bifundamentals to arbitrary gauge nodes $G_I$ and with $F$
fundamental flavors.
We then give one specific example involving a total of
four gauge nodes.

\subsection{$G-[F]-\prod_I G_I$}
\label{sec_genstar}

Consider a gauge node with gauge group $G$ which can be a unitary, orthogonal
or symplectic gauge group. We allow fundamental ($Q$) and bifundamental
($B_I$)~\footnote{We also have the conjugate representations for unitary gauge
groups.} matter with R-charges $r_Q$ and $r_{B_I}$ for
$G$, with the bifundamental chirals coupling to other gauge nodes with gauge
groups $G_I$. Each of these chirals is charged under a separate axial $U(1)$
group~\footnote{For the cases involving a unitary group the chirals are paired,
although for unitary $G$ we can also consider 
conjugate representations and the two members of each pair of chirals have
the same charge under the same axial $U(1)$. For unitary $G$ we can also
consider the case with different numbers of fundamental and anti-fundamental
flavors, but they still have the same axial charge.}. Again we assume each
$G_I$ is unitary, orthogonal or symplectic.
This gives a star-shaped quiver with centre $G$. However, this may be embedded
in a larger quiver diagram where we allow the gauge nodes $G_I$ to have
arbitrary matter content and couplings to any gauge nodes other than $G$,
including other $G_I$ nodes.

The Seiberg-like duality of $G$ then produces a dual gauge node $\tilde{G}$
determined in the usual way by the gauge group $G$ and the total number of
fundamental and bifundamental chirals for $G$. The node $\tilde{G}$ is at the
centre of the quiver with the same star-shape but now the fundamental ($q$)
and bifundamental ($b_I$) chirals have R-charges $1 - r_Q$ and $1 - r_{B_I}$
and the signs of the axial charges are reversed. In addition all the gauge
nodes $G_I$ now gain (in addition to any flavors they already had) the same
number of flavors ($p_I$) as $G$, a rank-$2$ representation chiral ($\phi_I)$
and a bifundamental chiral ($b_{IJ} \equiv b_{JI}$) for each pair $G_I, G_J$. There
is also a rank-2 flavor symmetry chiral $M$. If $G$ is orthogonal/symplectic,
$M$ and all $\phi_I$ are symmetric/anti-symmetric. Finally, there is also a
singlet chiral $\sigma$, or two singlets $\sigma^{\pm}$ if $G$ is unitary.

\begin{figure}
\centering
\scalebox{0.7}{
\begin{tikzpicture}
\path 
(-3,2) node[circle, minimum size=64,  fill=olive!10, draw](AG1) {$G_1$}
(0,0) node[circle, minimum size=64,  fill=pink!20, draw](AG2) {$G$}
(-3,-2) node[circle, minimum size=64,  fill=olive!10, draw](AG5) {$G_2$}
(0,-4) node[circle, minimum size=64,  fill=olive!10, draw](AG3) {$G_3$}
(3,-2) node[circle, minimum size=64, fill=olive!10, draw](AG6) {$G_4$}
(3,2) node[circle, minimum size=64, fill=olive!10, draw](AG4) {$G_k$}
(0,4) node[minimum size=50, fill=cyan!10, draw](AF2) {$F$};
\draw (AG1) -- (AG2) node [midway, above] {$B_{1}$};
\draw (AG3) -- (AG2) node [midway, right] {$B_{3}$};
\draw (AG5) -- (AG2) node [midway, right] {$B_{2}$};
\draw (AG6) -- (AG2) node [midway, right] {$B_{4}$};
\draw (AG4) -- (AG2) node [midway, above] {$B_{k}$};
\draw (AG2) -- (AF2) node [midway, right] {$Q$};
\draw[ultra thick, dashed,-]  (4,1) to[bend left] (4,-1);
\node[circle, minimum size=50, fill=olive!10, draw] (DG1) at (9,-3) {$G_1$};
\node[circle, minimum size=50, fill=olive!10, draw] (DG4) at (15,-3) {$G_2$};
\node[circle, minimum size=40,fill=olive!10, draw] (DG2) at (16,0) {$G_3$};
\node[circle, minimum size=30, fill=olive!10, draw] (DG5) at (12,1.5) {$G_4$};
\node[circle, minimum size=40, fill=olive!10, draw] (DG3) at (8,0) {$G_k$};
\node[circle, minimum size=50, fill=pink!20, draw] (CG) at (12,5) {$\tilde{G}$};
\node[minimum size=40, fill=cyan!10, draw] (F1) at (12,-7) {$F$};

\draw (CG) -- (DG1) node [midway, right] {$b_{1}$}; 
\draw (CG) -- (DG4) node [midway, left] {$b_{2}$}; 
\draw (CG) -- (DG2) node [midway, right] {$b_3$}; 
\draw[dashed] (CG) -- (DG5) node [midway, right] {$b_4$}; 
\draw (CG) -- (DG3) node [midway, left] {$b_k$}; 
\draw[dashed] (CG) -- (F1) node [midway, right] {$q$}; 
\draw (DG1) -- (F1) node [midway, left] {$p_1$}; 
\draw (DG1) -- (DG4) node [midway, below] {$b_1$}; 
\draw[dashed] (DG1) -- (DG2) node [midway, below] {$b_{1,2}$}; 
\draw[dashed] (DG1) -- (DG5) node [midway, below] {$b_{1,4}$}; 
\draw (DG1) -- (DG3) node [midway, left] {$b_{1,k}$}; 
\draw (DG4) -- (F1) node [midway, right] {$p_2$}; 
\draw (DG4) -- (DG2) node [midway, right] {$b_{2,3}$}; 
\draw[dashed] (DG4) -- (DG5) node [midway, left] {$b_{2,4}$}; 
\draw[dashed] (DG4) -- (DG3) node [midway, left] {$b_{2,k}$}; 
\draw[dashed] (DG2) -- (F1) node [midway, left] {$p_3$}; 
\draw[dashed] (DG2) -- (DG5) node [midway, right] {$b_{3,4}$}; 
\draw[dashed] (DG2) -- (DG3) node [midway, above] {$b_{3,k}$}; 
\draw[dashed] (DG5) -- (F1) node [midway, left] {$p_4$}; 
\draw[dashed] (DG5) -- (DG3) node [midway, left] {$b_{4,k}$}; 
\draw[dashed] (DG3) -- (F1) node [midway, right] {$p_k$}; 
\draw[ultra thick, dashed,-]  (11,1.5) to[bend right] (9,1);
\draw (F1.south west) .. controls ++(-1, -2) and ++(1, -2) .. node[below]{$M$} (F1.south east);
\draw (DG1.west) .. controls ++(-2,-1) and ++(-1,-2) .. node[left]{$\phi_1$} (DG1.south);
\draw (DG4.east) .. controls ++(2,-1) and ++(1,-2) .. node[right]{$\phi_2$} (DG4.south);
\draw (DG2.north) .. controls ++(1,0.5) and ++(2,0.5) .. node[right]{$\phi_3$} (DG2.east);
\draw (DG5.north east)[dashed] .. controls ++(1, 0.5) and ++(-1,0.5) .. node[above]{$\phi_4$} (DG5.north west);
\draw (DG3.north) .. controls ++(-1,0.5) and ++(-2,0.5) .. node[above]{$\phi_k$} (DG3.west);
\end{tikzpicture}
}
\caption{A star-shaped quiver with a single $G$ gauge node coupled to $k$ adjacent gauge nodes and its dual $k$-gonal bipyramid quiver. } \label{fig:starG}
\end{figure}
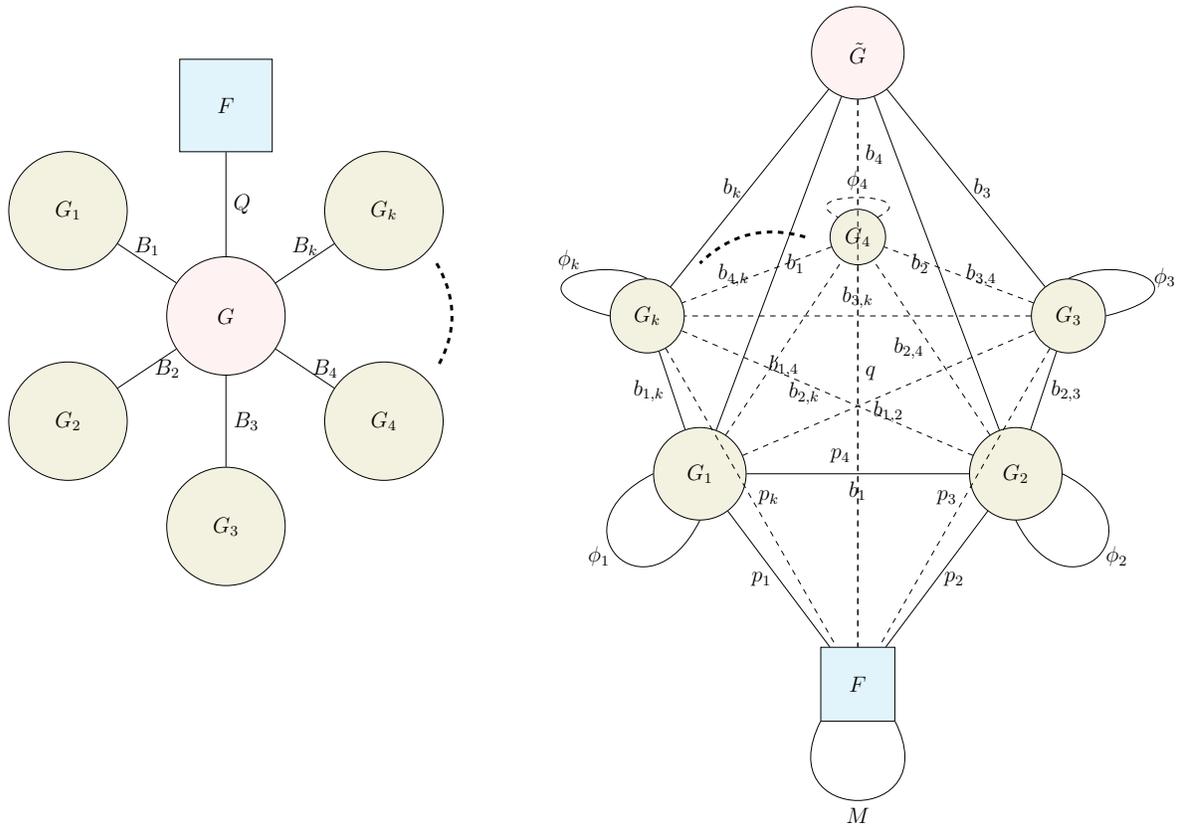

The general structure of this duality is to map the star-shaped quiver to a
bi-pyramidal shaped quiver as illustrated in Figure~\ref{fig:starG}. 

Clearly
the dual quiver now has additional closed triangles, and for each of them we
get a cubic contribution to the superpotential. Also each rank-$2$ chiral
together with the (bi)fundamental chirals for $G$ gives a cubic term in the
superpotential. The final ingredient in the superpotential is a quadratic
term coupling $\sigma$ to the monopole $v$
with minimal flux for
$\tilde{G}$ and no flux for any other gauge node -- for the unitary case
$\sigma^{\pm}$ couples to $v^{\pm}$ which are the minimal positive and
negative flux monopoles. The resulting superpotential is of the form
\begin{align}
\mathcal{W} = & \sigma v + \Tr(qMq)
 + \sum_I \left( \Tr(b_I \phi_I b_I) + q b_I p_I \right)
 + \sum_{I < J} \Tr(b_I b_{IJ} b_J)
\end{align}
with obvious modification to the case with unitary groups to include
appropriate additional or replacement terms with anti-fundamental operators and
replacing $\sigma v$ with $\sigma^{\pm} v^{\pm}$. This fully
determines the charges of the chirals $M$, $\phi_I$ and $b_{IJ}$ 
as follows
\begin{align}
\label{Gen_charges}
\begin{array}{c|c|c|c|c}
& U(1)_{a_I} & U(1)_{a_J} & U(1)_{a_F} & U(1)_R \\ \hline
B_I & 1 & 0 & 0 & r_{B_I} \\
Q & 0 & 0 & 1 & r_Q \\ \hline
b_I & -1 & 0 & 0 & 1 - r_{B_I} \\
b_{IJ} & 1 & 1 & 0 & r_{B_I} + r_{B_J} \\
\phi_I & 2 & 0 & 0 & 2r_{B_I} \\
p_I & 1 & 0 & 1 & r_{B_I} + r_Q \\
M & 0 & 0 & 2 & 2r_Q \\
q & 0 & 0 & -1 & 1 - r_Q \\
\end{array}
\end{align}
with the charges of $\sigma$ similarly determined by the charges of the monopole $v$.

In the case of unitary gauge groups we also need to map the topological
fugacities. We expect that when dualizing a gauge group $G = U(N)$ with
topological fugacity $z$ and a
non-zero number of flavors, the dual theory has topological fugacity $1/z$
for gauge node $\tilde{G}$. Also, for all the gauge nodes $G_I = U(N_I)$
connected to $G$ in the quiver having topological fugacities $z_I$, in the dual
theory they have topological fugacities $z z_I$. However, if $G$ has no
flavors, the mapping of topological fugacities is trivial.

In terms of matching indices, the conjectures here, allowing arbitrary
extensions of the basic star-shaped quiver, are equivalent to conjecturing that
the duality holds if we take the minimal star-shaped quiver (i.e.\ not
including any fields not defined above) and also remove the vector multiplets for every $G_I$ (i.e.\ consider these as global symmetries) but still keep the
dependence on the magnetic charges $m_i^{(I)}$ in the contributions from the
chirals coupled to $G_I$. In the case of a boundary, the conjecture for the
half-indices is similarly
equivalent to an identity without the vector multiplet and any 2d matter
contributions associated to the $G_I$ nodes. In this sense it is possible to
check the matching of half-indices without knowing the details of the matter
coupled to the $G_I$ nodes. However, while this would be the approach for any
attempt to analytically demonstrate the exact matching of indices and
half-indices, it is rather challenging for direct checking of the expansion to
a given order in $q$ due to the very large number of terms in the expansion.
Indeed, in the examples we have presented we have set the flavor fugacities to one for precisely this reason.

\subsection{Chern-Simons levels}
\label{sec_genstar_cs}
It is also possible to derive dualities with non-vanishing Chern-Simons levels
for some or all gauge nodes. We do this simply by considering the duality with
Chern-Simons levels zero but including additional fundamental chirals for
each gauge node. Turning on masses for these additional flavors and integrating
them out gives rise to a Chern-Simons level for each gauge node as in the case of a single gauge node \cite{Willett:2011gp, Giveon:2008zn}. For a unitary
or symplectic gauge node we shift the Chern-Simons level by $1/2$ for each
fundamental (or anti-fundamental) chiral integrated out, with the sign
correlated with taking the mass to $\pm \infty$. For orthogonal gauge groups the shift is by $1$ for each fundamental flavor. If we consider such a limit for some
chirals $Q$ to generate Chern-Simons level $k$ for gauge node $G$, then in the
dual theory we need to remove the same number of chirals $q$ which are flavors
for $\tilde{G}$. However, in the dual theory this also requires removing the
same number of flavors $p_I$ for each gauge node $G_I$ which is connected to
$\tilde{G}$ in the dual quiver (equivalently, to $G$ in the original quiver).
In the indices and half-indices this limit can be taken by send a combination of axial and flavor fugacities to either $0$ or $\infty$. As the chirals $q$
have opposite charges to $Q$ this means the dual gauge node $\tilde{G}$ will
have Chern-Simons level $-k$. However, the gauge nodes $G_I$ will get a shift
in the Chern-Simons level by $k$ (or $2k$ if $G_I$ is orthogonal and $G$ is
not, or $k/2$ if $G$ is orthogonal and $G_I$ is not) since $p_I$ have opposite
relevant charges to $q$ due to the superpotential term $\Tr(q b_I p_I)$. In
all cases the singlet(s) $\sigma$ ($\sigma^{\pm}$) are removed in this limit.

\subsection{Boundary conditions}
\label{sec_genstar_bc}
We further conjecture that with a boundary these theories are dual with the
boundary conditions being all Neumann in the original theory, while in the dual
all boundary conditions are Neumann except for the $\tilde{G}$ vector multiplet
and the chirals $b_I$, $q$ and $\sigma$ which all have Dirichlet boundary
conditions. For anomaly matching and to cancel
gauge anomalies we need to include for each gauge node in the original theory
a Fermi which is bifundamental under the gauge group and its Seiberg-like dual.
Note that to specify these bifundamental Fermis for gauge nodes $G_I$ requires
details of the matter content of these gauge nodes, i.e.\ knowledge of
additional parts of the quiver diagram. In the dual theory we need the same
Fermis for each $G_I$ but no Fermis for $\tilde{G}$. In the case of unitary
gauge groups we also need bideterminant Fermis for each unitary bifundamental Fermi, and
for all unitary $G_I$ (but not $\tilde{G}$) in the dual theory a determinant 2d chiral -- however, we remind the reader that we
have not been able to check examples of dualities with boundaries for unitary
groups due to the presence of these 2d chirals.

Alternatively we expect the duality with boundaries to hold if we swap all
Neumann and Dirichlet boundary conditions and then we need a bifundamental
Fermi only for $\tilde{G}$. Again, for unitary groups we will need additional
2d matter in determinant representations. 

We now present one example to illustrate the generalization to a star-shaped
quiver with the central node connected to three other gauge nodes.

\subsection{$USp(4)-[1] \times SO(2)^{\otimes 3}$}
\begin{figure}
\centering
\scalebox{0.7}{
\begin{tikzpicture}
\path 
(-4,0) node[circle, minimum size=64, fill=green!10, draw](AG1) {$SO(2)$}
(0,0) node[circle, minimum size=64, fill=yellow!20, draw](AG2) {$USp(4)$}
(0,-4) node[circle, minimum size=64,  fill=green!10, draw](AG3) {$SO(2)$}
(4,0) node[circle, minimum size=64,  fill=green!10, draw](AG4) {$SO(2)$}
(0,4) node[minimum size=50,  fill=cyan!10, draw](AF2) {$2$};
\draw (AG1) -- (AG2) node [midway, above] {$B_1$};
\draw (AG3) -- (AG2) node [midway, right] {$B_2$};
\draw (AG4) -- (AG2) node [midway, above] {$B_3$};
\draw (AG2) -- (AF2) node [midway, right] {$Q$};
\node[circle, minimum size=60, fill=green!10, draw] (DG1) at (9,-3) {$SO(2)$};
\node[circle, minimum size=25, fill=green!10, draw] (DG2) at (15,-1) {$SO(2)$};
\node[circle, minimum size=10, fill=green!10, draw] (DG3) at (9,0) {$SO(2)$};
\node[circle, minimum size=30,  fill=yellow!20, draw] (CG) at (12,5) {$USp(2)$};
\node[minimum size=40,  fill=cyan!10, draw] (F1) at (12,-7) {$2$};

\draw (CG) -- (DG1) node [midway, right] {$b_1$}; 
\draw (CG) -- (DG2) node [midway, right] {$b_2$};
\draw (CG) -- (DG3) node [midway, left] {$b_3$};
\draw[dashed] (CG) -- (F1) node [midway, right] {$q$}; 
\draw (DG1) -- (F1) node [midway, left] {$p_1$};
\draw (DG1) -- (DG2) node [midway, below] {$c_1$}; 
\draw (DG1) -- (DG3) node [midway, left] {$c_2$};
\draw (DG2) -- (F1) node [midway, right] {$p_2$};
\draw[dashed] (DG2) -- (DG3) node [midway, above] {$c_3$};
\draw[dashed] (DG3) -- (F1) node [midway, right] {$p_3$}; 
\draw (F1.south west) .. controls ++(-1, -2) and ++(1, -2) .. node[below]{$M$} (F1.south east);
\draw (DG1.west) .. controls ++(-2,-1) and ++(-1,-2) .. node[left]{$\phi_1$} (DG1.south);
\draw (DG2.north east) .. controls ++(2, 1) and ++(2,-1) .. node[right]{$\phi_2$} (DG2.south east);
\draw (DG3.north) .. controls ++(-1,0.5) and ++(-2,0.5) .. node[above]{$\phi_3$} (DG3.west);
\end{tikzpicture}
}
\caption{A star-shaped quiver with a single $USp(4)$ gauge node coupled to $SO(2)^{\otimes 3}$ gauge nodes and its dual triangular bipyramid quiver with a single $USp(2)$ gauge node coupled to $SO(2)^{\otimes 3}$. } \label{fig:starUSp(4)SO(2)3}
\end{figure}
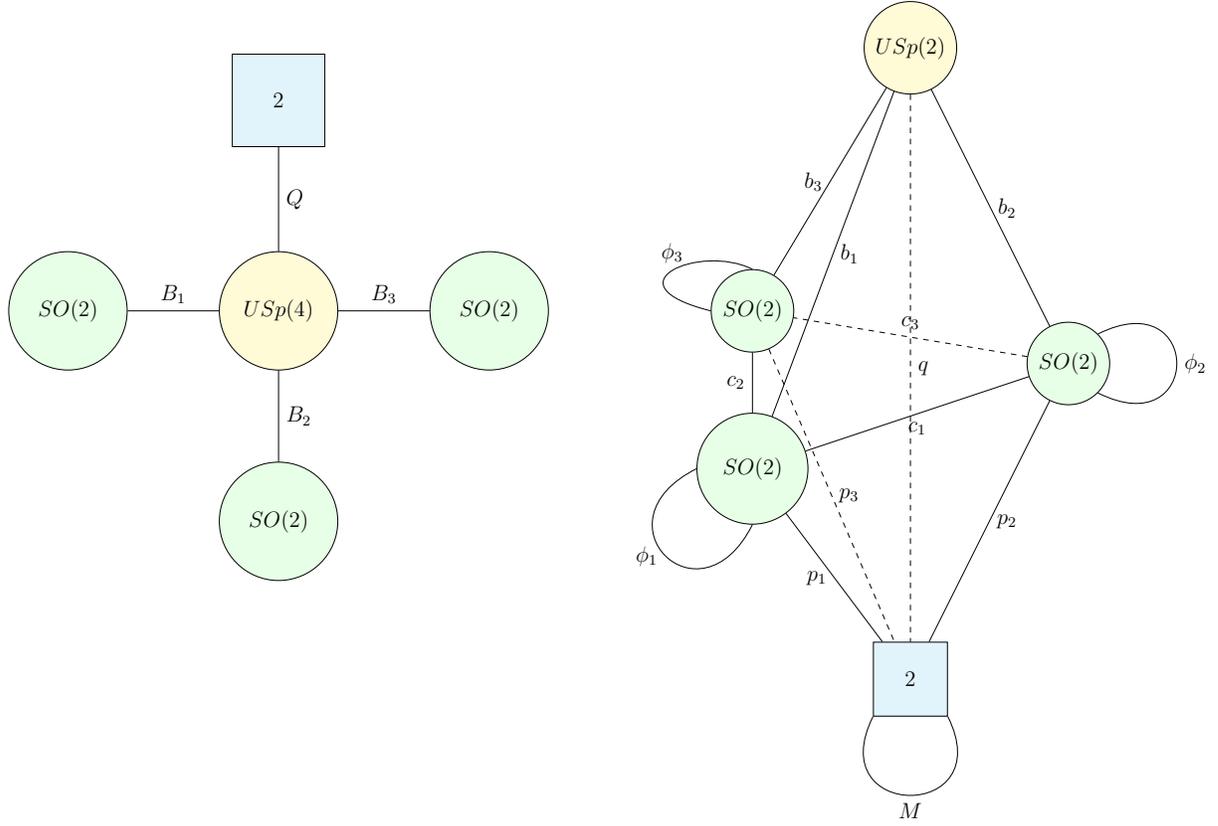

Let us consider a star-shaped quiver with gauge group $USp(4)\times SO(2)^{\otimes 3}$ and bifundamental chiral multiplets between $USp(4)$ and each of the $SO(2)$ gauge nodes while there are two $USp(4)$ flavors. Taking the Seiberg-like dual of the $USp(4)$ gauge node gives a theory with gauge group $USp(2)\times SO(2)^{\otimes 3}$, bifundamental chirals for each pair of gauge nodes and 2 flavors for each gauge node. 
The dual theory can be described as a triangular bipyramid quiver (see Figure \ref{fig:starUSp(4)SO(2)3}).

As we conjecture, 
the full-indices of the proposed dual theories beautifully agree with each other! 
In the following we show the indices for $r_Q = 1/14$ and all $r_{B_I} = 1/5$
\begingroup
\begin{align}
&I ^{\textrm{star $USp(4)-SO(2)^{\otimes 3}$}}= I^{\textrm{bipyramid $USp(2)-SO(2)^{\otimes 3}$}}
\nonumber\\
=&1+a_4^2 q^{1/7}+a_4^4 q^{2/7}+3 a^2 q^{2/5}+a_4^6 q^{3/7}+15 a^2 a_4^2 q^{19/35}
+a_4^8 q^{4/7}+\frac{q^{23/35}}{a^6 a_4^2}+15 a^2 a_4^4 q^{24/35}
\nonumber\\
&
+a_4^{10} q^{5/7}+\left(12 a^4+\frac{1}{a^6}\right) q^{4/5}+15 a^2 a_4^6 q^{29/35}+a_4^{12} q^{6/7}+\frac{\left(93 a^{10}+1\right)
   a_4^2 q^{33/35}}{a^6}+15 a^2 a_4^8 q^{34/35}
   \nonumber\\
   &
   +\left(a_4^{14}-10\right) q+\frac{3 q^{37/35}}{a^4 a_4^2}+\frac{\left(168 a^{10}+1\right) a_4^4 q^{38/35}}{a^6}
   +15 a^2 a_4^{10} q^{39/35}+a_4^2 \left(a_4^{14}-30\right) q^{8/7}
      \nonumber\\
&
   +\left(35 a^6+\frac{18}{a^4}\right) q^{6/5}+\frac{\left(168 a^{10}+1\right)
   a_4^6 q^{43/35}}{a^6}
   +q^{44/35} \left(15 a^2 a_4^{12}+\frac{6}{a^8 a_4^2}\right)
      \nonumber\\
&
   +a_4^4 \left(a_4^{14}-30\right) q^{9/7}+\frac{q^{46/35}}{a^{12} a_4^4}+\cdots
   \end{align}

\subsection*{Acknowledgements}
The authors would like to thank Stefano Cremonesi and Kimyeong Lee for useful discussions and comments. 
The research of T.O. was supported by KIAS Individual Grants (PG084301) at Korea Institute for Advanced Study. 
This work was supported in part by STFC Consolidated Grants ST/P000371/1 and ST/T000708/1.

\bibliographystyle{utphys}
\bibliography{ref}

\end{document}